\documentclass[a4paper,10pt]{article}

\pdfoutput=1

\usepackage{jheppub}

\usepackage{graphicx}    % needed for figures
\usepackage{dcolumn}    % needed for some tables
\usepackage{bm}             % for math
\usepackage{amssymb}   % for math
\usepackage{subfigure}
\usepackage{verbatim}
\usepackage{environ}

\usepackage[english]{babel}
\usepackage{xcolor}
\usepackage{graphicx}
\usepackage{amsmath}
\usepackage{dsfont}
\usepackage[makeroom]{cancel}
\usepackage{url}
\usepackage{hyperref}
\usepackage{slashed}

\newcommand{\secn}[1]{Section~\ref{#1}}
\newcommand{\appn}[1]{Appendix~\ref{#1}}
\newcommand{\beq}{\begin{eqnarray}}
\newcommand{\eeq}{\end{eqnarray}}

\newcommand{\as}{\alpha_{\mbox{\tiny{S}}}}
\newcommand{\eps}{\epsilon}

\newcommand{\Gam}{\Gamma}

\newcommand{\npo}{{n+1}}
\newcommand{\npt}{{n+2}}

\newcommand{\mc}{\mathcal}

\newcommand{\Li}{\mbox{Li}}
\newcommand{\dt}{\!\cdot\!}

\newcommand{\nnb}{\nonumber}

\newcommand{\kb}{\bar{k}}

\newcommand{\kt}{\tilde{k}}

\newcommand{\tk}{\tilde k}

\def\eq#1{Eq.~(\ref{#1})}
\newcommand\eqs[2]{Eqs.~(\ref{#1}-\ref{#2})}

\newcommand{\bS}[1]{{\bf S}_{#1}}
\newcommand{\bC}[1]{{\bf C}_{#1}}

\newcommand{\bbS}[1]{\overline{\bf S}_{#1}}
\newcommand{\bbC}[1]{\overline{\bf C}_{#1}}

\newcommand{\kkl}[1]{\{\bar k\}^{(#1)}}
\newcommand{\kk}[2]{\bar k_{#1}^{(#2)}}
\newcommand{\sk}[2]{\bar s_{#1}^{(#2)}}

\newcommand{\Norm}{\mc{N}_1}

\newcommand{\Bn}{B}

\newcommand{\euler}{\gamma_{\!_{_E}}}

\newcommand{\aA}{a}
\newcommand{\bB}{b}
\newcommand{\cC}{c}
\newcommand{\dD}{d}
\newcommand{\eE}{e}
\newcommand{\fF}{f}

\newcommand{\zg}{({\rm 0g})}
\newcommand{\og}{({\rm 1g})}
\newcommand{\tg}{({\rm 2g})}

\preprint{\\  \rightline{P3H-20-061} \\ 
                  \rightline{TTP20-036} \\ 
                  \rightline{CERN-TH-2020-178}}

\title{
Analytic integration of soft and collinear radiation \\ in factorised QCD cross 
sections at NNLO
}

\author[a,b]{Lorenzo Magnea,}
\author[c]{Giovanni Pelliccioli,}
\author[b,d]{Chiara Signorile-Signorile,}
\author[b]{Paolo Torrielli,}
\author[b]{and Sandro Uccirati}

\affiliation[a]{Theoretical Physics Department, CERN, CH-1211 Geneva 23, Switzerland}
\affiliation[b]{Dipartimento di Fisica and Arnold-Regge Center, Universit\`a di Torino,\\
                 and INFN, Sezione di Torino, Via P. Giuria 1, I-10125 Torino, Italy}
\affiliation[c]{Institut f\"ur Theoretische Physik und Astrophysik, Universit\"at W\"urzburg,\\
             Emil-Hilb-Weg 22, 97074 W\"urzburg, Germany}
\affiliation[d]{Institut f\"ur Theoretische Teilchenphysik,
                Karlsruher Institut f\"ur Technologie (KIT), \\
                76128 Karlsruhe, Germany\\}

\emailAdd{lorenzo.magnea@unito.it}
\emailAdd{giovanni.pelliccioli@physik.uni-wuerzburg.de}
\emailAdd{chiara.signorile-signorile@kit.edu}
\emailAdd{torriell@to.infn.it}
\emailAdd{uccirati@to.infn.it}

\abstract{
  Within the framework of local analytic sector subtraction,
  we present the full analytic integration of double-real and real-virtual 
  local infrared counterterms that enter NNLO QCD computations with any number 
  of massless final-state partons. We show that a careful choice of phase-space
  mappings leads to simple analytic results, including non-singular terms, that
  can be obtained with conventional integration techniques.
}

\begin{document}
\maketitle

%%%%%%%%%%%%%%%%%%%%%%%%%%%%%%%%%%%%%%%%%%%%%

\section{Introduction}

%%%%%%%%%%%%%%%%%%%%%%%%%%%%%%%%%%%%%%%%%%%%%

\noindent
Computing QCD cross-sections at next-to-next-to-leading order (NNLO) 
in the strong coupling is becoming mandatory to provide sufficiently precise 
fixed-order predictions for many processes of interest at high-energy colliders. 
This precision goal has led to the development of a host of new techniques 
in perturbative quantum field theory, ranging from the determination of parton 
distributions, to jet algorithms and of course to the calculation of high-order 
scattering amplitudes (for a recent review, see~\cite{Heinrich:2020ybq}). 

One of the problems that need to be efficiently tackled in order to perform 
multi-parton NNLO QCD calculations is the cancellation of infrared singularities.
Indeed, it is well known that, beyond leading order (LO) in QCD, both virtual 
corrections and real-radiation corrections contribute to any infrared-safe 
cross section: while these contributions are separately infrared (IR) singular, 
their sum (after UV renormalisation of virtual corrections) gives finite predictions 
for physical observables \cite{Kinoshita:1962ur,Lee:1964is}. This 
cancellation is well-understood in principle, but the increasing complexity of 
scattering amplitudes at high orders, and the intricate dependence of many
collider observables on experimental cuts and jet algorithms, lead to
significant difficulties in the practical implementation of the cancellation.

Subtraction algorithms form a class of proposed solutions to this problem.
The basic ingredient of subtraction is the construction of universal infrared 
counterterms, defined locally in the radiative phase spaces. Such counterterms
are required to mimic the behaviour of the radiative squared matrix element 
in all singular phase-space regions; on the other hand, they must be simple 
enough to be integrated over unresolved degrees of freedom in $d=4-2\eps$ 
dimensions, in order to analytically cancel the poles in $\eps$ arising from 
virtual corrections. Given such a set of counterterms, one proceeds by 
subtracting the counterterms from the radiative squared matrix element, so 
that the resulting expression can be numerically integrated without encountering 
singular contributions. One then adds to virtual corrections the integral of the 
counterterm over the radiative degrees of freedom, thus cancelling all infrared 
poles, and without having introduced any approximations in the distribution
of the chosen infrared-safe observable.

At next-to-leading order (NLO), subtraction is well understood and successfully 
applied to a vast ensemble of observable multi-parton distributions. The most used 
subtraction methods at NLO are the Frixione-Kunszt-Signer (FKS)~\cite{Frixione:1995ms} 
scheme and the Catani-Seymour (CS)~\cite{Catani:1996vz} algorithm.
One order higher in the perturbative expansion (at NNLO), the development 
of a fully general and efficient subtraction method has been the subject of
active research by many groups for several years. The literature is too vast 
to be comprehensively cited, but the main characteristics and some 
important applications of the most developed methods can be found 
in Refs.~\cite{Frixione:2004is,GehrmannDeRidder:2005cm,Somogyi:2005xz,
GehrmannDeRidder:2008ug,Czakon:2010td,Czakon:2013goa,Czakon:2014oma,
Czakon:2015owf,DelDuca:2015zqa,Ridder:2015dxa,DelDuca:2016csb,
DelDuca:2016ily,Caola:2017dug,Currie:2017eqf,Caola:2019pfz,Caola:2019nzf,
Chawdhry:2019bji,Czakon:2019tmo,Gehrmann-DeRidder:2019ibf}.
It must also be mentioned that subtraction is not the only possible approach to 
the problem: an alternative viewpoint is provided by slicing methods, where
an infrared cutoff is introduced to isolate the singular regions of the radiative 
phase space, and approximate expressions for the matrix elements are employed
below the cutoff scale. Such methods were successfully used already 
at NLO~\cite{Giele:1991vf,Giele:1993dj}, and have been applied at NNLO 
to a number of important processes~\cite{Catani:2007vq,Catani:2009sm,
Boughezal:2011jf,Bonciani:2015sha,Boughezal:2015dva,Boughezal:2015dra,
Gaunt:2015pea,Grazzini:2017mhc,Catani:2019hip,Kallweit:2020gcp}. Furthermore, new ideas 
have been recently proposed~\cite{Cacciari:2015jma,Herzog:2018ily,
Dreyer:2018rfu}, including theoretical developments concerning infrared 
factorisation~\cite{Magnea:2018ebr}, and the analysis of the infrared structure 
of Feynman diagrams~\cite{Anastasiou:2018rib,Ma:2019hjq,Anastasiou:2020sdt}, 
as well as purely numerical methods based on the cancellation of singular 
contributions at the integrand level, before loop and phase-space integrals 
are performed~\cite{Catani:2008xa,Bierenbaum:2010cy,Sborlini:2016hat,
Sborlini:2016gbr,Capatti:2019edf,Capatti:2020xjc}. Finally, the first 
developments for the extension of some of these tools to N$^3$LO have 
been presented~\cite{Dreyer:2016oyx,Cieri:2018oms,Currie:2018fgr,
Dreyer:2018qbw}. This vast activity bears witness to the fact that the 
problem of subtraction (or more generally the problem of local cancellation 
of IR divergences) at NNLO is very intricate: available NNLO schemes are 
often characterised by a remarkable degree of complexity if compared with 
the NLO ones, and do not always feature desirable aspects such as universality, 
analytic control, and full locality in phase space. We believe that there is
much room for further understanding, especially in view of future extensions 
to several (possibly massive) partons in the final state, and to higher perturbative 
orders.

In the present work, building on the results of Ref.~\cite{Magnea:2018hab},
we tackle the problem of analytic integration of local subtraction counterterms;
in the context of other NNLO subtraction schemes, this issue was addressed in
Refs.~\cite{Gehrmann-DeRidder:phdthesis,GehrmannDeRidder:1997wx,Gehrmann-DeRidder:2003pne,
  Anastasiou:2003gr,Aglietti:2008fe,DelDuca:2013kw,Caola:2018pxp,Delto:2019asp}.
To be more precise, we note that the definition of a set of infrared counterterms
has two main ingredients. On the one hand, these local functions in the radiative 
phase space must, in all unresolved limits, reproduce the factorised soft and
collinear kernels which emerge in QCD at leading power in the soft-parton energies 
and in the collinear-parton transverse momenta. On the other hand, phase space
itself must be factorised and parametrised so that the integration over the 
radiative degrees of freedom can be completely decoupled from the integration
over the Born configurations: only when this step has been successfully performed
can one claim the universality of the resulting subtraction algorithm. The
necessary mappings of phase space have been  extensively discussed in 
Ref.~\cite{DelDuca:2019ctm}: many choices are possible, and this choice is a 
crucial ingredient of any subtraction procedure.

Let us consider more carefully the interplay between the choice of infrared
counterterms and the choice of phase-space mapping. Any QCD (squared)
amplitude with the emission of one or more unresolved partons can be written 
as a product (to be understood as a matrix product in the colour and helicity 
spaces) of the (squared) amplitude for the process without the emissions, 
times a soft or a collinear kernel, containing all dependence on the momenta 
of the unresolved radiated particle(s). Any definition of subtraction counterterms must 
have the same factorised structure, and the kernels defined by the counterterms 
must reproduce the kernels of the QCD factorisation formulae, in all singular 
regions. Quite naturally, therefore, the first na\"ive choice is to use in the
counterterms the kernels of the QCD factorisation formulae 
themselves. This is, for example, the case for FKS subtraction~\cite{Frixione:1995ms} 
at NLO, and for the Colourful subtraction scheme~\cite{Somogyi:2005xz} at 
NNLO. Other well-known choices are the 
dipoles in the CS subtraction scheme~\cite{Catani:1996vz} at NLO, and the 
antennas in the Antenna subtraction method~\cite{GehrmannDeRidder:2005cm} 
at NNLO. The CS and antenna kernels have expressions that are more 
involved than the ones of the QCD factorisation formulae, but still reduce 
to the QCD soft or collinear kernels in all singular limits. 
When it comes to the choice of phase-space mappings, the FKS and Colourful 
methods essentially involve the momenta of all outgoing particles of the 
radiative process, producing rather complicated expressions in the phase-space of the 
radiated particles, which then need to be integrated in $d$-dimensions. 
As a consequence, in the Colourful 
approach, these expressions in some cases can be integrated just numerically 
(this is not the case in FKS, because of the simplicity of NLO kernels). 
An easier solution for the phase-space mappings is the one adopted in the 
CS and antenna subtractions, where the only momenta involved in the 
parametrisation are the ones contained in the kernels. 
This choice overcomes the complexity of the latter (which in these 
subtraction procedures are more complicated than the QCD soft and 
collinear kernels), and allows for their analytical integration in the
radiative phase space. 

In what  follows, we pursue a different approach, recently proposed in 
Ref.~\cite{Magnea:2018hab}, which combines a definition of the counterterm 
kernels as close as possible to the QCD soft and collinear kernels (as is the case
for the FKS and Colourful methods), together with phase-space mappings 
involving only the particles present in the particular kernel being integrated 
(as is the case for the CS and Antenna subtraction methods). 
As was shown already in preliminary 
tests performed in Ref.~\cite{Magnea:2018hab}, this approach leads to simpler 
integrals, that can readily be computed analytically with conventional methods.
The goal of this paper is thus to present the analytic integration in $d$-dimensions 
of the soft and collinear kernels of QCD factorisation formulae at NLO and 
NNLO, once a specific choice of phase-space mappings, along the lines of 
Ref.~\cite{Magnea:2018hab}, is adopted. We emphasise that the results we 
present have a universal aspect: the full integration of NNLO QCD kernels 
with an exact factorisation of the radiation phase space, such that the
on-shellness of the underlying Born configuration is ensured, and momentum
conservation  is properly enforced.
On the other hand, these integrals are essential building blocks for 
the subtraction procedure of Ref.~\cite{Magnea:2018hab}: indeed, all required
integrals for a complete subtraction algorithm for massless final-state partons
are either contained in the results presented here, or are significantly simpler
than the ones we perform.

The structure of the paper is as follows: in \secn{treeonereal}, for clarity and 
completeness, we present the exact integration of NLO soft and collinear kernels,
which was discussed already in Ref.~\cite{Magnea:2018hab}. In \secn{treetworeal},
we turn to the integration of tree-level kernels for double-unresolved radiation,
considering explicitly double-soft emission and  the case of three partons becoming
collinear (which we describe  as `double-collinear' limit). In both cases, we consider
un-ordered emissions, where both partons involved become unresolved at the same 
rate. We emphasise that this is the most intricate configuration in view of integration:
hierarchical limits, with one of the two partons becoming unresolved at a higher rate 
than the other one, lead to a subset of the integrals considered here; similarly,
nested soft-collinear limits lead to simpler integrals. In \secn{olokore}, we tackle 
the problem of real-virtual corrections, and integrate the QCD kernels for single
real radiation at one-loop. In the process, we display the non-trivial cancellation
of all singularities proportional to colour tripoles, which is an essential consistency
check, given the absence of such singularities in double-virtual and double-real
contributions. Finally, in \secn{conclu}, we summarise our results and present 
perspectives for future work. A number of technical details, including a thorough
discussion of the phase-space mappings that we employ, and the treatment
of integrals with non-trivial azimuthal dependence, are discussed in the
Appendices.

%%%%%%%%%%%%%%%%%%%%%%%%%%%%%%%%%%%%%%%%%%%%%

\section{Tree-level infrared kernels with one real emission}
\label{treeonereal}

%%%%%%%%%%%%%%%%%%%%%%%%%%%%%%%%%%%%%%%%%%%%%

\noindent
In this section we recall methods and results for the integration of the tree-level 
factorisation kernels with a single unresolved real emission, as performed in 
Ref.~\cite{Magnea:2018hab}, and we introduce notations that we will use in the 
rest of the paper. We consider a generic process with a colour-singlet initial state,
producing $n$ massless coloured particles in the final state at lowest order.
We will therefore be interested in scattering amplitudes involving up to $n, \npo$
and $\npt$ final-state coloured particles at LO, NLO and NNLO, respectively. We 
will denote the sets of momenta of coloured particles by $\{k\}$, where the number 
of particles involved will be clear from the context. Furthermore, we will adopt the 
notation $\{ k \}_{\slashed i}$ for the set obtained from $\{k\}$ by omitting the 
$i$-th particle, and $\{k\}_{[ij]}$ for the set obtained from $\{k\}$ by removing 
particles $i$ and $j$, and introducing in their stead a single particle with momentum 
$k_i + k_j$. We note from the outset that, if the set $\{k\}$ involves $\npo$ on-shell
momenta $k_i$ satisfying $k_i^2 = 0$ and $\sum_i k_i^\mu  = q^\mu$, then
the set $\{ k \}_{\slashed i}$ does not satisfy the same momentum sum
outside the strict soft limit $k_i = 0$, while in the set $\{k\}_{[ij]}$ the
momentum $k_i + k_j$ is off-shell outside the strict collinear limit $k_i^\mu 
\propto k_j^\mu$. A crucial concern in what follows will be, therefore, to
choose a parametrisation of the radiative phase space factorising a lowest-order
parton configuration with $n$ on-shell partons and enforcing momentum 
conservation.

We expand perturbatively the amplitude for the emission of $n$ partons as
\beq
  {\cal A}_n  \, = \, {\cal A}_n^{(0)}  \, + \, 
  {\cal A}_n^{(1)}  \, + \, {\cal A}_n^{(2)}  \, + \, \ldots \, ,
\label{pertexpA}
\eeq
and we will use the notation $B\left(\{k\}\right)$ for the Born-level squared 
matrix element, $B(\{k\}) = |{\cal A}_n^{(0)}|^2$. At NLO, we will also need 
the {\it colour-connected} Born squared matrix elements, $B_{lm} = {\cal A}_n^{(0)*} 
({\bf T}_l \cdot {\bf T}_m) {\cal A}_n^{(0)}$, where we use the standard 
notation~\cite{Bassetto:1984ik,Catani:1996vz} for the colour-insertion 
operators ${\bf T}_i$, responsible for the radiation of a gluon from Born-level 
parton $i$, and the {\it spin-connected} Born squared matrix elements, 
$B_{\mu\nu}$, obtained by stripping the spin polarisation vector of a selected 
parton from the Born amplitude and from its complex conjugate. In this 
language, the virtual correction at NLO is given by $V(\{k\}) = 2 \, {\rm Re}
({\cal A}_n^{(0) *} {\cal A}_n^{(1)})$, and the real radiation contribution is
$R(\{k\}) = |{\cal A}_\npo^{(0)}|^2$.

With these definitions, we can write the well-known factorised expressions for 
$R(\{k\})$ in the limits where one particle becomes unresolved, as follows. Defining
the Mandelstam invariants of the process as $s_{ab} = (k_a + k_b)^2 = 2 k_a \cdot k_b$, 
we can introduce a soft-limit operator $\bS{i}$, extracting the leading 
power of $R(\{k\})$ as $s_{im} \to 0$, uniformly for all $m \neq i$, {\it i.e.} 
taking all ratios of the form $s_{il}/s_{im}$ to be of order one; similarly, the 
collinear-limit operator $\bC{ij}$ extracts the leading power of $R(\{k\})$ as 
$s_{ij} \to 0$, with all ratios $s_{im}/s_{jm}$, for $m \neq i,j$, taken to be 
independent of $m$ in the limit. Under these limits, $R(\{k\})$ factorises as 
\beq
\bS{i} \, R 
& = & 
- \, \Norm
\sum_{\substack{l \neq i\\ m \neq i}} \, \mc I_{lm}^{(i)} \, 
B_{lm}(\{ k \}_{\slashed i}) 
\, ,
\quad\;
\bC{ij} \, R 
= 
\frac{\Norm}{s_{ij}} \, \Big[
P_{ij}\,B(\{k\}_{[ij]}) 
+  
Q_{ij}^{\mu \nu}\,B_{\mu \nu}(\{k\}_{[ij]}) 
\Big] 
\, ,
\label{SandC}
\eeq
where the normalisation factor $\Norm$ is given by
\beq
\Norm 
=
8 \pi \as \! \left(\! \frac{\mu^2 e^{\euler}}{4 \pi} \!\right)^{\!\!\eps} 
\! ,
\qquad
\label{calN1}
\eeq
with $\mu$ the renormalisation scale and $\euler$ the Euler-Mascheroni constant. 

In order to express the infrared kernels in a compact and flavour-symmetric way,
we introduce flavour Kronecker delta functions: if $f_i$ is the flavour of parton $i$,
we define for example $\delta_{f_i g}$ as $\delta_{f_i g} = 1$ if parton $i$ is a gluon, 
and $\delta_{f_i g} = 0$ otherwise; in similar vein, we define $\delta_{f \{q,\bar q\}} \equiv
\delta_{fq} + \delta_{f \bar q}$,  and $\delta_{\{f_i f_j\} \{q \bar q\}} \equiv \delta_{f_i q} 
\delta_{f_j \bar q} + \delta_{f_i \bar q} \delta_{f_i q}$. The soft limit is then 
expressed in terms of the eikonal kernel $\mc I_{lm}^{(i)}$, which is given by
\beq
\mc I_{lm}^{(i)} & = & 
\delta_{f_i g} \, \frac{s_{lm}}{s_{il}\,s_{im}} 
\, .
\label{eq:eikonal}
\eeq
In order to characterise precisely the collinear limit for partons $i$ and $j$, on the
other hand, we select a massless reference vector $k_r$, which is conveniently chosen 
among the momenta $\{k\}$ of the outgoing particles; we then introduce ratios
of Mandelstam invariants, that can be interpreted as longitudinal momentum 
fractions along the collinear  direction, as
\beq
x_i \, = \, \frac{s_{ir}}{s_{ir}+s_{jr}}
\, ,
\qquad
x_j \, = \, \frac{s_{jr}}{s_{ir}+s_{jr}}
\, ,
\qquad
x_i + x_j \, = \, 1 \, ,
\label{xixj}
\eeq 
and a transverse-momentum vector
\beq
\label{ktdef1}
\kt^\mu_{ij} \, = \,  
x_j\,k_i^\mu - x_i\,k_j^\mu - ( x_j - x_i )\,\frac{s_{ij}}{s_{ir}+s_{jr}}\,k_r^\mu \, = \,
- \kt^\mu_{ji} \, .
\eeq 
We can now write the Altarelli-Parisi kernels $P_{ij}$, for collinear emissions in a 
generic flavour configuration, in the form
\beq
P_{ij} 
& = & 
P_{ij}^{\zg} \,\delta_{ \{f_i f_j\} \{q \bar q\} }
+ 
P_{ij}^{\og}\,\delta_{f_i g} \delta_{f_j \{q, \bar q\}} 
+  
P_{ji}^{\og}\,\delta_{f_j g} \delta_{f_i \{q, \bar q\}}
+
P_{ij}^{\tg}\,\delta_{f_i g} \delta_{f_j g} 
\, ,
\eeq
where $P^{(k {\rm g})}_{ij}$ represents the flavour contribution with $k$ radiated 
collinear gluons ($k = 0, 1, 2$), and can be written explicitly as
\beq
\hspace{-3mm}
P_{ij}^{\zg}
=
T_R \bigg( 1 - \frac{2 x_i x_j}{1 - \eps} \bigg) 
\, ,
\quad
P_{ij}^{\og}
=
C_F \bigg[ 2\,\frac{x_j}{x_i} + (1-\eps)x_i \bigg]
\, ,
\quad
P_{ij}^{\tg}
= 
2 \, C_A \bigg( \!\frac{x_i}{x_j} + \frac{x_j}{x_i} + x_i x_j \!\bigg) 
\, .
\eeq
The azimuthal tensor kernel $Q_{ij}^{\mu \nu}$, on the other hand, is 
\beq
Q_{ij}^{\mu\nu} 
& = & 
\left(
- \, 
\delta_{f_ig} \, \delta_{f_jg} \, 2 \, C_A \,  x_i x_j 
+ 
\delta_{\{f_i f_j\}\{q \bar q\}} \, T_R \, \frac{2 x_i x_j}{1 - \eps} 
\right)
\left[ - \, g^{\mu\nu} + (d - 2) \, \frac{\kt^\mu_{ij} \kt^\nu_{ij}}{\kt^2_{ij}}   \right] 
\, .
\eeq
The task is now to introduce a parametrisation of the ($\npo$)-particle phase space
in terms of $n$ on-shell massless momenta, carrying the same total momentum as
the original set of $\npo$ partons, and factorising the integration over the degrees 
of freedom of the unresolved parton. A broad set of solutions to this problem, inspired
from \cite{Catani:1996vz}, is described
in~\appn{app:map1}, and we apply it below, with the goal of simplifying 
as much as possible the subsequent integration.

%%%%%%%%%%%%%%%%%%%%%%%%%

\subsection{Phase-space mappings and integration for the soft kernel}
\label{softolo}

%%%%%%%%%%%%%%%%%%%%%%%%%

\noindent
For the eikonal kernel $\mc I_{lm}^{(i)}$, we perform the mapping described in~\appn{app:map1}, choosing the momenta $\{k_{\aA},k_{\bB},k_{\cC}\}$ 
differently for each term in the sum in \eq{SandC}, as
\beq
k_{\aA} \to k_i \, ,
\qquad\qquad
k_{\bB} \to k_l \, ,
\qquad\qquad
k_{\cC} \to k_m \, .
\eeq
Promoting the set $\{k\}_{\slashed i}$ (which preserves momentum conservation 
just in the soft limit) to the momentum-conserving set $\kkl{ilm}$ of 
\appn{app:map1}, we define the mapped soft limit of $R(\{k\})$ as
\beq
\bbS{i} \, R 
\; \equiv \; 
- \, \Norm \,
\sum_{\substack{l \neq i\\ m \neq i}} \, \mc I_{lm}^{(i)} \, 
B_{lm} \! \left( \kkl{ilm} \right) 
\, ,
\label{eq:bSi R} 
\eeq
which manifestly satisfies the condition ${\bf S}_i \, \bbS{i} \, R = {\bf S}_i \, R$, 
necessary to ensure a local cancellation. \eq{eq:bSi R} can be exactly integrated in 
$d = 4 - 2 \epsilon$ dimensions over the radiative phase space. One writes
\beq
&&
\int d\Phi_\npo \,
\bbS{i} \, R 
\; = \;
- \, 
\sum_{\substack{l \neq i\\ m \neq i}} \, 
\int d \Phi_n \big(\kkl{ilm}\big) \, 
J_{\rm s}^{ilm} \,
B_{lm} \! \left( \kkl{ilm} \right)
\, ,
\label{eq:int bSi R} 
\eeq
where the soft integral
\beq
\label{genJs}
&&
J_{\rm s}^{ilm}
\; \equiv \;
\Norm
\int d\Phi_{\rm rad}^{(ilm)}\,
\mc I_{lm}^{(i)}
=
\delta_{f_i g} \,
\Norm
\int d\Phi_{\rm rad}^{(ilm)} \, \frac{s_{lm}}{s_{il}\,s_{im}} 
\equiv
\delta_{f_i g} \,J_{\rm s} \left(\sk{lm}{ilm}\right) 
\eeq
depends on the kinematics of particles $i$, $l$, $m$ only through the  
\emph{radiative soft function} $J_{\rm s}$, with argument $\sk{lm}{ilm}$.
Substituting the expression for the Mandelstam invariants given in \eq{eq:sij NLO}, 
$J_{\rm s}$ can be trivially calculated to all orders in $\epsilon$, with the result
\beq
J_{\rm s}(s) 
& = & 
\Norm
N (\eps) \, s^{-\eps} \!\!
\int_0^\pi\!\!\!d\phi \, \sin^{- 2 \eps}\!\phi \!
\int_0^1 \!\! dy \int_0^1 \!\! dz
\Big[ y (1 - y)^2 z (1 - z) \Big]^{- \eps} \! (1 - y)
\frac{1-z}{y\,z} 
\nnb \\
& = &
\frac{\as}{2\pi}
\left( \frac{s}{e^{\euler\!}\mu^2} \right)^{\!\!-\eps} 
\frac{\Gam(1 - \eps) \Gam(2 - \eps)}{\eps^2 \, \Gam(2 - 3 \eps)}
\nnb \\
& = &
\frac{\as}{2\pi}
\left( \frac{s}{\mu^2} \right)^{\!\!-\eps} 
\left[
\frac1{\eps^2}
+
\frac2{\eps}
+
6
-
\frac{7}{12}\,\pi^2
+
\mc O(\eps)
\right]
\, .
\label{eq:Js}
\eeq

%%%%%%%%%%%%%%%%%%%%%%%%%

\subsection{Phase-space mappings and integration for the collinear kernels}
\label{collolo}

%%%%%%%%%%%%%%%%%%%%%%%%%

\noindent
For the collinear kernels, we choose the momenta $\{k_{\aA},k_{\bB},k_{\cC}\}$ 
of the mapping of \appn{app:map1} in the most natural way as 
\beq
k_{\aA} \to k_i \, ,
\qquad\qquad
k_{\bB} \to k_j \, ,
\qquad\qquad
k_{\cC} \to k_r \, .
\eeq
We promote the set $\{k\}_{[ij]}$ (where the momentum $k_i+k_j$ is on-shell 
only in the collinear limit) to the set of on-shell momenta $\kkl{ijr}$ of 
\appn{app:map1}, and we define the mapped collinear limit of $R(\{k\})$ as
\beq
\bbC{ij} \, R 
\; \equiv \;
\frac{\Norm}{s_{ij}} \, \Big[
P_{ij}\,B\left( \kkl{ijr} \right)
+  
Q_{ij}^{\mu \nu}\,B_{\mu \nu}\left( \kkl{ijr} \right)
\Big] 
\, ,
\label{eq:bCij R} 
\eeq
which can easily be shown to satisfy the locality condition 
${\bf C}_{ij}\,\bbC{ij}\,R={\bf C}_{ij}\,R$. 
Proceeding with the integration, we first notice that the azimuthal kernel 
$Q_{ij}^{\mu\nu}$ integrates to zero \cite{Catani:1996vz}, because of its tensor
structure, taking into account that $\kt_{ij}\cdot \kk{j}{ijr} = 0$.
The remaining terms, involving the $P_{ij}$ kernels, can again be
integrated exactly in the radiation phase space. We write
\beq
&&
\int d\Phi_\npo \,
\bbC{ij} \, R 
\; = \;
\int d \Phi_n \big(\kkl{ijr}\big) \,
J_{\rm c}^{ijr} \,
B \! \left( \kkl{ijr} \right)
\, ,
\label{eq:int bCij R} 
\eeq
where the collinear integral
\beq
\label{Jcdef}
J_{\rm c}^{ijr}
& \; \equiv \; &
\Norm 
\int d\Phi_{\rm rad}^{(ijr)} \,
\frac{P_{ij}}{s_{ij}} 
\nnb\\
& \; \equiv \; &
\delta_{ \{f_i f_j\} \{q \bar q\} } \, J_{\rm c}^{\zg} \left(\sk{jr}{ijr}\right)
+ 
\Big( 
\delta_{f_i g} \delta_{f_j \{q, \bar q\}} 
+ 
\delta_{f_j g} \delta_{f_i \{q, \bar q\}} 
\Big)
J_{\rm c}^{\og}\left(\sk{jr}{ijr}\right)
\nnb\\
&&
+ \,
\delta_{f_i g} \delta_{f_j g} \, J_{\rm c}^{\tg} \left(\sk{jr}{ijr}\right)
\eeq
depends on the kinematics of particles $i$, $j$, $r$ only through the  
\emph{radiative collinear functions} $J_{\rm c}^{\zg}$, $J_{\rm c}^{\og}$, 
$J_{\rm c}^{\tg}$ with argument $\sk{jr}{ijr}$.  Using again the expression 
for the Mandelstam invariants given in \eq{eq:sij NLO} one finds the following
results. The radiation of a collinear quark-antiquark pair gives 
\beq
\label{eq:Jc}
J_{\rm c}^{\zg}(s)
& = & 
\Norm \,
N (\eps) \, s^{-\eps} \!\!
\int_0^\pi\!\!\!d\phi \, \sin^{- 2 \eps}\!\phi \!
\int_0^1 \!\! dy \int_0^1 \!\! dz
\Big[ y (1 - y)^2 z (1 - z) \Big]^{- \eps} \, \frac{1-y}{y} \,
T_R \bigg( 1 - \frac{2z(1-z)}{1 - \eps} \bigg)
\nnb \\
& = &
\frac{\as}{2\pi}
\left( \frac{s}{e^{\euler\!}\mu^2} \right)^{\!\!-\eps} 
\frac{\Gam(1-\eps)\Gam(2-\eps)}{\eps\,\Gam(2 - 3\eps)} \,
T_R \,
\frac{-2}{3-2\eps}
\nnb \\
& = &
\frac{\as}{2\pi}
\left( \frac{s}{\mu^2} \right)^{\!\!-\eps} 
T_R 
\left[
-
\frac23\,\frac1{\eps}
-
\frac{16}{9}
+
\mc O (\eps)
\right]
\, ;
\eeq
the radiation of a collinear gluon from a quark or an antiquark, on the other hand, 
yields 
\beq
J_{\rm c}^{\og}(s)
& = & 
\Norm \,
N (\eps) \, s^{-\eps} \!\!
\int_0^\pi\!\!\!d\phi \, \sin^{- 2 \eps}\!\phi \!
\int_0^1 \!\! dy \! \int_0^1 \!\! dz
\Big[ y (1 - y)^2 z (1 - z) \Big]^{- \eps} \, \frac{1-y}{y} \,
C_F \bigg( \! 2\frac{1-z}{z} + (1-\eps)z \! \bigg)
\nnb \\
& = &
\frac{\as}{2\pi}
\left( \frac{s}{e^{\euler\!}\mu^2} \right)^{\!\!-\eps} 
\frac{\Gam(1-\eps)\Gam(2-\eps)}{\eps\,\Gam(2-3\eps)} \,
C_F \,
\bigg(
\frac2{\eps}
-
\frac{1}{2}
\bigg)
\nnb \\
& = &
\frac{\as}{2\pi}
\left( \frac{s}{\mu^2} \right)^{\!\!-\eps} 
C_F 
\left[
\frac2{\eps^2}
+
\frac72\,\frac1{\eps}
+
11
-
\frac{7}{6}\,\pi^2
+
\mc O (\eps)
\right]
\, ;
\eeq
finally, the radiation of two collinear gluons from a gluon yields
\beq
J_{\rm c}^{\tg}(s)
& = & 
\Norm \,
N (\eps) \, s^{-\eps} \!\!
\int_0^\pi\!\!\!d\phi \, \sin^{- 2 \eps}\!\phi \!
\int_0^1 \!\! dy \int_0^1 \!\! dz
\Big[ y (1 - y)^2 z (1 - z) \Big]^{- \eps} \, \frac{1-y}{y} \,
\nnb\\
&&
\hspace{50mm}
\times \,\, 
2 \, C_A \bigg[ \frac{z}{1-z} + \frac{1-z}{z} + z(1-z) \bigg]
\nnb \\
& = &
\frac{\as}{2\pi}
\left( \frac{s}{e^{\euler\!}\mu^2} \right)^{\!\!-\eps} 
\frac{\Gam(1-\eps)\Gam(2-\eps)}{\eps \,\Gam(2-3\eps)} \,
C_A \,
\bigg(
\frac{4}{\eps}
-
\frac{1}{3-2\eps}
\bigg)
\nnb \\
& = &
\frac{\as}{2\pi} \!
\left( \frac{s}{\mu^2} \right)^{\!\!-\eps} \!\!
C_A 
\left[
\frac4{\eps^2}
+
\frac{23}{3}\,\frac1{\eps}
+
\frac{208}{9}
-
\frac{7}{3}\,\pi^2
+
\mc O (\eps)
\right]
\, ,
\eeq
which completes the required NLO calculations. To be precise, in order to build
a complete NLO subtraction procedure one also needs to introduce and integrate
the soft-collinear kernel, extracted from the combined limits ${\bf S}_i {\bf C}_{ij} R
= {\bf C}_{ij} {\bf S}_i R$, after introducing an appropriate mapping. This presents 
no further difficulties, as discussed in detail in Ref.~\cite{Magnea:2018hab}.

%%%%%%%%%%%%%%%%%%%%%%%%%%%%%%%%%%%%%%%%%%%%%

\section{Tree-level infrared kernels with two real emissions}
\label{treetworeal}

%%%%%%%%%%%%%%%%%%%%%%%%%%%%%%%%%%%%%%%%%%%%%

\noindent
In this section we consider the integration of tree-level infrared kernels with 
two real emissions. We first rewrite the factorisation formulae derived in 
Ref.~\cite{Catani:1998nv,Catani:1999ss} for the emission of two soft particles and three 
collinear particles. Indicating with $RR(\{k\}) = |{\cal A}_\npt^{(0)}|^2$ the 
tree-level squared matrix element for the emission of two extra partons, 
the general structure of the double-soft limit $\bS{ij}$, where 
both particles $i$ and $j$ become uniformly soft, can be written as
\beq
\label{eq:SSCC1}
\bS{ij} \, RR 
&=&
\frac{\Norm^{\,2}}{2} \!
\sum_{\substack{c \neq i,j \\ d \neq i,j}} 
\bigg[ 
\sum_{\substack{e\neq i,j\\f\neq i,j}} 
\mc I_{cd}^{(i)} \, \mc I_{ef}^{(j)} \,B_{cdef}(\{k\}_{\slashed i \slashed j})
+
\mc I_{cd}^{(ij)} \, B_{cd}(\{k\}_{\slashed i\slashed j})
\bigg] 
\\
&=&
\frac{\Norm^{\,2}}{2} \!
\sum_{\substack{c \neq i,j \\ d \neq i,j,c}} \!
\bigg[
\sum_{\substack{e \neq i,j,c,d \\ f \neq i,j,c,d,e}} \!\!\!\!
\mc I_{cd}^{(i)} \,
\mc I_{ef}^{(j)} 
B_{cdef}(\{k\}_{\slashed i\slashed j})
+
4 \!\!\!
\sum_{\substack{e \neq i,j,c,d}} \!\!
\mc I_{cd}^{(i)} \,
\mc I_{ed}^{(j)} 
B_{cded}(\{k\}_{\slashed i\slashed j})
\nnb\\
&&
\qquad\qquad\quad
+ \,\,
2 \,
\mc I_{cd}^{(i)} \, 
\mc I_{cd}^{(j)}
B_{cdcd}(\{k\}_{\slashed i\slashed j})
+
\left( 
\mc I_{cd}^{(ij)} - \frac12\,\mc I_{cc}^{(ij)} - \frac12\,\mc I_{dd}^{(ij)} 
\right) \!
B_{cd}(\{k\}_{\slashed i\slashed j})
\bigg]
\, .
\nnb
\eeq
On the other  hand, in the collinear limit $\bC{ijk}$, where particles $i$, $j$ and $k$ 
become uniformly collinear, we have the general structure
\beq
\label{eq:SSCC2}
\bC{ijk} \, RR 
& = & 
\frac{\Norm^{\,2}}{s_{ijk}^2} \, 
\bigg[
P_{ijk} \, B(\{k\}_{[ijk]})
+
Q_{ijk}^{\mu\nu} \,B_{\mu\nu}(\{k\}_{[ijk]})
\bigg]
\, .
\eeq
In \eqs{eq:SSCC1}{eq:SSCC2}, $\Norm$ is given by \eq{calN1}, while the momentum 
sets $\{k\}_{\slashed i \slashed j}$ and $\{k\}_{[ijk]}$ are obtained from $\{k\}$ by 
removing $k_i,k_j$, and by combining $k_i,k_j,k_k$ into $k=k_i+k_j+k_k$, respectively. 
In \eq{eq:SSCC1}, furthermore, we have introduced the doubly-colour-connected 
Born squared matrix element 
$B_{cdef} = 
{\cal A}_n^{(0)*} 
\{ {\bf T}_c\cdot{\bf T}_d , {\bf T}_e\cdot {\bf T}_f \} 
{\cal A}_n^{(0)}$, 
which is multiplied times the product of two eikonal factors, defined in 
\eq{eq:eikonal}. 
In the latter expression of \eq{eq:SSCC1}, we have rearranged all sums in such a way 
that each term features only unequal colour indices. The (singly-)colour-connected 
squared amplitude $B_{cd}$, on the other hand, multiplies the pure NNLO soft kernel, 
which can be written as 
\beq
\mc I_{cd}^{(ij)}
& = &
\delta_{\{f_if_j\}\{q \bar q\}}\,2\,T_R\,
\mc I_{cd}^{(q\bar q)(ij)}
-
\delta_{f_i g}\,\delta_{f_j g} \, 2\,C_A\,
\mc I_{cd}^{(gg)(ij)} \, ,
\eeq
with the explicit expressions~\cite{Catani:1999ss}
\beq
\mc I_{cd}^{(q\bar q)(ij)}
& = &
\frac{s_{ic}s_{jd}+s_{id}s_{jc}-s_{ij}s_{cd}}{s_{ij}^2(s_{ic}+s_{jc})(s_{id}+s_{jd})}
\, ,
\\[3mm]
\mc I_{cd}^{(gg)(ij)}
& = &
\frac{(1-\eps)(s_{ic}s_{jd}+s_{id}s_{jc})-2s_{ij}s_{cd}}
     {s_{ij}^2(s_{ic}+s_{jc})(s_{id}+s_{jd})}
\nnb\\
&&
\, + \,
s_{cd} \,\,
\frac{s_{ic}s_{jd}+s_{id}s_{jc}-s_{ij}s_{cd}}{s_{ij}s_{ic}s_{jd}s_{id}s_{jc}} \,
\bigg[
1
-
\frac12 \,
\frac{s_{ic}s_{jd}+s_{id}s_{jc}}{(s_{ic}\!+\!s_{jc})(s_{id}\!+\!s_{jd})}
\bigg]
\, .
\nnb
\eeq
In the collinear factorisation formula, \eq{eq:SSCC2}, the collinear kernels 
can be organised by flavour structure as
\beq
\label{eq:Pijk-Qijk-1}
P_{ijk} 
& = & 
  P_{ijk}^{\zg}\,\delta_{\{f_if_j\}\{q \bar q\}}\,\delta_{f_k \{q',\bar q'\}}
+ P_{jki}^{\zg}\,\delta_{\{f_jf_k\}\{q \bar q\}}\,\delta_{f_i \{q',\bar q'\}}
+ P_{kij}^{\zg}\,\delta_{\{f_kf_i\}\{q \bar q\}}\,\delta_{f_j \{q',\bar q'\}}
\nnb\\
&&
+ \; P_{ijk}^{(\rm{0g},\rm id)}\,\delta_{\{\{f_if_j\}f_k\}\{q \bar q\}}
+    P_{jki}^{(\rm{0g},\rm id)}\,\delta_{\{\{f_jf_k\}f_i\}\{q \bar q\}}
+    P_{kij}^{(\rm{0g},\rm id)}\,\delta_{\{\{f_kf_i\}f_j\}\{q \bar q\}}
\nnb\\
&&
+ \; P_{ijk}^{\og}\,\delta_{\{f_if_j\}\{q \bar q\}}\,\delta_{f_k g}
+    P_{jki}^{\og}\,\delta_{\{f_jf_k\}\{q \bar q\}}\,\delta_{f_i g}
+    P_{kij}^{\og}\,\delta_{\{f_kf_i\}\{q \bar q\}}\,\delta_{f_j g}
\nnb\\
&&
+ \; P_{ijk}^{\tg}\,\delta_{f_i g}\,\delta_{f_j g}\,\delta_{f_k \{q,\bar q\}}
+    P_{jki}^{\tg}\,\delta_{f_j g}\,\delta_{f_k g}\,\delta_{f_i \{q,\bar q\}}
+    P_{kij}^{\tg}\,\delta_{f_k g}\,\delta_{f_i g}\,\delta_{f_j \{q,\bar q\}}
\nnb\\
&&
+ \; P_{ijk}^{({\rm 3g})}\,\delta_{f_i g}\,\delta_{f_j g}\,\delta_{f_k g} 
\, ,
\eeq
where $q'$ is a quark of flavour equal to or different from that of $q$; 
similarly, the azimuthal tensor kernel can be written as
\beq
\label{eq:Pijk-Qijk-2}
Q_{ijk}^{\mu\nu}
& = &
  Q_{ijk}^{\og\mu\nu}\,\delta_{\{f_if_j\}\{q \bar q\}}\,\delta_{f_k g}
+ Q_{jki}^{\og\mu\nu}\,\delta_{\{f_jf_k\}\{q \bar q\}}\,\delta_{f_i g}
+ Q_{kij}^{\og\mu\nu}\,\delta_{\{f_kf_i\}\{q \bar q\}}\,\delta_{f_j g}
\nnb\\
&&
+ \; Q_{ijk}^{(\rm{3g})\mu\nu}\,\delta_{f_i g}\,\delta_{f_j g}\,\delta_{f_k g} 
\, .
\eeq
In \eqs{eq:Pijk-Qijk-1}{eq:Pijk-Qijk-2} we introduced
$
\delta_{\{\{f_af_b\}f_c\}\{q \bar q\}} = 
\delta_{f_a q}\,\delta_{f_b q}\,\delta_{f_c \bar q}
+ 
\delta_{f_a \bar q}\,\delta_{f_b \bar q}\,\delta_{f_c q}
$,
and, as before, the superscripts ($k \rm{g}$) refer to the number of final-state 
gluons featuring in the various kernels.

The expressions for $P_{ijk}^{\zg}$, $P_{ijk}^{(\rm{0g},\rm id)}$, $P_{ijk}^{\og}$, 
$P_{ijk}^{\tg}$, and $P_{ijk}^{(\rm 3g)}$ can be extracted from 
Ref.~\cite{Catani:1999ss}, and can be written as
\beq
\label{Pijk0g}
P_{ijk}^{\zg}
& = &
C_F T_R \,
\Bigg\{
\!\!
- 
\frac{s_{ijk}^2}{2s_{ij}^2}\,
\bigg(
\frac{s_{jk}}{s_{ijk}} - \frac{s_{ik}}{s_{ijk}} + \frac{z_i\!-\!z_j}{z_{ij}}
\bigg)^2
\!\!
+
\frac{s_{ijk}}{s_{ij}}\,
\bigg[
2 \, \frac{z_k \!-\! z_iz_j}{z_{ij}}
+
(1-\eps) z_{ij}
\bigg]
\!
-
\!\frac12
+
\eps
\Bigg\}
\, ,
\qquad\quad
\eeq
\beq
\label{Pijk0gid}
P_{ijk}^{(\rm{0g},\rm id)}
& = &
C_F(2C_F\!-\!C_A)\,
\Bigg\{ \!
- 
\frac{s_{ijk}^2\, z_k}{2s_{jk}s_{ik}}  \,
\bigg[
\frac{1+z_k^2}{z_{jk}z_{ik}}
-
\eps\,
\bigg(
\frac{z_{ik}}{z_{jk}}
+
\frac{z_{jk}}{z_{ik}}
+
1
+
\eps
\bigg)
\bigg]
+ \,
(1-\eps)\bigg[ \frac{s_{ij}}{s_{jk}} + \frac{s_{ij}}{s_{ik}} - \eps \bigg]
\nnb \\
&&
\hspace{26mm}
+ \,
\frac{s_{ijk}}{2s_{jk}}\,
\bigg[
\frac{1 + z_k^2 - \eps z_{jk}^2}{z_{ik}}
-
2(1-\eps)\frac{z_j}{z_{jk}}
-
\eps(1+z_k)
-
\eps^2\,z_{jk}
\bigg]
\nnb \\
&&
\hspace{26mm}
+ \,
\frac{s_{ijk}}{2s_{ik}}\,
\bigg[
\frac{1 + z_k^2 - \eps z_{ik}^2}{z_{jk}}
-
2(1-\eps)\frac{z_i}{z_{ik}}
-
\eps(1+z_k)
-
\eps^2\,z_{ik}
\bigg]
\Bigg\}
\, ,
\qquad
\\[2mm]
\label{Pijk1g}
P_{ijk}^{\og}
& = &
C_F T_R \,
\Bigg\{
\frac{2 s_{ijk}^2}{s_{ik}s_{jk}}  \,
\bigg[
1 + z_k^2 - \frac{z_k+2z_iz_j}{1-\eps}
\bigg]
- 
(1-\eps)\bigg[ \frac{s_{ij}}{s_{jk}} + \frac{s_{ij}}{s_{ik}} \bigg]
-
2
\nnb \\
&&
\hspace{13.5mm}
- \,
\frac{s_{ijk}}{s_{jk}}\,
\bigg[
1 + 2z_k + \eps - \frac{2z_{jk}}{1-\eps}
\bigg]
-
\frac{s_{ijk}}{s_{ik}}\,
\bigg[
1 + 2z_k + \eps - \frac{2z_{ik}}{1-\eps}
\bigg]
\Bigg\}
\nnb \\
&&
\hspace{-4.2mm}
+ \,\,
C_A T_R \,\,
\Bigg\{
- 
\frac{s_{ijk}^2}{2s_{ij}^2}\,
\bigg(
\frac{s_{jk}}{s_{ijk}} - \frac{s_{ik}}{s_{ijk}} + \frac{z_i-z_j}{z_{ij}}
\bigg)^2
-
\frac{s_{ijk}^2}{s_{ik}s_{jk}}  \,
\bigg[
1 + z_k^2 - \frac{z_k+2z_iz_j}{1-\eps}
\bigg]
\nnb \\
&&
\hspace{13.5mm}
+ \,
\frac{s_{ijk}^2}{2s_{ij}s_{ik}}\,\frac{z_i}{z_k z_{ij}}
\bigg[
z_{ij}^3 - z_k^3 - \frac{2z_i(z_{jk} - 2z_jz_k)}{1-\eps}
\bigg]
\nnb \\
&&
\hspace{13.5mm}
+ \, 
\frac{s_{ijk}^2}{2s_{ij}s_{jk}}\,\frac{z_j}{z_k z_{ij}}
\bigg[
z_{ij}^3 - z_k^3 - \frac{2z_j(z_{ik} - 2z_iz_k)}{1-\eps}
\bigg]
\nnb \\
&&
\hspace{13.5mm}
+ \,
\frac{s_{ijk}}{2s_{ik}}\,\frac{z_{ik}}{z_kz_{ij}}
\bigg[
1 + z_kz_{ij} - \frac{2z_jz_{ik}}{1-\eps}
\bigg]
+ 
\frac{s_{ijk}}{2s_{jk}}\,\frac{z_{jk}}{z_kz_{ij}}
\bigg[
1 + z_kz_{ij} - \frac{2z_iz_{jk}}{1-\eps}
\bigg]
\nnb \\
&&
\hspace{13.5mm}
+ \,
\frac{s_{ijk}}{s_{ij}}\,\frac{1}{z_kz_{ij}}
\bigg[
1 + z_k^3 + \frac{z_k(z_i-z_j)^2-2 z_iz_j(1+z_k)}{1-\eps}
\bigg]
-
\frac12
+
\eps
\Bigg\}
\, ,
\\[2mm]
\label{Pijk2g}
P_{ijk}^{\tg}
& = &
C_F^2\,
\Bigg\{
\frac{s_{ijk}^2\,z_k}{2s_{ik}s_{jk}}\,
\bigg[
\frac{1+z_k^2-\eps z_{ij}^2}{z_iz_j}
+
\eps(1-\eps)
\bigg]
-
(1-\eps)^2 \, \frac{s_{jk}}{s_{ik}}
+
\eps(1-\eps)
\nnb \\
&&
\hspace{10mm}
+ \,
\frac{s_{ijk}}{s_{ik}}\,
\bigg[
\frac{z_k z_{jk} + z_{ik}^2 - \eps z_{ik} z_{ij}^2}{z_iz_j}
+
\eps\,z_{ik}
+
\eps^2\,(1+z_k)
\bigg]
\Bigg\}
\nnb \\
&& \hspace{-3.5mm}
+ \,\,
C_F C_A\,\,
\Bigg\{
(1-\eps)\frac{s_{ijk}^2}{4s_{ij}^2}
\bigg(
\frac{s_{jk}}{s_{ijk}} - \frac{s_{ik}}{s_{ijk}} + \frac{z_i-z_j}{z_{ij}}
\bigg)^2
- 
\frac{s_{ijk}^2\,z_k}{4s_{ik}s_{jk}}\,
\bigg[
\frac{z_{ij}^2(1-\eps) + 2z_k}{z_iz_j}
+
\eps(1-\eps)
\bigg]
\nnb \\
&&
\hspace{15mm}
+ \,
\frac{s_{ijk}^2}{2s_{ij}s_{ik}}\,
\bigg[
\frac{z_{ij}^2(1-\eps) + 2z_k}{z_j}
+
\frac{z_j^2(1-\eps) + 2z_{ik}}{z_{ij}}
\bigg]
+
\frac14(1-\eps)(1-2\eps)
\nnb \\
&&
\hspace{15mm}
+ \,
\frac{s_{ijk}}{2s_{ik}}\,
\bigg[
(1-\eps)
\frac{z_{ik}^3+z_k^2-z_j}{z_j z_{ij}}
-
2\eps \,
\frac{z_{ik}(z_j-z_k)}{z_j z_{ij}}
\nnb\\
&&
\hspace{28mm}
- \,
\frac{z_k z_{jk}+z_{ik}^3}{z_iz_j}
+ 
\eps\,z_{ik}
\frac{z_{ij}^2}{z_iz_j}
-
\eps(1+z_k)
-
\eps^2 z_{ik}
\bigg]
\nnb \\
&&
\hspace{15mm}
+ \,
\frac{s_{ijk}}{2s_{ij}}\,
\bigg[
(1-\eps)
\frac{z_i(2z_{jk}+z_i^2)-z_j(6z_{ik}+z_j^2)}{z_j z_{ij}}
+
2\eps \,
\frac{z_k(z_i-2z_j)-z_j}{z_j z_{ij}}
\bigg]
\Bigg\}
\nnb \\
&& \hspace{-3.5mm}
+ \,\,
( i \leftrightarrow j)
\, ,
\\[2mm]
\label{Pijk3g}
P_{ijk}^{(\rm 3g)}
& = &
C_A^2 \,
\Bigg\{
(1-\eps)\frac{s_{ijk}^2}{4s_{ij}^2}
\bigg(
\frac{s_{jk}}{s_{ijk}} - \frac{s_{ik}}{s_{ijk}} + \frac{z_i-z_j}{z_{ij}}
\bigg)^2
+
\frac34(1-\eps)
\nnb \\
&&
\hspace{8mm}
+ \,
\frac{s_{ijk}^2}{2s_{ij}s_{ik}}
\bigg[
\frac{2 z_i z_j z_{ik}(1\!-\!2z_k)}{z_k z_{ij}}
+
\frac{1\!+\!2z_i\!+\!2z_i^2}{z_{ik} z_{ij}}
+ 
\frac{1\!-\!2z_iz_{jk}}{z_j z_k}
+
2 z_j z_k
+
z_i(1\!+\!2z_i)
-
4
\bigg]
\nnb\\
&&
\hspace{8mm}
+ \, 
\frac{s_{ijk}}{s_{ij}}\,
\bigg[
4\,\frac{z_iz_j-1}{z_{ij}}
+
\frac{z_iz_j-2}{z_k}
+
\frac{(1-z_kz_{ij})^2}{z_iz_k z_{jk}}
+
\frac52\,z_k
+
\frac32
\bigg]
\Bigg\}
\nnb\\
&&
+ \, \,
\mbox{( 5 permutations )}
\, ,
\eeq
where we defined
\beq
\label{zdefijk}
z_a
& = &
\frac{s_{ar}}{s_{ir}\!+\!s_{jr}\!+\!s_{kr}}
\, ,
\qquad
\qquad
z_{ab}
\, = \,
z_a + z_b
\, ,
\qquad
\qquad
a, b
\, = \,
i, j, k
\, ,
\eeq
and $k_r$, as before, is a massless reference vector, which can be chosen among 
the momenta of the outgoing particles.

From Ref.~\cite{Catani:1999ss} we can also obtain the expressions for the azimuthal 
tensor kernels $Q_{ijk}^{\og\mu\nu}$ and $Q_{ijk}^{(3g)\mu\nu}$, which we report
in our notation for  completeness. They are
\beq
Q_{ijk}^{\og\mu\nu}
& = &
- \, C_F T_R \,\,
\frac{2}{1-\eps}
\frac{s_{ijk}}{s_{ik}s_{jk}} \,
\Bigg[
\tk_i^2
\, q_i^{\mu\nu}
+
\tk_j^2
\, q_j^{\mu\nu}
-
\eps \, 
\tk_k^2
\, q_k^{\mu\nu}
\Bigg]
\\
&&
+ \,
C_A T_R \,\,
\frac{s_{ijk}}{2(1-\eps)} \,
\Bigg\{
\nnb\\
&&
\hspace{7mm}
\bigg[
\frac{2}{s_{ik}s_{jk}}
-
\frac{4z_j}{z_k}
\frac{s_{ij}+2s_{jk}}{s_{ij}^2s_{jk}}
+
\frac{2(z_is_{jk}+z_js_{ik})}{z_{ij}s_{ij}s_{ik}s_{jk}}
+
\bigg(
\frac{2z_iz_j}{z_k z_{ij}} - 1 + \eps
\bigg)
\frac{s_{ik}-s_{jk}}{s_{ij}s_{ik}s_{jk}}
\bigg]
\,
\tk_i^2 \, q_i^{\mu\nu}
\nnb\\
&&
\hspace{4mm}
+ \,
\bigg[
\frac{2}{s_{ik}s_{jk}}
-
\frac{4z_i}{z_k}
\frac{s_{ij}+2s_{ik}}{s_{ij}^2s_{ik}}
+
\frac{2(z_is_{jk}+z_js_{ik})}{z_{ij}s_{ij}s_{ik}s_{jk}}
+
\bigg(
\frac{2z_iz_j}{z_k z_{ij}} - 1 + \eps
\bigg)
\frac{s_{jk}-s_{ik}}{s_{ij}s_{ik}s_{jk}}
\bigg]
\,
\tk_j^2 \, q_j^{\mu\nu}
\nnb \\
&&
\hspace{4mm}
+ \,
\bigg[ 
\frac{2z_iz_j}{z_k z_{ij}}
\frac{1}{s_{ij}}
\bigg(
\frac{4}{s_{ij}}
+
\frac{1}{s_{ik}}
+
\frac{1}{s_{jk}}
\bigg)
+
\frac{z_i-z_j}{z_{ij}}\frac{s_{ik}-s_{jk}}{s_{ij}s_{ik}s_{jk}}
-
\eps \,
\frac{s_{ijk}+s_{ij}}{s_{ij}s_{ik}s_{jk}}
\bigg]
\,
\tk_k^2 \, q_k^{\mu\nu}
\Bigg\}
\, ,
\nnb\\
Q_{ijk}^{(3g)\mu\nu}
& = &
C_A^2 \,\,
\frac{s_{ijk}}{s_{ij}} \,
\Bigg\{
\bigg[
\frac{2z_j}{z_k}
\frac{1}{s_{ij}}
+
\bigg(\! \frac{z_jz_{ik}}{z_k z_{ij}} - \frac32 \!\bigg)
\frac{1}{s_{ik}}
\bigg]
\,
\tk_i^2\,q_i^{\mu\nu}
\nnb\\
&&
\hspace{12.5mm}
+ \,
\bigg[
\frac{2z_i}{z_k}
\frac{1}{s_{ij}}
-
\bigg(\! 
\frac{z_jz_{ik}}{z_k z_{ij}} - \frac32 - \frac{z_i}{z_k} + \frac{z_i}{z_{ij}} 
\!\bigg)
\frac{1}{s_{ik}}
\bigg]
\,
\tk_j^2 \,q_j^{\mu\nu}
\nnb \\
&&
\hspace{12.5mm}
- \, 
\bigg[
\frac{2z_iz_j}{z_{ij}z_k}
\frac{1}{s_{ij}}
+
\bigg(\! 
\frac{z_jz_{ik}}{z_k z_{ij}} - \frac32 - \frac{z_i}{z_j} + \frac{z_i}{z_{ik}} 
\!\bigg)
\frac{1}{s_{ik}}
\bigg]
\,
\tk_k^2 \,q_k^{\mu\nu}
\Bigg\} \,
+ \,
\mbox{( 5 permutations )}
\, ,
\nnb
\eeq
where all terms are proportional to azimuthal tensors of the form
\beq
&&
q_a^{\mu\nu} 
\, = \, 
- g^{\mu\nu} + (d-2)\,\frac{\tk_a^\mu\tk_a^\nu}{\tk_a^2}
\, ,
\eeq
and, in analogy with \eq{ktdef1}, we defined a transverse-momentum vector
\beq
\label{ktdef2}
\qquad
&& \kt_{a}^\mu 
\, = \, 
k_a^\mu - z_a k^\mu - ( k \dt k_a - z_a k^2 ) \frac{k_r^\mu}{k \dt k_r}
\, ,
\qquad\qquad
a,b,c = i,j,k
\, ,
\nnb\\
&&
\tk_a^2 
\, = \, 
z_a( z_a k^2 - 2 \, k \dt k_a )
\, = \,
z_a( s_{bc} - z_{bc} s_{ijk}) 
\, ,
\eeq
where the momentum $k^\mu = k_i^\mu + k_j^\mu + k_k^\mu$ is the parent 
momentum of the three collinear particles. We stress that the symmetry of 
$P_{ijk}^{(3g)}$, $Q_{ijk}^{(3g)\mu\nu}$ under exchange of $i$, $j$, $k$, and 
of all other kernels under exchange of $i$, $j$, guarantees that the kernels 
$P_{ijk}$ and $Q_{ijk}^{\mu\nu}$ defined in \eqs{eq:Pijk-Qijk-1}{eq:Pijk-Qijk-2} 
are totally symmetric under permutations of $i$, $j$, and $k$. 

%%%%%%%%%%%%%%%%%%%%%%%%%

\subsection{Phase-space mappings and integration}
\label{sec:mappings2}

%%%%%%%%%%%%%%%%%%%%%%%%%

\subsubsection{Double-soft kernel}
\label{2radsoft}

%%%%%%%%%%%%

\noindent
In order to integrate the double-soft kernel in \eq{eq:SSCC1} we introduce different 
phase-space mappings according to the number of different momenta involved in 
the various contributions to the kernel. For the terms containing $B_{cd}$ and 
$B_{cdcd}$, where only the four particles $i,j,c,d$ are present, we use the mapping 
described in \appn{app:map2 4}, with the identifications
\beq
k_{\aA} \to k_i \, ,
\quad\qquad
k_{\bB} \to k_j \, ,
\quad\qquad
k_{\cC} \to k_c \, ,
\quad\qquad
k_{\dD} \to k_d \, .
\eeq
For the terms with $B_{cded}$ involving the five particles $i,j,c,d,e$, 
we use the mapping given in \appn{app:map2 5}, with
\beq
k_{\aA} \to k_i \, ,
\quad\qquad
k_{\bB} \to k_j \, ,
\quad\qquad
k_{\cC} \to k_c \, ,
\quad\qquad
k_{\dD} \to k_d \, ,
\quad\qquad
k_{\eE} \to k_e \, .
\eeq
Finally, for the terms proportional to $B_{cdef}$, we use the 
mapping of \appn{app:map2 6}, with
\beq
\hspace{-5mm}
k_{\aA} \to k_i \, ,
\quad\quad
k_{\bB} \to k_j \, ,
\quad\quad
k_{\cC} \to k_c \, ,
\quad\quad
k_{\dD} \to k_d \, ,
\quad\quad
k_{\eE} \to k_e \, ,
\quad\quad
k_{\fF} \to k_f \, .
\eeq
The mapped double-soft limit of $RR(\{k\})$ is then 
\beq
\label{eq:bSS RR}
\bbS{ij} \, RR 
&=&
\frac{\Norm^{\,2}}{2} \!
\sum_{\substack{c \neq i,j \\ d \neq i,j,c}} \!
\Bigg[
\sum_{\substack{e \neq i,j,c,d \\ f \neq i,j,c,d,e}} \!\!\!\!
\mc I_{cd}^{(i)} \,
\mc I_{ef}^{(j)} 
B_{cdef} \Big( \! \kkl{icd,jef} \! \Big) 
\nnb\\
&&
\hspace{18mm}
+ \,
4 \!\!\!
\sum_{\substack{e \neq i,j,c,d}} \!\!
\mc I_{cd}^{(i)} \,
\mc I_{ed}^{(j)} 
B_{cded} \Big( \! \kkl{icd,jed} \! \Big) 
+ 
2\,
\mc I_{cd}^{(i)} \, 
\mc I_{cd}^{(j)}
B_{cdcd} \Big( \! \kkl{ijcd} \! \Big) 
\nnb\\
&&
\hspace{18mm}
+ \,
\left( 
\mc I_{cd}^{(ij)} - \frac12\,\mc I_{cc}^{(ij)} - \frac12\,\mc I_{dd}^{(ij)} 
\right) \!
B_{cd} \Big( \! \kkl{ijcd} \! \Big) 
\Bigg]
\, ,
\eeq
and its integral in the $\npt$ phase-space is given by
\beq
\label{eq:int bSS RR}
\int d\Phi_\npt \,
\bbS{ij} \, RR 
& = &
\frac{1}{2} \!
\sum_{\substack{c \neq i,j \\ d \neq i,j,c}} \!
\Bigg\{
\sum_{\substack{e \neq i,j,c,d \\ f \neq i,j,c,d,e}} \!\!
\int\!d \Phi_n \big(\kkl{icd,jef}\big)  \,
J_{{\rm s} \otimes {\rm s}}^{ijcdef} \,
B_{cdef} \Big( \! \kkl{icd,jef} \! \Big)
\nnb \\
&&
\hspace{15mm}
+ \,
4 \!\!\!
\sum_{\substack{e \neq i,j,c,d}}
\int\!d \Phi_n \big( \kkl{icd,jed} \big)  \,
J_{\rm s \otimes s}^{ijcde} \,
B_{cded} \Big( \! \kkl{icd,jed} \! \Big)
\nnb\\
&&
\hspace{15mm}
+ \,
2 
\int\!d \Phi_n \big(\kkl{ijcd}\big)  \,
J_{\rm s \otimes s}^{ijcd} \,
B_{cdcd} \Big( \! \kkl{ijcd} \! \Big)
\nnb\\
&&
\hspace{15mm}
+
\int\!d \Phi_n \big(\kkl{ijcd}\big) \,
J_{\rm ss}^{ijcd} \,
B_{cd} \Big( \! \kkl{ijcd} \! \Big)
\;
\Bigg\}
\, ,
\eeq
where the radiative integrals of products of eikonal kernels are defined by
\beq
J_{\rm s \otimes s}^{ijcdef}
& \equiv &
\Norm^{\,2}
\int d\Phi_{\rm rad,2}^{(icd,jef)}\,
\mc I_{cd}^{(i)} \, \mc I_{ef}^{(j)}
\, \equiv \,
\delta_{f_i g} \,\delta_{f_j g} \,
J_{\rm s \otimes s}^{(4)} \Big(\sk{cd}{icd,jef},\sk{ef}{icd,jef} \Big) 
\, ,
\nnb \\
J_{\rm s \otimes s}^{ijcde}
& \equiv &
\Norm^{\,2}
\int d\Phi_{\rm rad,2}^{(icd,jed)}\,
\mc I_{cd}^{(i)} \, \mc I_{ed}^{(j)}
\, \equiv \, 
\delta_{f_i g} \,\delta_{f_j g} \,
J_{\rm s \otimes s}^{(3)} \Big(\sk{cd}{icd,jed},\sk{ed}{icd,jed} \Big) 
\, ,
\nnb \\
J_{\rm s \otimes s}^{ijcd}
& \equiv &
\Norm^{\,2}
\int d\Phi_{\rm rad,2}^{(ijcd)}\,
\mc I_{cd}^{(i)} \, \mc I_{cd}^{(j)}
\, \equiv \, 
\delta_{f_i g} \,\delta_{f_j g} \,
J_{\rm s \otimes s}^{(2)} \Big(\sk{cd}{ijcd} \Big) 
\, ,
\eeq
while the radiative integral of the pure double-soft kernel is
\beq
\label{eq:Jss def}
J_{\rm ss}^{ijcd}
& \equiv &
\Norm^{\,2}
\int d\Phi_{\rm rad,2}^{(ijcd)}\,
\left( 
\mc I_{cd}^{(ij)} - \frac12\,\mc I_{cc}^{(ij)} - \frac12\,\mc I_{dd}^{(ij)} 
\right)
\nnb \\
& \equiv &
\delta_{\{f_if_j\}\{q \bar q\}} \, 2\,T_R\,
J_{\rm ss}^{(\rm q \bar q)}\Big(\sk{cd}{ijcd}\Big) 
-
\delta_{f_i g}\,\delta_{f_j g} \, 2\,C_A\,
J_{\rm ss}^{(\rm gg)}\Big(\sk{cd}{ijcd}\Big) 
\, .
\eeq
The kinematic dependence of these integrals is described by the five  
\emph{radiative double-soft functions} $J_{\rm s \otimes s}^{(4)}$, 
$J_{\rm s \otimes s}^{(3)}$, $J_{\rm s \otimes s}^{(2)}$, $J_{\rm ss}^{(\rm q \bar q)}$ 
and $J_{\rm ss}^{(\rm gg)}$. The integrals defining $J_{\rm s \otimes s}^{(4)}$ 
and $J_{\rm s \otimes s}^{(3)}$ are factorised, so their calculation is trivial, and 
can be performed to all orders in $\epsilon$, analogously to the case with 
one emission. We find 
\beq
\label{J4sos}
J_{\rm s \otimes s}^{(4)}(s, s')
& = &
\Norm^{\,2} \,
N^2(\eps) \, 
(s\,s')^{\!-\eps}
\!\!
\int_0^\pi \!\!\! d \phi' \, (\sin\phi')^{- 2 \eps} \!
\int_0^1 \!\!\! d y' \!
\int_0^1 \!\!\! d z' \!
\int_0^\pi \!\!\! d \phi \, (\sin\phi)^{- 2 \eps} \!
\int_0^1 \!\!\! d y \!
\int_0^1 \!\!\! d z \, 
\nnb \\
& & 
\qquad
\times\, 
\Big[ y'(1-y')^2\,z'(1-z')\,y(1-y)^2\,z(1-z) \Big]^{- \eps} 
(1-y')(1-y) 
\frac{1-z'}{y'z'}\,
\frac{1-z}{yz}
\nnb \\
& = &
J_{\rm s}(s)\,J_{\rm s}(s')
=
\left( \frac{\as}{2\pi} \right)^{\!2}
\left( \frac{ss'}{e^{2\euler\!}\mu^4} \right)^{\!\!-\eps} \, 
\bigg[
\frac{\Gam(1 - \eps) \Gam(2 - \eps)}{\eps^2 \, \Gam(2 - 3 \eps)}
\bigg]^2
\nnb \\
& = &
\!\bigg(\! \frac{\as}{2\pi} \!\bigg)^{\!\!2}
\left(\! \frac{ss'}{\mu^4} \!\right)^{\!\!\!-\eps}
\Bigg[
\frac1{\eps^4}
+ 
\frac4{\eps^3}
+
\bigg( 16 - \frac{7}{6}\pi^2 \bigg) \frac1{\eps^2}
+
\bigg( 60 - \frac{14}{3}\pi^2 - \frac{50}{3}\zeta_3 \bigg) 
\frac1{\eps}
\nnb \\
&& \hspace{27.7mm}
+ \, 
216
-
\frac{56}{3}\pi^2 
-
\frac{200}{3}\zeta_3 
+
\frac{29}{120}\pi^4
\!
+
\mc O (\eps)
\Bigg] \, ,
\eeq
and
\beq
\label{J3sos}
J_{\rm s \otimes s}^{(3)}(s, s')
& = &
\Norm^{\,2} \,
N^2(\eps) \, 
(s\,s')^{\!-\eps}
\!\!
\int_0^\pi \!\!\! d \phi' \, (\sin\phi')^{- 2 \eps} \!
\int_0^1 \!\!\! d y' \!
\int_0^1 \!\!\! d z' \!
\int_0^\pi \!\!\! d \phi \, (\sin\phi)^{- 2 \eps} \!
\int_0^1 \!\!\! d y \!
\int_0^1 \!\!\! d z \, 
\nnb \\
& & 
\qquad
\times\, 
\Big[ y'(1-y')^2\,z'(1-z')\,y(1-y)^3\,z(1-z) \Big]^{- \eps} 
(1-y')(1-y)
\frac{1-z'}{y'z'}\,
\frac{1-z}{yz}
\nnb \\
& = &
\left( \frac{\as}{2\pi} \right)^{\!2}
\left( \frac{ss'}{e^{2\euler\!}\mu^4} \right)^{\!\!-\eps} 
\frac{\Gam(1 - \eps) \Gam(2 - \eps)}{\eps^2 \, \Gam(2 - 3 \eps)}
\,
\frac{\Gam(1 - \eps) \Gam(2 - \eps)\Gam(2 - 3\eps)}
     {\eps^2 \, \Gam(2 - 2 \eps)\Gam(2 - 4 \eps)}
\nnb \\
& = &
\!\bigg(\! \frac{\as}{2\pi} \!\bigg)^{\!\!2}
\left(\! \frac{ss'}{\mu^4} \!\right)^{\!\!\!-\eps}
\Bigg[
\frac1{\eps^4}
+ 
\frac4{\eps^3}
+
\!\left(\! 17 - \frac{4}{3}\pi^2 \!\right) \!\! \frac1{\eps^2}
+ 
\!\left(\! 70 - \frac{16}{3}\pi^2 - \frac{68}{3}\zeta_3 \!\right) \!\!
\frac1{\eps} 
\nnb \\
&&  \hspace{27.9mm}
+ \,  
284
-
\frac{68}{3}\pi^2 
-
\frac{272}{3}\zeta_3
+
\frac{13}{90}\pi^4
+
\mc O (\eps)
\Bigg] \, .
\eeq
The integrals defining $J_{\rm s \otimes s}^{(2)}$, $J_{\rm ss}^{(gg)}$, 
$J_{\rm ss}^{(q \bar q)}$ are not factorised, and are thus more involved. 
They have been performed following the procedure described in \secn{sec:int}, 
with the results
\beq
J_{\rm s \otimes s}^{(2)}(s)
& = & 
\bigg(\! \frac{\as}{2\pi} \!\bigg)^{\!\!2} \!\!
\left(\! \frac{s}{\mu^2} \!\right)^{\!\!\!-2\eps} \! 
\bigg[
\frac1{\eps^4}
+ 
\frac4{\eps^3}
+
\bigg( \! 18 - \frac{3}{2}\pi^2 \! \bigg) \frac1{\eps^2}
+
\bigg( \!76 - 6\pi^2 - \frac{74}{3}\zeta_3 \!\bigg) \frac1{\eps}
\nnb\\
&&
\qquad\qquad\qquad
+ \,
312
-
27\pi^2 
-
\frac{308}{3}\zeta_3 
+
\frac{49}{120}\pi^4
\!
+
\mc O (\eps)
\bigg]
,
\nnb \\[3mm]
J_{\rm ss}^{({\rm q\bar q})}(s)
& = & 
\bigg(\! \frac{\as}{2\pi} \!\bigg)^{\!\!2} \!\!
\left(\! \frac{s}{\mu^2} \!\right)^{\!\!\!-2\eps} \!
\bigg[
\;
\frac{1}{6}  \, \frac1{\eps^3}
+ 
\frac{17}{18} \, \frac1{\eps^2}
+ 
\left( \frac{116}{27} - \frac{7}{36}\,\pi^2 \right) \frac1{\eps}
+ 
\frac{1474}{81} 
- 
\frac{131}{108}\,\pi^2 
- 
\frac{19}{9}\,\zeta_3
+
\mc O (\eps)
\bigg]
\, ,
\nnb\\[3mm]
J_{\rm ss}^{({\rm gg})}(s)
& = &
\bigg(\! \frac{\as}{2\pi} \!\bigg)^{\!\!2} \!\!
\left(\! \frac{s}{\mu^2} \!\right)^{\!\!\!-2\eps} \!
\bigg[
\;\;
\frac12 \, \frac1{\eps^4}
+ 
\frac{35}{12} \, \frac1{\eps^3}
+
\left( \frac{487}{36} - \frac{2}{3}\,\pi^2 \right) \frac1{\eps^2}
+
\left( \frac{1562}{27} - \frac{269}{72}\,\pi^2 - \frac{77}{6}\,\zeta_3 \right) 
\frac1{\eps}
\nnb\\
&&
\qquad\qquad\qquad
+ \,
\frac{19351}{81} 
-
\frac{3829}{216}\,\pi^2 
-
\frac{1025}{18}\,\zeta_3 
-
\frac{23}{240}\,\pi^4
+
\mc O (\eps)
\;
\bigg]
\, .
\eeq

%%%%%%%%%%%%

\subsubsection{Double-collinear kernel}
\label{2radcoll}

%%%%%%%%%%%%

\noindent 
In order to integrate the double-collinear kernel, we perform the phase-space
mappings described in \appn{app:map2 4}, with the choices
\beq
k_a \to k_i \, ,
\qquad\qquad
k_b \to k_j \, ,
\qquad\qquad
k_c \to k_k \, ,
\qquad\qquad
k_d \to k_r \, .
\eeq
The mapped double-collinear limit of $RR(\{k\})$ is then 
\beq
\label{eq:bCC RR}
\bbC{ijk} \, RR 
\, = \,  
\frac{\Norm^{\,2}}{s_{ijk}^2} \, \bigg[
P_{ijk} \,
B \Big( \! \kkl{ijkr} \! \Big)
+
Q_{ijk}^{\mu\nu} \,
B_{\mu\nu} \Big( \! \kkl{ijkr} \! \Big)
\bigg]
\, .
\eeq
As was the case for the single-collinear limit at  NLO, the integrals of the azimuthal
tensor kernel $Q_{ijk}^{\mu\nu}$ vanish because of its Lorentz structure:
\beq
\int d\Phi_\npt \,q_a^{\mu\nu}
\, = \,
\int d\Phi_n \big(\kkl{ijkr}\big) \int d\Phi_{\rm rad,2}^{(ijkr)} \,q_a^{\mu\nu}
\, = \,
0,
\qquad
\mbox{for}
\quad
a = i,j,k
\, ,
\eeq
which relies on the fact that $\kt_{a}\cdot \kk{k}{ijkr} = 0$ for $a=i,j,k$.
The remaining terms, featuring the $P_{ijk}$ kernels, can be integrated in 
the ($\npt$)-particle phase-space, and the result can be written as
\beq
\label{eq:int bCC RR}
&&
\int d \Phi_\npt \,
\bbC{ijk} \, RR
\; = \,
\int d \Phi_n \big(\kkl{ijrk}\big) \,
J_{\rm cc}^{ijkr} \,
B \! \left( \kkl{ijkr} \right) 
\, ,
\eeq
where the radiative integral
\beq
J_{\rm cc}^{ijkr}
& \; \equiv \; &
\Norm^{\,2}
\int d\Phi_{\rm rad,2}^{(ijkr)} \,
\frac{P_{ijk}}{s_{ijk}^2} 
\\ 
& \; \equiv \; &
\delta_{ \{f_i f_j f_k\} \{q \bar q q',q \bar q \bar q'\} } \, 
J_{\rm cc}^{\zg} \Big( \sk{kr}{ijkr} \Big)
+ 
\delta_{ \{f_i f_j f_k\} \{q q \bar q,\bar q \bar q q\} } \, 
J_{\rm cc}^{(\rm 0g, id)} \Big( \sk{kr}{ijkr} \Big)
\nnb \\
&&
\hspace{-4mm}
+ \,
\delta_{ \{f_i f_j f_k\} \{q \bar q g\} } \, J_{\rm cc}^{\og} \Big( \sk{kr}{ijkr} \Big)
+ 
\delta_{ \{f_i f_j f_k\} \{g g q, g g \bar q\} } \, J_{\rm cc}^{\tg} \Big( \sk{kr}{ijkr} \Big)
+ 
\delta_{f_i g} \delta_{f_j g} \delta_{f_k g} \, J_{\rm cc}^{(\rm 3g)} \Big( \sk{kr}{ijkr} \Big)
\nnb
\eeq
admits a flavour decomposition following from \eq{eq:Pijk-Qijk-1}, 
and has a kinematic dependence described by the five \emph{radiative 
double-collinear functions} $J_{\rm cc}^{\zg}$, $J_{\rm cc}^{\rm (0g, id)}$, 
$J_{\rm cc}^{\og}$, $J_{\rm cc}^{\tg}$, $J_{\rm cc}^{\rm (3g)}$ with argument 
$\sk{kr}{ijkr}$. Here we have introduced symmetrised flavour delta functions, 
according to
\beq
\delta_{ \{f_i f_j f_k\} \{q \bar q q',q \bar q \bar q'\} }
& = &
  \delta_{\{f_if_j\}\{q \bar q\}}\,\delta_{f_k \{q',\bar q'\}}
+ \delta_{\{f_jf_k\}\{q \bar q\}}\,\delta_{f_i \{q',\bar q'\}}
+ \delta_{\{f_kf_i\}\{q \bar q\}}\,\delta_{f_j \{q',\bar q'\}} \, ,
\nnb \\
\delta_{ \{f_i f_j f_k\} \{q q \bar q,\bar q \bar q q\} }
& = &
  \delta_{\{\{f_if_j\}f_k\}\{q \bar q\}}
+ \delta_{\{\{f_jf_k\}f_i\}\{q \bar q\}}
+ \delta_{\{\{f_kf_i\}f_j\}\{q \bar q\}} \, ,
\nnb \\
\delta_{ \{f_i f_j f_k\} \{q \bar q g\} }
& = &
  \delta_{\{f_if_j\}\{q \bar q\}}\,\delta_{f_k g}
+ \delta_{\{f_jf_k\}\{q \bar q\}}\,\delta_{f_i g}
+ \delta_{\{f_kf_i\}\{q \bar q\}}\,\delta_{f_j g} \, ,
\nnb \\
\delta_{ \{f_i f_j f_k\} \{g g q, g g \bar q\} }
& = &
  \delta_{f_i g}\,\delta_{f_j g}\,\delta_{f_k \{q,\bar q\}}
+ \delta_{f_j g}\,\delta_{f_k g}\,\delta_{f_i \{q,\bar q\}}
+ \delta_{f_k g}\,\delta_{f_i g}\,\delta_{f_j \{q,\bar q\}} \, ,
\eeq
where again $q'$ is a quark of flavour equal to or different from that of $q$.
The integration 
of $J_{\rm cc}^{\zg}$, $J_{\rm cc}^{(\rm 0g, id)}$, $J_{\rm cc}^{\og}$, 
$J_{\rm cc}^{\tg}$, $J_{\rm cc}^{(\rm 3g)}$ is the computation of the highest 
complexity among those presented in this paper. It can however be performed 
analytically following the procedure described in \secn{sec:int}.  The results are
\beq
J_{\rm cc}^{\zg}(s)
& = &
\bigg( \frac{\as}{2\pi} \bigg)^{\!2} 
\left( \frac{s}{\mu^2} \right)^{\!\!-2\eps} 
C_F T_R \,
\nnb\\
&&
\times \,
\bigg[
-
\frac{1}{3}  \, \frac1{\eps^3}
-
\frac{31}{18} \, \frac1{\eps^2}
-
\left( \frac{889}{108} - \frac{\pi^2}{2} \right) \frac1{\eps}
- \,
\frac{23941}{648} 
+
\frac{31}{12}\,\pi^2 
+ 
\frac{80}{9}\,\zeta_3
+
{\cal O}(\eps)
\bigg]
\, ,
\nnb\\[2mm]
J_{\rm cc}^{(\rm 0g, id)}(s)
& = &
\bigg( \frac{\as}{2\pi} \bigg)^{\!2} 
\left( \frac{s}{\mu^2} \right)^{\!\!-2\eps} 
C_F \, \big( 2 C_F -  C_A \big) \,
\nnb\\
&&
\times \,
\bigg[
- 
\left( \frac{13}{8} - \frac{1}{4}\,\pi^2 + \zeta_3 \right) 
\frac1{\eps}
- \,
\frac{227}{16}
+
\pi^2 
+ 
\frac{17}{2}\,\zeta_3
-
\frac{11}{120}\,\pi^4
+
{\cal O}(\eps)
\bigg]
\, ,
\nnb
\\[2mm]
J_{\rm cc}^{\og}(s)
& = &
\bigg( \frac{\as}{2\pi} \bigg)^{\!2} 
\left( \frac{s}{\mu^2} \right)^{\!\!-2\eps} 
\nnb\\
&&
\hspace{-2mm}
\times \,
\Bigg\{
C_F T_R \,
\bigg[
-
\frac{2}{3}  \, \frac1{\eps^3}
-
\frac{31}{9} \, \frac1{\eps^2}
-
\left( \frac{889}{54} - \pi^2 \right) \frac1{\eps}
- \,
\frac{23833}{324} 
+
\frac{31}{6}\,\pi^2 
+ 
\frac{160}{9}\,\zeta_3
+
{\cal O}(\eps)
\bigg]
\nnb\\
&&
+ \,
C_{\!A} T_R \,
\bigg[
-
\frac{4}{3}  \frac1{\eps^3}
- 
\frac{41}{6} \frac1{\eps^2}
-
\left( \frac{1675}{54} \! - \! \frac{17}{9}\pi^2 \right) \! \frac1{\eps}
-
\frac{10808}{81} 
+
\frac{1063}{108}\pi^2 
+ 
\frac{278}{9}\zeta_3
+
{\cal O}(\eps)
\bigg]
\Bigg\}
\, ,
\nnb
\eeq
\beq
\label{collresults}
J_{\rm cc}^{\tg}(s)
& = &
\bigg(\! \frac{\as}{2\pi} \!\bigg)^{\!\!2} \!\!
\left(\! \frac{s}{\mu^2} \!\right)^{\!\!\!-2\eps} \!
\Bigg\{
C_F^2\,
\bigg[
\;\;
\frac4{\eps^4}
+
\frac{14}{\eps^3}
+
\left( \frac{251}{4} - 6\,\pi^2 \right) \frac1{\eps^2}
+
\left( 
\frac{2125}{8} - 21\,\pi^2 - \frac{308}{3}\,\zeta_3 
\right) \frac1{\eps}
\nnb\\
&&
\hspace{30mm}
+ \,
\frac{17607}{16} 
-
\frac{753}{8}\,\pi^2 
- 
\frac{1096}{3}\,\zeta_3
+
\frac{13}{10}\,\pi^4 
+
{\cal O}(\eps)
\;
\bigg]
\nnb\\
&&
\hspace{10mm}
+ \,
C_F C_A\,
\bigg[
\;\;
\frac1{\eps^4}
+
\frac{16}{3} \, \frac1{\eps^3}
+
\left( \frac{905}{36} - \frac{4}{3}\,\pi^2 \right) \frac1{\eps^2}
+
\left( 
\frac{11773}{108} - \frac{89}{12}\,\pi^2 - \frac{65}{3}\,\zeta_3 
\right) 
\frac1{\eps}
\nnb\\
&&
\hspace{30mm}
+ \,
\frac{295789}{648} 
-
\frac{845}{24}\,\pi^2 
-
\frac{2191}{18}\,\zeta_3
+
\frac{19}{120}\,\pi^4
+
{\cal O}(\eps)
\;
\bigg]
\;
\Bigg\}
\, ,
\nnb\\[3mm]
J_{\rm cc}^{(\rm 3g)}(s)
& = &
\bigg(\! \frac{\as}{2\pi} \!\bigg)^{\!\!2} \!\!
\left(\! \frac{s}{\mu^2} \!\right)^{\!\!\!-2\eps} \!
C_A^2 \,
\bigg[
\;\;
\frac{15}{\eps^4}
+
\frac{63}{\eps^3}
+
\left( \frac{853}{3} - 22\,\pi^2 \right) \frac1{\eps^2}
+
\left( 
\frac{10900}{9} - \frac{275}{3}\,\pi^2 - 376\,\zeta_3 
\right) 
\frac1{\eps}
\nnb\\
&&
\hspace{26mm}
+ \,
\frac{180739}{36} 
-
\frac{3736}{9}\,\pi^2 
-
1555\,\zeta_3
+
\frac{41}{10}\,\pi^4
+
{\cal O}(\eps)
\;
\bigg]
\, .
\eeq
This completes the integration of the factorised kernels for tree-level double-unresolved
radiation. Once again, in order to build a complete subtraction procedure at NNLO,
one needs to consider both strongly-ordered and composite limits, mixing soft and 
collinear configurations. For such limits, it is important to find a consistent set of
phase-space mappings, which need to be mutually consistent when the relevant limits are
taken, in order to guarantee a local cancellation of singularities: a procedure to do so
is described in Ref.~\cite{Magnea:2018hab}. When it comes to the phase-space integration,
however, all the composite and strongly-ordered limits are either contained in 
the results we just stated, or lead to significantly simpler integrals. We have thus 
provided all the key ingredients necessary for the integration of local counterterms
(for massless final-state partons) at NNLO.

%%%%%%%%%%%%%%%%%%%%%%%%%

\subsection{Details of the integration procedure}
\label{sec:int}

%%%%%%%%%%%%%%%%%%%%%%%%%

\noindent
In this section we describe in detail how the integration of the kernels 
$J_{\rm s \otimes s}^{(2)}$, $J_{\rm ss}^{(\rm gg)}$, $J_{\rm ss}^{(\rm q \bar 
q)}$, $J_{\rm cc}^{\zg}$, $J_{\rm cc}^{(\rm 0g, id)}$, $J_{\rm cc}^{\og}$, 
$J_{\rm cc}^{\tg}$, $J_{\rm cc}^{(\rm 3g)}$ has been performed. We note
that the procedure we follow, while certainly non-trivial, does not require
the deployment of advanced techniques such as integration by parts
or the use of differential equations (see, for example,~\cite{Chetyrkin:1981qh,
Laporta:2001dd,Smirnov:2006ry,Henn:2013pwa,Henn:2014qga,Anastasiou:2002yz,Anastasiou:2003yy}): in this 
sense our method, at NNLO, allows for a complete analytic integration
of all subtraction counterterms, by means of relatively simple tools.

The integration procedure is simplified by a careful analysis of
the symmetries of the relevant integrals under exchanges of particle
labels. When integrating in the two-body radiative phase space 
$d \Phi_{\rm rad, 2}^{\,(abcd)}$, the freedom in choosing $k_a$, $k_b$, 
$k_c$, $k_d$ does not stem from the symmetries of the kernel itself, but 
from those of the four-body phase space. In particular, following 
Ref.~\cite{Gehrmann-DeRidder:2003pne}, we note that the four-body 
phase space for momenta $k_a$, $k_b$, $k_c$, $k_d$ is symmetric 
under the permutation of the four momenta, as well as under the following 
permutations of Mandelstam invariants: 
\beq
s_{ab} \leftrightarrow s_{cd} \, ,
\qquad \,
s_{ac} \leftrightarrow s_{bd} \, ,
\qquad \,
s_{ad} \leftrightarrow s_{bc} \, .
\label{eq:symmetry sij}
\eeq
These symmetries are reflected in our parametrisations of phase space, 
in particular when moving from the set $\{k_a, k_b, k_c, k_d \}$ to the set
$\{\kk{c}{abcd}, \kk{d}{abcd}, y,z,\phi,y',z',w'\}$, and this is crucial to 
simplify the analytic integration. 

In order to exploit these symmetries for the integration of the soft and collinear 
kernels, after assigning the momenta $k_a$, $k_b$, $k_c$, $k_d$ according to 
the discussion of \secn{sec:mappings2}, we apply the following transformations: 
\begin{itemize}
\item
in the terms containing $1/(s_{ad}+s_{bd})/(s_{ad}+s_{cd})$, all 
permutations of the invariants
$
s_{ab} \leftrightarrow s_{cd},
\, 
s_{ac} \leftrightarrow s_{bd},
\,
s_{ad} \leftrightarrow s_{bc}
$ 
are performed;
\item 
in the terms containing $1/(s_{ad}+s_{cd})$ (but not $1/(s_{ad}+s_{bd})$), 
the permutation $k_b \leftrightarrow k_c$ is performed;
\item 
in the terms containing $1/(s_{bd}+s_{cd})$ (but not $1/(s_{ad}+s_{bd})$), 
the permutation $k_a \leftrightarrow k_c$ is performed;
\item
in the terms containing $1/(s_{ad} \, s_{bd})$, the partial fractioning
\beq
\frac1{s_{ad} \, s_{bd}} \, = \,
\frac1{s_{ad}+s_{bd}}\,\left( \frac1{s_{ad}} + \frac1{s_{bd}} \right)
\eeq
is performed, and in the first term the permutation $k_a \leftrightarrow k_b$ 
is applied. 
\item
in the terms containing $1/s_{ad}$ (but not $1/s_{bd}$), the permutation 
$k_a \leftrightarrow k_b$ is performed. 
\end{itemize}
After these transformations, the denominators of all integrands feature only the 
following combinations of invariants:
\beq
\label{eq:denom_list}
s_{ab} \, , \quad s_{ac} \, , \quad s_{bc} \, , \quad s_{cd} \, , \quad s_{bd} \, , 
\quad 
s_{ac}+s_{bc} \, , \quad s_{ad}+s_{bd} \, , \quad s_{ab}+s_{bc} \, ,
\eeq
and they can be parametrised as (see \eq{invaproco}) 
\beq
s_{ab} & = & y' \, y \, s_{abcd} \, ,
\nnb \\
s_{ac} & = & z' ( 1 - y' ) \, y \, s_{abcd} \, ,
\nnb \\
s_{bc} & = & ( 1 - y' ) ( 1  - z' ) \, y \, s_{abcd} \, ,
\nnb \\
s_{cd} & = & ( 1 - y' ) ( 1 - y ) ( 1 - z ) \, s_{abcd} \, ,
\nnb \\
s_{bd} 
& = & 
(1-y) \left[\,
y'z'(1-z) + (1-z')z + 2(1-2w')\sqrt{y'z'(1-z')z(1-z)} 
\,\right] s_{abcd} \, ,
\nnb\\
s_{ac}+s_{bc} &=& (1-y')\,y\,s_{abcd} \, , 
\nnb\\
s_{ad}+s_{bd} &=& (y'+z-y'\,z)\,(1-y)\,s_{abcd} \, , 
\nnb\\ 
s_{ab}+s_{bc} &=& (1-z'+z'y')\,y\,s_{abcd} \, .
\label{invaproco_text}
\eeq
We now detail the integration procedure, focusing on one variable at a time. 
In \secn{subsec:yxp} we analyse the trivial integration over $y$, and the 
first non-trivial structure arising from the $w'$ integration. Then, the subsequent 
integrations over $z$ and $z'$ are detailed in \secn{subsec:zzp}, including a 
discussion on how we linearise the argument of the resulting hypergeometric 
functions. Finally, \secn{subsec:ypt} concerns the $\eps$-expansion 
of intermediate results, and the last integration step. 

%%%%%%%%%%%%

\subsubsection{Integration on $y$ and on the azimuthal variable $w'$}
\label{subsec:yxp}

%%%%%%%%%%%%

\noindent
Since in all denominators in the list (\ref{eq:denom_list})
the dependence on $y$ is factorised, the integration in 
the $y$ variable is always of the form 
\beq\label{eq:integy}
\int_0^1 \!\!\! d y \,
\Big[ y\,(1-y) \Big]^{1-2\eps} \, y^n\,(1-y)^m \, ,
\qquad
n,m \in \mathbb{Z} \, ,
\eeq
which clearly gives $B\!\left(n+2-2\eps,m+2-2\eps\right)$. 

We now switch to the integration over the azimuthal variable $w'$. According to 
\eq{invaproco_text}, the only denominator containing the azimuthal variable $w'$ 
is $s_{bd}$, while the presence of the $w'$ in the numerator uniquely stems from 
linear combinations of $s_{ad}$ and $s_{bd}$, see \eq{invaproco}. Thus, terms 
without $s_{bd}$ in the denominator are of the form 
\beq
\label{altint}
\int_0^1 \!\!\! d w' \, 
\big[ w'(1-w')\big]^{-\frac12-\eps}
(1-2w')^n
\, = \,
2^{- 1+ 2 \eps} \left( 1 + (-1)^n \right) 
B \!\left( \frac{1}{2} - \eps, \frac{1 + n}{2} \right) \, , 
\qquad
n \in \mathbb{N} \, .
\quad
\eeq
Terms containing the ratio $s_{ad}/s_{bd}$ can be simplified according to
\beq
\frac{s_{ad}}{s_{bd}} \, = \, 
\frac{s_{ad}+s_{bd}}{s_{bd}} - 1 = 
(y'+z-y'\,z)\,(1-y)\,\frac{s_{abcd}}{s_{bd}} - 1 \, ;
\eeq
therefore, no dependence on $w'$ in the numerator is left in the presence of 
the denominator $s_{bd}$. The only non-trivial integration involving the azimuthal 
variable $w'$ is then
\beq
\int_0^1 \!\!\! d w' \, 
\big[ w'(1-w')\big]^{-\frac12-\eps}\,
\frac{s_{abcd}}{s_{bd}}
& = &
\frac{1}{1-y}
\int_0^1 \!\!\! d w' \, 
\big[ w'(1-w')\big]^{-\frac12-\eps} \, 
\frac{1}{(A+B)^2-4ABw'}
\nnb\\
& \equiv & 
\frac{1}{1-y} \, I_{w'} \, ,
\eeq
with $A = \sqrt{y'z'(1-z)}$ and $B = \sqrt{z(1-z')}$. 
Note that, as already discussed at the beginning of this section (see 
\eq{eq:integy}), the $y$ dependence is trivially factorised. Therefore, from 
now on, we understand the $y$ dependence to be integrated out. 

The integral $I_{w'}$ is of the type described in \appn{app:Iw1}, with $a=1$ and 
$b=1+\eps$. From \eq{eq:I1bAB} we get then
\beq
\hspace{-5mm}
I_{w'}
& = & 
I_{1+\eps} \Big( \sqrt{y'z'(1-z)},  \sqrt{z(1-z')} \Big)
\nnb\\
&=& 
\frac{\Gamma^2(1/2-\eps)}{\Gamma(1-2\eps)}\,
\Bigg[
\;
\frac{1}{z(1-z')} \,\, 
{}_2F_1\bigg(1,1+\eps,1-\eps,\frac{y'z'(1-z)}{z(1-z')}\bigg)\,
\Theta\!\left(1-\frac{y'z'(1-z)}{z(1-z')}\right)
\nnb\\
&&
\hspace{18mm}
+ \,
\frac{1}{y'z'(1-z)} \,\,
{}_2F_1\bigg(1,1+\eps,1-\eps,\frac{z(1-z')}{y'z'(1-z)}\bigg)\,
\Theta\!\left(\frac{y'z'(1-z)}{z(1-z')}-1\right)
\Bigg].
\label{eq:Iw'}
\eeq

%%%%%%%%%%%%

\subsubsection{Integration of the variables $z$ and $z'$}
\label{subsec:zzp}

%%%%%%%%%%%%

\noindent
After integrating over $y$ and $w'$, one is left with three integrations over the 
variables $z, \, z'$ and $y'$. We now analyse the $z$ and the $z'$ integrations. 
While all numerators have a polynomial dependence on $z, z'$, the  denominators
manifest a richer structure. In particular,
\begin{itemize}
\item 
the invariants $s_{ab}$, $s_{ac}$, $s_{bc}$, $s_{cd}$, $s_{ac}+s_{bc}$ feature 
a trivial dependence on $z'$ and $z$, as they are just products of powers of 
$z'$, $(1-z')$, $z$ and $(1-z)$;
\item 
the structure $s_{ad}+s_{bd}$ does not depend on $z'$, while it depends on 
$z$ through the factor $y'+z-y' z$; analogously, $s_{ab}+s_{bc}$ depends only 
on $z'$, through the factor $1-z'+z'y'$; 
\item 
when the denominator is $s_{bd}$, the $z,z'$ dependence is confined to the 
arguments and the prefactors of the hypergeometric functions of in \eq{eq:Iw'}, 
as well as in the accompanying $\Theta$ functions; the latter are to be
understood as constraints on the integration region for either $z$ or $z'$. 
\end{itemize}
The soft and collinear kernels feature products of the invariant structures
described above. Among them, a non-trivial dependence on $z$ and $z'$ arises
from the following building blocks:
\beq
&& 
\frac{1}{y'+z-y'\,z} 
\, , 
\qquad
\frac{1}{1-z'+z'y'} 
\, , 
\qquad
\frac{I_{w'}}{y'+z-y'\,z}
\, ,
\qquad
\frac{I_{w'}}{1-z'+z'y'}
\, , 
\qquad
I_{w'}
\, .
\label{eq:zz' structures}
\eeq
In contributions proportional to the first structure in \eq{eq:zz' structures}, the $z'$ 
integration gives Beta functions, while the $z$ integration takes the form
\beq
\int_0^1 d z \, 
\frac{z^{n-\eps} (1-z)^{m-\eps}}{y'+z-y'\,z}
\, = \, 
B(n+1-\eps,m+1-\eps)\,
{}_2F_1\left(1,m+1-\eps,n+m+2-2\eps,1-y'\right)
\, ,
\nnb\\
\eeq
where we used ${}_2F_1(a,b,c,x) = (1-x)^{-a}{}_2F_1(a,c-b,c,-x/(1-x))$. Note that 
$m,n$ stand for generic powers of $z$, arising from the numerators. Similarly, in 
terms that embed the second structure in \eq{eq:zz' structures}, the $z$ integration 
is trivial (Beta functions), while the $z'$ integration takes the form 
\beq
\int_0^1 d z' \, 
\frac{ (z')^{n-\eps} (1-z')^{m-\eps}}{1-z'+z'y'}
\, = \, 
B(n+1-\eps,m+1-\eps)\,
{}_2F_1\left(1,n+1-\eps,n+m+2-2\eps,1-y'\right)
\, .
\nnb\\
\eeq
In the third (fourth) structure of \eq{eq:zz' structures} the whole $z'$ ($z$) 
dependence is contained in $I_{w'}$, and this variable is integrated first.
Finally, in the fifth structure of \eq{eq:zz' structures}, where no denominator 
depends on $z$ nor on $z'$, the order of integration of $z$ and  $z'$ is irrelevant. 
Accounting for generic numerators, whose dependence upon $z$ and $z'$ is 
polynomial, we can cast all integrals to be performed as combinations of the 
following building blocks\footnote{
In some cases it is necessary to apply the partial fractioning
\[
\frac{1}{z(1-z)} = \frac1z + \frac1{1-z}\,.
\]}:
\beq
&&
I_{w'z}^{(n)} 
\, = \, 
\int_0^1 \!\!\! d z \, 
\Big[ z(1-z) \Big]^{- \eps} (1-z)^n\,I_{w'}
\, ,
\qquad
J_{w'z}^{(n)}
\, = \, 
\int_0^1 \!\!\! d z \, 
\Big[ z(1-z) \Big]^{- \eps} z^n\,I_{w'}
\, ,
\label{eq:w'z}
\\
&&
I_{w'z'}^{(n)}
\, = \, 
\int_0^1 \!\!\! d z' \, 
\Big[ z'(1-z') \Big]^{- \eps} (z')^n\,I_{w'}
\, ,
\qquad
J_{w'z'}^{(n)} 
\, = \, 
\int_0^1 \!\!\! d z' \, 
\Big[ z'(1-z') \Big]^{- \eps} (1-z')^n\,I_{w'}
\, ,\label{eq:w'z'}
\eeq
where $n$ is an integer such that $n \ge -1$.

Because of the symmetries of $I_{w'}$ upon $z\leftrightarrow 1-z'$,
the results for $I_{w'z'}^{(n)}$ and $J_{w'z'}^{(n)}$ can be inferred
from those for $I_{w'z}^{(n)}$ and $J_{w'z}^{(n)}$, respectively. 
We then proceed with the computation of the latter two integrals,
which are  of the type described in \eq{eq:appIv} of \appn{app:Iab2}
with $b=1+\eps$. Specifically
\beq
I_{w'z}^{(n)} 
& = &
\int_0^1 \!\!\! d z \, 
(z)^{-\eps}(1-z)^{n-\eps} \,
I_{1+\eps}(A,B)
\,\, = \,\, 
I_{1+\eps,-\eps,n-\eps}(1-z',y'z')
\, ,
\nnb\\
J_{w'z}^{(n)}
& = &
\int_0^1 \!\!\! d z \, 
(z)^{n-\eps}(1-z)^{- \eps} \,
I_{1+\eps}(A,B)
\,\, = \,\,
I_{1+\eps,n-\eps,-\eps}(1-z',y'z')
\, .
\eeq
We see that  the integral $I_{1+\eps,-\eps,n-\eps}(1-z',y'z')$ is of the 
special type $I_{b,1-b,\gamma}(C,D)$ described in \eq{eq:Iv s1}, 
while the integral $I_{1+\eps,n-\eps,-\eps}(1-z',y'z')$ is of the special 
type $I_{b,\beta,1-b}(C,D)$ described in \eq{eq:Iv s2}. 
Using the results derived there we have
\beq
I_{w'z}^{(n)} 
& = &
\frac{1}{1-z'} \,
\frac{\Gamma^2(1/2-\eps)}{\Gamma(1-2\eps)}\,
\frac{\Gamma(-\eps)\Gamma(n\!+\!1\!-\!\eps)}{\Gamma(n+1-2\eps)} \,\,
{}_2F_1\bigg(1,n\!+\!1\!-\!\eps,1\!-\!\eps,-\frac{y'z'}{1\!-\!z'}\bigg)
\, ,
\nnb\\
J_{w'z}^{(n)} 
& = &
\frac{1}{y'z'} \,
\frac{\Gamma^2(1/2-\eps)}{\Gamma(1-2\eps)}\,
\frac{\Gamma(-\eps)\Gamma(n\!+\!1\!-\!\eps)}{\Gamma(n+1-2\eps)} \,\,
{}_2F_1\bigg(1,n\!+\!1\!-\!\eps,1\!-\!\eps,-\frac{1\!-\!z'}{y'z'}\bigg)
\, .
\quad\label{eq:Iwpzp1}
\eeq
We now show the result for specific values of $n$, and in particular we distinguish 
between $n = -1$ and $n \ge 0$. For $n = -1$, \eq{eq:Iwpzp1} reads
\beq
\label{eq:Iwpzp2}
I_{w'z}^{(-1)} 
& = &
\frac{1}{1-z'}\,
\frac{\Gamma^2(1/2-\eps)}{\Gamma(1-2\eps)}\,
\frac{\Gamma^2(-\eps)}{\Gamma(-2\eps)} \, \,
{}_2F_1\bigg(1,-\eps,1-\eps,-\frac{y'z'}{1-z'}\bigg)
\, ,
\\
J_{w'z}^{(-1)} 
& = &
\frac{1}{y'z'}\,
\frac{\Gamma^2(1/2-\eps)}{\Gamma(1-2\eps)}\,
\frac{\Gamma^2(-\eps)}{\Gamma(-2\eps)} \,\, 
{}_2F_1\bigg(1,-\eps,1-\eps,-\frac{1-z'}{y'z'}\bigg)
\nnb\\
& = &
- \,
\frac{\Gamma^2(1/2-\eps)}{\Gamma(1-2\eps)}\,
\frac{\Gamma(-\eps)\Gamma(1-\eps)}{\Gamma(-2\eps)}\,
\bigg[
\frac{1}{1+\eps}\,
\frac{1}{1\!-\!z'} \,\,
{}_2F_1\bigg(1,1+\eps,2+\eps,-\frac{y'z'}{1\!-\!z'}\bigg)
\nnb\\
&&
\hspace{47mm}
- \,
\Gamma(1+\eps)\Gamma(-\eps) \, \frac{(1-z')^{\eps}}{(y'z')^{1+\eps}}
\bigg]
\, ,
\nnb
\eeq
where in the second integral of \eq{eq:Iwpzp2}, we have inverted the 
argument of the hypergeometric function by means of \eq{eq:invF}. 
For $n \ge 0$ the hypergeometric functions are of the class 
${}_2F_1(1,c+n,c,x)$, with $c=1-\eps$, and can therefore be written
as a finite sum in the form
\beq
{}_2F_1(1,c+n,c,x) 
& = & 
(c-1)
\sum_{k=0}^{n}
\frac{\Gamma(n+1)\Gamma(c+n-k-1)}{\Gamma(n-k+1)\Gamma(c+n)}\,
\frac{1}{(1-x)^{k+1}}
\, ,
\qquad
n \ge 0
\, .
\label{eq:F(1,c+n,c,x}
\eeq
The integrals of \eq{eq:Iwpzp1} for $n\geq 0$ are then given by 
\beq
I_{w'z}^{(n)} 
& = &
\frac{\Gamma^2(1/2-\eps)}{\Gamma(1-2\eps)}\,
\frac{\Gamma(1\!-\!\eps)\Gamma(n\!+\!1)}{\Gamma(n+1-2\eps)}\,
\sum_{k=0}^{n}\,
\frac{\Gamma(n\!-\!k\!-\!\eps)}{\Gamma(n\!-\!k\!+\!1)}\,
\frac{(1-z')^k}{(1-z'+z'y')^{k+1}}
\, ,
\quad
n \ge 0
\, ,
\nnb\\
J_{w'z}^{(n)}
& = &
\frac{\Gamma^2(1/2-\eps)}{\Gamma(1-2\eps)}\,
\frac{\Gamma(1\!-\!\eps)\Gamma(n\!+\!1)}{\Gamma(n+1-2\eps)}\,
\sum_{k=0}^{n}\,
\frac{\Gamma(n\!-\!k\!-\!\eps)}{\Gamma(n\!-\!k\!+\!1)}\,
\frac{(y')^k(z')^k}{(1-z'+z'y')^{k+1}}
\, ,
\quad
n \ge 0
\, .
\label{eq:n>0}
\eeq
Notice that the two integrals coincide for $n=0$, evaluating  to
\beq
I_{w'z}^{(0)} 
& = &
J_{w'z}^{(0)} 
\; = \;
\frac12 \,
\frac{\Gamma^2(1/2-\eps)}{\Gamma(1-2\eps)}\,
\frac{\Gamma^2(-\eps)}{\Gamma(-2\eps)}\,
\frac1{1-z'+z'y'}
\, .
\quad
\eeq
After the first $z$ ($z'$) integration has been performed, all non-trivial dependence 
on the remaining $z'$ ($z$) variable is encoded in one of the following structures: 
\beq
I_{w'zz'}^{(n,p,q,m)} 
& \equiv &
\int_0^1 \! d z' \, 
\frac{(1\!-\!z')^{p-\eps} (z')^{q-\eps}}{(1\!-\!z'\!+\!z'y')^m} \, I_{w'z}^{(n)}
\nnb\\
& = &
\int_0^1 \! d z \, 
\int_0^1 \! d z' \, 
\frac{(1\!-\!z')^{p-\eps} (z')^{q-\eps}}{(1\!-\!z'\!+\!z'y')^m} \, 
\Big[ z(1\!-\!z) \Big]^{- \eps} (1\!-\!z)^n \,
I_{w'}
\, ,
\nnb\\[2mm]
J_{w'zz'}^{(n,p,q,m)} 
& \equiv &
\int_0^1 \! d z' \, 
\frac{(1\!-\!z')^{p-\eps} (z')^{q-\eps}}{(1\!-\!z'\!+\!z'y')^m} \, J_{w'z}^{(n)}
\nnb\\
& = &
\int_0^1 \! d z \, 
\int_0^1 \! d z' \, 
\frac{(1\!-\!z')^{p-\eps} (z')^{q-\eps}}{(1\!-\!z'\!+\!z'y')^m} \,
\Big[ z(1\!-\!z) \Big]^{- \eps} z^n \,
I_{w'}
\, ,
\label{eq:w'zz'}
\eeq
where the integers $n,p,q,m$ are such that $n,p,q\ge-1$, while $m=0,1$. 
For later convenience, we recursively use the following partial fractioning
\beq
z^{p-\eps} (1-z)^{q-\eps}
& = &
z^{p+1-\eps} (1-z)^{q-\eps} + z^{p-\eps} (1-z)^{q+1-\eps}
\, ,
\nnb\\
(z')^{p-\eps} (1-z')^{q-\eps}
& = &
(z')^{p+1-\eps} (1-z')^{q-\eps} + (z')^{p-\eps} (1-z')^{q+1-\eps}
\, ,
\label{eq:p+q>m-1}
\eeq
until the condition $p+q\ge m$ is satisfied. 

Using the symmetry under the exchange $z \leftrightarrow 1-z'$, the 
integrals $I_{w'zz'}^{(n,p,q,m)}$, $J_{w'zz'}^{(n,p,q,m)} $ can equivalently 
be written in terms of $I_{w'z'}^{(n)}$ and $J_{w'z'}^{(n)}$, as 
\beq
I_{w'zz'}^{(n,p,q,m)} 
& \equiv &
\int_0^1 \! d z \, 
\frac{z^{p-\eps} (1\!-\!z)^{q-\eps}}{(y'\!+\!z\!-\!y'z)^m} \, I_{w'z'}^{(n)}
\nnb\\
& = &
\int_0^1 \! d z \, 
\int_0^1 \!  d z' \, 
\frac{z^{p-\eps} (1\!-\!z)^{q-\eps}}{(y'\!+\!z\!-\!y'z)^m} \, 
\Big[ z'(1\!-\!z') \Big]^{- \eps} (z')^n \,
I_{w'}
\, ,
\nnb\\[2mm]
J_{w'zz'}^{(n,p,q,m)} 
& \equiv &
\int_0^1 \! d z \, 
\frac{z^{p-\eps} (1\!-\!z)^{q-\eps}}{(y'\!+\!z\!-\!y'z)^m} \, J_{w'z'}^{(n)}
\nnb\\
& = &
\int_0^1 \! d z \, 
\int_0^1 \! d z' \, 
\frac{z^{p-\eps} (1\!-\!z)^{q-\eps}}{(y'\!+\!z\!-\!y'z)^m} \, 
\Big[ z'(1\!-\!z') \Big]^{- \eps} (1\!-\!z')^n \,
I_{w'}
\, .
\eeq
To proceed with the computation, we choose the representation of 
$I_{w'zz'}^{(n,p,q,m)}$, $J_{w'zz'}^{(n,p,q,m)}$ in terms of $I_{w'z}^{(n)}$ 
and $J_{w'z}^{(n)}$, according to \eq{eq:w'zz'}. Thanks to the results in
\eq{eq:n>0}, the case $n\ge0$ is trivial, and yields
\beq
I_{w'zz'}^{(n,p,q,m)} 
& = &
\frac{\Gamma^2(1/2-\eps)}{\Gamma(1-2\eps)}\,
\frac{\Gamma(1\!-\!\eps)\Gamma(n\!+\!1)}{\Gamma(n+1-2\eps)}\,
\sum_{k=0}^{n}\,
\frac{\Gamma(n\!-\!k\!-\!\eps)}{\Gamma(n\!-\!k\!+\!1)}\,
\frac{\Gamma(p\!+\!k\!+\!1\!-\!\eps)\Gamma(q\!+\!1\!-\!\eps)}
     {\Gamma(p\!+\!q\!+\!k\!+\!2\!-\!2\eps)}\,
\nnb\\
&&
\,\times\,\,\,{}_2F_1\left(
       m\!+\!k\!+\!1,
       q\!+\!1\!-\!\eps,
       p\!+\!q\!+\!k\!+\!2\!-\!2\eps,
       1\!-\!y'
       \right)
\, ,
\hspace{18mm}
n \ge 0
\, ,
\nnb\\[0.3cm]
J_{w'zz'}^{(n,p,q,m)} 
& = &
\frac{\Gamma^2(1/2-\eps)}{\Gamma(1-2\eps)}\,
\frac{\Gamma(1\!-\!\eps)\Gamma(n\!+\!1)}{\Gamma(n+1-2\eps)}\,
\sum_{k=0}^{n}\,
\frac{\Gamma(n\!-\!k\!-\!\eps)}{\Gamma(n\!-\!k\!+\!1)}\,
\frac{\Gamma(p\!+\!1\!-\!\eps)\Gamma(q\!+\!k\!+\!1\!-\!\eps)}
     {\Gamma(p\!+\!q\!+\!k\!+\!2\!-\!2\eps)}\,
\nnb\\
&&
\,\times\,\,\,(y')^{k}\,
{}_2F_1\left(
       m\!+\!k\!+\!1,
       q\!+\!k\!+\!1\!-\!\eps,
       p\!+\!q\!+\!k\!+\!2\!-\!2\eps,
       1\!-\!y'
       \right)
\, ,
\hspace{5mm}
n \ge 0
\, .
\eeq
For $n=-1$, we exploit the integral representation of the hypergeometric 
functions in \eq{eq:Iwpzp2}, introducing the auxiliary integration variable 
$t$, and write
\beq
I_{w'zz'}^{(-1,p,q,m)} 
& = &
\frac{\Gamma^2(1/2-\eps)}{\Gamma(1-2\eps)}\,
\frac{\Gamma(-\eps)\Gamma(1-\eps)}{\Gamma(-2\eps)}\,
\int_0^1 \!\!\! d z' \! 
\int_0^1 \!\!\! d t \, 
\frac{(1-z')^{p-\eps}(z')^{q-\eps}}{(1-z'+z'y')^m} \, 
\frac{t^{-1-\eps}}{1-z'+tz'y'}
\, ,
\nnb\\
J_{w'zz'}^{(-1,p,q,m)} 
& = &
- \,
\frac{\Gamma^2(1/2-\eps)}{\Gamma(1-2\eps)}\,
\frac{\Gamma(-\eps)\Gamma(1-\eps)}{\Gamma(-2\eps)}\,
\int_0^1 \!\!\! d z' \, \frac{(1-z')^{p} (z')^{q-\eps}}{(1\!-\!z'\!+\!z'y')^m}
\nnb\\
&&
\hspace{35mm}
\times \, 
\Bigg[
\int_0^1 \! d t \,\,
\frac{(1-z')^{-\eps} \, t^{\eps}}{1\!-\!z'\!+\!tz'y'} -
\frac{\Gamma(1+\eps)\Gamma(-\eps)}{(z' y')^{1+\eps}} \, 
\Bigg]
\, .
\label{eq:neqminusone}
\eeq
The second expression makes sense only if $p\ge0$, which is the case in 
all soft and collinear kernels. For $m=0$, the $z'$ integration gives 
\beq
I_{w'zz'}^{(-1,p,q,0)}
& = &
\frac{\Gamma^2(1/2-\eps)}{\Gamma(1-2\eps)}\,
\frac{\Gamma(-\eps)\Gamma(1-\eps)}{\Gamma(-2\eps)}\,
\frac{\Gamma(p\!+\!1\!-\!\eps)\Gamma(q\!+\!1\!-\!\eps)}
     {\Gamma(p\!+\!q\!+\!2\!-\!2\eps)}
\\
&&
\hspace{25mm}
\times \, 
\int_0^1 \!\!\! d t \, 
t^{-1-\eps} \,
{}_2F_1\left(1,q+1-\eps,p+q+2-2\eps,1-ty'\right)
\, ,
\nnb\\
J_{w'zz'}^{(-1,p,q,0)}
& = &
- \,
\frac{\Gamma^2(1/2-\eps)}{\Gamma(1-2\eps)}\,
\frac{\Gamma(-\eps)\Gamma(1-\eps)}{\Gamma(-2\eps)}\,
\bigg[
  - \,
\frac{\Gamma(1+\eps)\Gamma(-\eps)}{(y')^{1+\eps}} \,
\frac{\Gamma(p+1)\Gamma(q-2\eps)}{\Gamma(p+q+1-2\eps)}\,
\quad
\nnb\\
&&
+ \,
\frac{\Gamma(p+1-\eps)\Gamma(q+1-\eps)}{\Gamma(p+q+2-2\eps)}\,
\int_0^1 \!\!\! d t \, 
t^{\eps} \,
{}_2F_1\left(1,q\!+\!1\!-\eps,p+q+2\!-\!2\eps,1\!-ty'\right)
\bigg]
\, ,
\quad
p\ge0
\, .
\nnb
\eeq
For $m=1$, before performing the remaining $z'$ integration, we employ the 
partial fractioning
\beq
\frac{1}{1-z'+z'y'} \, \frac{1}{1-z'+tz'y'}
& = &
\frac{1}{1-t}\,\frac{1}{y'z'} \,
\bigg[ \frac{1}{1-z'+tz'y'} - \frac{1}{1-z'+z'y'} \bigg]
\, .
\eeq
with the results
\beq
I_{w'zz'}^{(-1,p,q,1)}
& = &
\frac{1}{y'}\,
\frac{\Gamma^2(1/2-\eps)}{\Gamma(1-2\eps)}\,
\frac{\Gamma(-\eps)\Gamma(1-\eps)}{\Gamma(-2\eps)}\,
\frac{\Gamma(p+1-\eps)\Gamma(q-\eps)}{\Gamma(p+q+1-2\eps)}\,
\int_0^1 \!\!\! d t \, 
\frac{t^{-1-\eps}}{1-t} \,
\\
&&
\times \bigg[
{}_2F_1\left(1,q\!-\!\eps,p\!+\!q\!+\!1\!-\!2\eps,1-ty'\right)
-
{}_2F_1\left(1,q\!-\!\eps,p\!+\!q\!+\!1\!-\!2\eps,1-y'\right)
\bigg]
\, ,
\nnb\\
J_{w'zz'}^{(-1,p,q,1)}
& = &
- \,
\frac{1}{y'}\,
\frac{\Gamma^2(1/2-\eps)}{\Gamma(1-2\eps)}\,
\frac{\Gamma(-\eps)\Gamma(1-\eps)}{\Gamma(-2\eps)}\,\,
\bigg\{
\frac{\Gamma(p+1-\eps)\Gamma(q-\eps)}{\Gamma(p+q+1-2\eps)} 
\nnb\\
&&
\times
\int_0^1 \! d t \, 
\frac{t^{\eps}}{1-t} \,
\bigg[
{}_2F_1\left(1,q\!-\!\eps,p\!+\!q\!+\!1\!-\!2\eps,1-ty'\right)
-
{}_2F_1\left(1,q\!-\!\eps,p\!+\!q\!+\!1\!-\!2\eps,1-y'\right)
\bigg]
\nnb\\
&&
- \,
\frac{\Gamma(1+\eps)\Gamma(-\eps)}{(y')^{\eps}} \,
\frac{\Gamma(p+1)\Gamma(q-2\eps)}{\Gamma(p\!+\!q\!+\!1\!-\!2\eps)}\,
{}_2F_1\left(1,q\!-\!2\eps,p\!+\!q\!+\!1\!-\!2\eps,1-y'\right)
\bigg\}
\, ,\quad p\ge0\,.
\nnb
\eeq
Summarising, for $n \ge 0$ we still have to perform the last integration over 
the $y'$ variable. Conversely, for the case $n=-1$, we are left with two integrations, 
one over the physical variable $y'$, the other over the auxiliary variable $t$ 
stemming from the integral representation of hypergeometric functions. Notice
that, so far, all our results are exact in $\epsilon$: only while performing these 
last steps we resort to an expansion\footnote{Formally, one could give exact 
results in terms of the hypergeometric functions ${}_3F_2$ and ${}_4F_3$ 
evaluated at unit argument.} in powers of $\epsilon$.

%%%%%%%%%%%%

\subsubsection{Expansion in $\eps$ and integration of the 
$y'$ and $t$ variables}
\label{subsec:ypt}

%%%%%%%%%%%%

\noindent
After the $y$, $w'$, $z$ and $z'$ integrations have been performed 
following the steps detailed in the previous sections, the integrations
over $y'$ and $t$ only involve monomials $y'$, $(1-y')$, $t$, $(1-t)$,
and hypergeometric functions of the types
\beq
\label{hyperone}
&&
\hspace{-3mm}
{}_2F_1(n_1,n_2-\eps,n_3-2\eps,1-\omega),
\qquad
n_1 \ge 1,
\quad
n_2 \ge 0,
\quad 
n_3 \ge n_1+1,n_2,
\qquad
\omega = ty', y'
\, ,
\nnb\\
&&
\hspace{-3mm}
{}_2F_1(1,n_2-2\eps,n_3-2\eps,1-\omega),
\qquad
n_2 \ge 0,
\quad 
n_3 \ge n_2 + 1,
\qquad
\omega = y'
\, .
\eeq
For the first type the constraint $n_3 \ge n_1+1$ is always achieved, 
thanks to the condition $p+q\ge m$, which comes from the partial 
fractioning described in \eq{eq:p+q>m-1}. 

We first manipulate these hypergeometric functions by means of the 
identity 
\beq
\hspace{-3mm}
{}_2F_1(a,b,c,x) = 
(1-x)^{c-b-a}\,{}_2F_1(c-a,c-b,c,x) =
(1-x)^{c-b-a}\,{}_2F_1(c-b,c-a,c,x)
\, ,
\eeq
to get 
\beq
\hspace{-3mm}
{}_2F_1(n_1,n_2\!-\!\eps,n_3\!-\!2\eps,1\!-\!\omega)
& = &
(\omega)^{n_3-n_2-n_1-\eps}\,\,
{}_2F_1(n_3\!-\!n_2\!-\!\eps,n_3\!-\!n_1\!-\!2\eps,n_3\!-\!2\eps,1\!-\!\omega)
\,,
\nnb\\
\hspace{-3mm}
{}_2F_1(1,n_2\!-\!2\eps,n_3\!-\!2\eps,1\!-\!\omega)
& = &
(\omega)^{n_3-n_2-1}\,\,
{}_2F_1(n_3\!-\!n_2,n_3\!-\!1\!-\!2\eps,n_3\!-\!2\eps,1\!-\!\omega)
\,.
\label{eq:intermediate1}
\quad
\eeq
Since $n_3\ge n_1+1$, the first hypergeometric function 
${}_2F_1(n_3\!-\!n_2\!-\!\eps,n_3\!-\!n_1\!-\!2\eps,n_3\!-\!2\eps,1\!-\!\omega)$ 
in \eq{eq:intermediate1} is of the type ${}_2F_1(a,b,b+n,x)$, and can be 
treated recursively using the relation 
\beq
\hspace{-3mm}
{}_2F_1(a,b,b+n,x)
& = &
\frac{1}{n-1}\,\Big[
(b+n-1) \, {}_2F_1(a,b,b+n-1,x)
-
b \, {}_2F_1(a,b+1,b+n,x)
\Big]
\, ,
\eeq
until it reduces to a hypergeometric function of the type 
${}_2F_1(a,b,b+1,x)$, with $a=m_1-\eps$, $b=m_2-2\eps$ ($m_1,m_2\ge0$). 
The other hypergeometric function 
${}_2F_1(n_3\!-\!n_2,n_3\!-\!1\!-\!2\eps,n_3\!-\!2\eps,1\!-\!\omega)$ 
in \eq{eq:intermediate1} is already of the type ${}_2F_1(a,b,b+1,x)$, 
but with $a=m_1+1$, $b=m_2-2\eps$ ($m_1,m_2\ge0$). 
We then make use of the relations (properly combined) 
\beq
{}_2F_1(a,b,b+1,x)
& = &
\frac{b}{a-1}\,\frac{1}{x}\,\Big[
(1-x)^{1-a} - {}_2F_1(a-1,b-1,b,x)
\Big]
\, ,
\nnb\\
{}_2F_1(a,b,b+1,x)
& = &
\frac{1}{a-1}\,\Big[
b\,(1-x)^{1-a} + (a-b-1)\,{}_2F_1(a-1,b,b+1,x)
\Big]
\, ,
\nnb\\
{}_2F_1(a,b,b+1,x)
& = &
\frac{b}{a-b}\,\frac{1}{x}\,\Big[
(1-x)^{1-a} - {}_2F_1(a,b-1,b,x)
\Big]
\, ,
\eeq
until all hypergeometric functions are reduced to the following forms 
\beq
{}_2F_1(-\eps,-2\eps,1-2\eps,1-\omega)
\, , 
\qquad\qquad
{}_2F_1(1,-2\eps,1-2\eps,1-\omega)
\, .
\eeq
Their expansions in $\eps$ is known to all orders and is given by 
\beq
{}_2F_1(-\eps,-2\eps,1-2\eps,1-\omega)
& = &
1 
-
\sum_{n=1}^{\infty}
\sum_{p=1}^{\infty}
(2\eps)^n(-\eps)^p \,
S_{n,p}(1-\omega)
\, ,
\nnb\\
{}_2F_1(1,-2\eps,1-2\eps,1-\omega)
& = &
1 
+
2\eps\ln \omega
-
\sum_{n=2}^{\infty}
(2\eps)^n \,
\Li_{n}(1-\omega)
\, ,
\label{magenteq}
\eeq
where the Spence functions $S_{n,p}(x)$ are defined by
\beq
S_{n,p}(x)
\, = \,
\frac{(-1)^{n+p-1}}{(n-1)!\,p!}\int_0^1\!\!dv \,
\frac{\ln^{n-1} v}{v}\,\ln^p(1-x\,v)
\, ,
\eeq
and reduce to standard polylogarithms for $p=1$, with $S_{n,1}(x) = \Li_{n+1}(x)$. 

At this point, all poles in $\eps$ can be extracted using the standard identities
\beq
\int_0^1 \!dx\, x^{-1+\alpha\eps}(1-x)^{-1+\beta\eps}\,f(x)
& = &
\int_0^1 \!dx\, x^{-1+\alpha\eps}(1-x)^{\beta\eps}\,f(x)
+
\int_0^1 \!dx\, x^{\alpha\eps}(1-x)^{-1+\beta\eps}\,f(x)
\, ,
\nnb\\
\int_0^1 \!dx\,x^{-1+\alpha\eps}\,f(x)
& = &
\frac{1}{\alpha\eps}\,f(0)
+
\int_0^1 \!dx\,x^{\alpha\eps}\,\frac{f(x)-f(0)}{x}
\, ,
\nnb\\
\int_0^1 \!dx\,(1-x)^{-1+\beta\eps}\,f(x)
& = &
\frac{1}{\beta\eps}\,f(1)
+
\int_0^1 \!dx\,(1-x)^{\beta\eps}\,\frac{f(x)-f(1)}{1-x}
\, ,
\eeq
where $x$ can be either $y'$ or $t$.  The remaining $\eps$ dependence does not 
generate any pole and can be safely expanded in Taylor series. Therefore, at this 
point, the remaining integrals (in $t$ or $y'$) can be easily performed using standard
techniques. Discarding terms that vanish in the $\eps\to 0$ limit, we obtain the final 
expressions given in \secn{sec:mappings2}.

%%%%%%%%%%%%%%%%%%%%%%%%%%%%%%%%%%%%%%%%%%%%%

\section{One-loop infrared kernels with one real emission}
\label{olokore}

%%%%%%%%%%%%%%%%%%%%%%%%%%%%%%%%%%%%%%%%%%%%%

\noindent
To complete the study of NNLO factorisation formulae we are left with the 
integration of one-loop infrared kernels involving the emission of one soft or 
two collinear particles at the one-loop level. These kernels are known from the 
literature \cite{Bern:1994zx,Bern:1998sc,Bern:1999ry,Kosower:1999xi,Kosower:1999rx,
Catani:2000pi,Somogyi:2006db}, and we rewrite them in the most suited form 
to perform their integration in the radiation phase-space in the context of our method. 

Indicating with $RV(\{k\})$ the renormalised one-loop squared matrix element for the emission 
of one unresolved parton $i$, the factorisation formulae for the soft limit $\bS{i}$ 
and for the collinear limit $\bC{ij}$ read 
\beq
\bS{i} \, RV
& = & 
- \Norm
\sum_{\substack{l \neq i \\ m \neq i}} 
\Bigg[
\mc I_{lm}^{(i)} \, 
V_{lm} \big(\{ k \}_{\slashed i} \big) 
-
\frac{\as}{2\pi}
\bigg(
\widetilde{\mc I}_{lm}^{(i)}
+
\mc I_{lm}^{(i)} \,
\frac{\beta_0}{2\eps}
\bigg)
B_{lm} \big(\{ k \}_{\slashed i} \big) 
+ 
\as \!\!\!
\sum_{\substack{p \neq i,l,m}} \!
\widetilde{\mc I}_{lmp}^{(i)}
B_{lmp} \big(\{ k \}_{\slashed i} \big)
\Bigg] ,
\nonumber \\
\bC{ij} RV
& = &
\frac{\Norm}{s_{ij}} \,
\Bigg\{
P_{ij}\,V \big(\{k\}_{[ij]} \big) 
+  
Q_{ij}^{\mu \nu}\,V_{\mu \nu} \big(\{k\}_{[ij]} \big) 
\nnb \\
&&
\hspace{2cm} + \, \frac{\as}{2\pi}
\Bigg[
\bigg(
\widetilde P_{ij}
- 
P_{ij}
\frac{\beta_0}{2\eps}
\bigg)
B \big(\{k\}_{[ij]} \big) 
+  
\bigg(
\widetilde Q_{ij}^{\mu \nu} \!
-
Q_{ij}^{\mu \nu}
\frac{\beta_0}{2\eps}
\bigg)
B_{\mu \nu} \big(\{k\}_{[ij]} \big) 
\Bigg]
\Bigg\}
\, ,
\eeq
where the symbols $\Norm$, $B$, $V$, $B_{lm}$, $B_{\mu\nu}$, $\{k\}_{\slashed i}$, 
$\{k\}_{[ij]}$, $\mc I_{lm}^{(i)}$, $P_{ij}$, and $Q_{ij}^{\mu \nu}$ were already 
introduced in \secn{treeonereal}. In addition, here we have introduced the completely 
antisymmetric tripole-colour-correlated Born squared matrix element 
\beq
\Bn_{lmp}
\, = \, 
\sum_{a,\,b,\,c}
f_{abc}\,
{\cal A}_n^{(0)*} 
\,
{\bf T}^a_l \, {\bf T}^b_m {\bf T}^c_p
\,
{\cal A}_n^{(0)}
\, ,
\eeq
as well as the colour-connected one-loop squared matrix element $V_{lm} \equiv
2 {\rm Re}\,{\cal A}_n^{(0)*}({\bf T}_l \cdot {\bf T}_m){\cal A}_n^{(1)}$, and the 
spin-connected one-loop squared matrix element $V_{\mu \nu}$, obtained 
by stripping the spin polarisation vectors of the particle with momentum 
$k_i+k_j$ from both the matrix element and its complex conjugate inside 
$V$. The one-loop soft kernels are 
\beq
\label{olosoftk}
\widetilde{\mc I}_{lm}^{(i)}
& = &
\delta_{f_i g} \,
C_A \,
\frac{\Gamma^3(1+\eps)\Gamma^4(1-\eps)}{\eps^2 \,
\Gamma(1+2\eps)\Gamma^2(1-2\eps)} \, 
\frac{s_{lm}}{s_{il}s_{im}}
\left(
\frac{e^{\euler}\mu^2 s_{lm}}{s_{il}s_{im}}
\right)^{\!\eps}
\, ,
\nnb\\
\widetilde{\mc I}_{lmp}^{(i)}
& = &
\delta_{f_i g} \,
\frac{\Gamma(1+\eps)\Gamma^2(1-\eps)}{\eps\,\Gamma(1-2\eps)} \,
\frac{s_{lm}}{s_{il}s_{im}} \,
\bigg(
\frac{e^{\euler}\mu^2\,s_{mp}}{s_{im}s_{ip}}
\bigg)^{\eps}
\, .
\eeq
Their collinear counterparts, on the other hand, can be written as
\beq
\label{olocollk}
\widetilde P_{ij}
& = &
\frac{\Gamma^2(1\!+\!\eps)\Gamma^3(1\!-\!\eps)}
     {\Gamma(1\!+\!2\eps)\Gamma^2(1\!-\!2\eps)}
\left(\! \frac{s_{ij}}{e^{\!\euler\!} \mu^2} \!\right)^{\!\!\!-\eps} 
\\
&& \hspace{2cm} \times \,
\bigg[
P_{ij} M_{ij}
+ 
N_{ij}^{\og\!} \delta_{f_i g} \delta_{f_j \{q, \bar q\}} 
+  
N_{ji}^{\og\!} \delta_{f_j g} \delta_{f_i \{q, \bar q\}}
+
N_{ij}^{\tg\!} \delta_{f_i g} \delta_{f_j g} 
\bigg]
,
\nnb\\
\widetilde Q_{ij}^{\mu \nu}
& = &
\frac{\Gamma^2(1\!+\!\eps)\Gamma^3(1\!-\!\eps)}
     {\Gamma(1\!+\!2\eps)\Gamma^2(1\!-\!2\eps)}
\!
\left(\! \frac{s_{ij}}{e^{\!\euler\!}\mu^2} \!\right)^{\!\!\!-\eps} 
\!
\bigg[
Q_{ij}^{\mu \nu} M_{ij}
-
N_{ij}^{\tg} \delta_{f_i g} \delta_{f_j g} 
\bigg(\!\!-g^{\mu\nu} + (d - 2) \, \frac{\kt_{ij}^\mu \kt_{ij}^\nu}{\kt_{ij}^2} \bigg)
\bigg]
\, , 
\nnb
\eeq
with
\beq
M_{ij}
& = &
\frac1{\eps^2}\,
\Big[
C_{f_{[ij]}\!} \!-\! C_{f_{i}} \!-\! C_{f_{j}}
+
\left(
C_{f_{[ij]}\!} \!+\! C_{f_{i}} \!-\! C_{f_{j}}
\right) \!
F(x_i)
+
\left(
C_{f_{[ij]}\!} \!+\! C_{f_{j}} \!-\! C_{f_{i}}
\right) \!
F(x_j)
\Big]
\\
&&
+ \,
\frac{1}{1\!-\!2\eps}
\Bigg[
\frac1\eps \big( \beta_0 - 3\,C_F \big)
+
C_A - 2C_F
+
\frac{C_A+4\,T_R\,N_f}{3(3-2\eps)}
\Bigg]
\,
\delta_{ \{f_i f_j\} \{q \bar q\} }
\, ,
\nnb\\
F(x) 
& = &
1 - {}_2F_1 \left(1,-\eps;1-\eps;\frac{x-1}{x} \right)
\, = \,
\eps\,\ln x
+
\sum_{n=2}^{+\infty} \, \eps^n \, \Li_n\left(\frac{x-1}{x} \right)
\, ,
\nnb
\eeq
and
\beq
N_{ij}^{\og}
\, = \,
C_F \,
\frac{C_A\!-\!C_F}
{1-2\eps}\,
\big(1-\eps x_i \big)
\, ,
\qquad
N_{ij}^{\tg}
\, = \,
4\,C_A\,
\frac{C_A(1-\eps)-2\,T_R\,N_f}{(1-2\eps)(2\!-\!2\eps)^2(3\!-\!2\eps)}\,
\big(1-2\eps x_i x_j \big)
\, .
\qquad
\eeq
While the one-loop kernels are rather intricate, there is only a single further 
unresolved radiation: the phase space mapping, to which we now turn, is
therefore simpler in this case.

%%%%%%%%%%%%%%%%%%%%%%%%%

\subsection{Phase-space mappings and integration}
\label{psmirv}

%%%%%%%%%%%%%%%%%%%%%%%%%

\subsubsection{One-loop soft kernel and cancellation of colour tripoles}
\label{softwolo}

%%%%%%%%%%%%

\noindent
As done for the tree-level infrared kernels with one real emission, for the soft 
kernels we perform the mapping described in~\appn{app:map1}, choosing 
the momenta $\{k_{\aA},k_{\bB},k_{\cC}\}$ as the momenta $\{k_i,k_l,k_m\}$ 
present in each term of the eikonal kernel, according to
\beq
k_{\aA} \to k_i \, ,
\qquad\qquad
k_{\bB} \to k_l \, ,
\qquad\qquad
k_{\cC} \to k_m \, .
\eeq
Promoting the set $\{k\}_{\slashed i}$ to the momentum conserving set 
$\kkl{ilm}$ of \appn{app:map1}, we define the mapped soft limit of 
$RV(\{k\})$ as
\beq
\bbS{i} \, RV 
& \equiv & 
- \, \Norm \,
\sum_{\substack{l \neq i\\ m \neq i}} \, 
\Bigg[
\;\;
\mc I_{lm}^{(i)} \, 
V_{lm} \!\left( \kkl{ilm} \right)
\\[-4mm]
&&
\hspace{18mm}
- \,
\frac{\as}{2\pi} 
\bigg(
\widetilde{\mc I}_{lm}^{(i)} \, 
+
\mc I_{lm}^{(i)} \,
\frac{\beta_0}{2\eps}
\bigg)
B_{lm} \!\left( \kkl{ilm} \right)
+
\as \!\!\!
\sum_{\substack{p \neq i,l,m}} \!
\widetilde{\mc I}_{lmp}^{(i)} \, 
B_{lmp} \!\left( \kkl{ilm} \right)
\;
\Bigg]
\, .
\nnb
\label{eq:bSi RV} 
\eeq
The integral of this function in the $(\npo)$-particle phase-space can be written as 
\beq
\label{eq:int bSi RV} 
\int d\Phi_\npo \,
\bbS{i} \, RV
& = &
- 
\sum_{\substack{l \neq i\\ m \neq i}} 
\int \! d \Phi_n(\kkl{ilm}) \, 
\Bigg[
\;\;
J_{\rm s}^{ilm} \,
V_{lm} \! \left( \kkl{ilm} \right) 
\\[-4mm]
&&
\hspace{7mm}
- \,
\frac{\as}{2\pi} 
\bigg(
\tilde J_{\rm s}^{ilm}
+
J_{\rm s}^{ilm} \,
\frac{\beta_0}{2\eps}
\bigg) \,
B_{lm} \! \left(\! \kkl{ilm} \!\right)
\, + \,
\as \!\!\!
\sum_{\substack{p \neq i,l,m}} \!\!\!
\tilde J_{\rm s}^{\, (i), lmp} \, 
B_{lmp} \! \left(\! \kkl{ilm} \!\right)
\Bigg]
\nnb
\, ,
\eeq
where $J_{\rm s}^{ilm}=\delta_{f_i g} \,J_{\rm s}$, defined and computed 
in \eqs{genJs}{eq:Js}, must here be expanded up to order $\mc O(\eps^2)$. 
One gets
\beq
J_{\rm s}(s) 
& = & 
\frac{\as}{2\pi}
\left(\! \frac{s}{\mu^2} \!\right)^{\!\!\!-\eps} 
\Bigg[
\frac1{\eps^2}
+
\frac2{\eps}
+
6
-
\frac{7}{12}\pi^2
+ 
\bigg( 18 - \frac{7}{6}\pi^2 - \frac{25}{3}\zeta_3 \bigg) \eps
\\ 
&& \hspace{3cm} + \,
\bigg(
54 - \frac{7}{2}\pi^2 - \frac{50}{3}\zeta_3 - \frac{71}{1440}\pi^4 
\bigg) 
\eps^2
+
\mc O(\eps^3)
\Bigg]
\, .
\nnb
\eeq
Also the integral $\tilde J_{\rm s}^{ilm}$, defined below, can be easily computed 
after substituting the expression for the Mandelstam invariants in our 
parametrisation, \eq{eq:sij NLO}. The result is
\beq
\tilde J_{\rm s}^{ilm}
& \equiv &
\Norm \!
\int \! d\Phi_{\rm rad}^{(ilm)}\,
\widetilde{\mc I}_{lm}^{(i)}
\, = \,
\delta_{f_i g} \,
\Norm \,
C_{\!A} \,
\frac{(e^{\euler\!}\mu^2)^{\eps}\,\Gamma^3(1\!+\!\eps)\Gamma^4(1\!-\!\eps)}
     {\eps^2\,\Gamma(1+2\eps)\Gamma^2(1-2\eps)}
\int \! d\Phi_{\rm rad}^{(ilm)}
\left(\! \frac{s_{lm}}{s_{il}s_{im}} \!\right)^{\!\!1+\eps}
\nnb \\[2mm]
& \equiv &
\delta_{f_i g} \,
C_{\!A} \,
\tilde J_{\rm s} \Big(\sk{lm}{ilm} \Big) 
\, ,
\eeq
whose kinematic dependence is described by the 
\emph{one-loop radiative soft function} $\tilde J_{\rm s}$ given by
\beq
\tilde J_{\rm s}(s) 
& = & 
\Norm \,
N (\eps) \,
\frac{(e^{\euler\!}\mu^2)^{\eps}\,\Gamma^3(1\!+\!\eps)\Gamma^4(1\!-\!\eps)}
     {\eps^2\,\Gamma(1+2\eps)\Gamma^2(1-2\eps)}\,
s^{-2 \eps} \!
\int_0^\pi\!\! d\phi \, \big( \! \sin \phi \big)^{- 2 \eps}
\int_0^1 \! dy  \int_0^1 \! dz \,
\nnb \\
&& \hspace{3cm} \times \, 
\Big[ y (1 - y)^2 z (1 - z) \Big]^{\!-\eps} (1 - y)
\left( \frac{1\!-\!z}{y\,z} \right)^{\!1+\eps}
\nnb
\\
& = &
\frac{\as}{2\pi}
\left( \frac{s}{e^{\euler\!}\mu^2} \right)^{\!\!-2\eps} 
\frac{\Gamma^3(1+\eps)\Gamma^3(1-\eps)}
     {4\,\eps^4\,\Gamma(1+2\eps)\Gam(2 - 4 \eps)}
\\
& = &
\frac{\as}{2\pi}
\left( \frac{s}{\mu^2} \right)^{\!\!-2\eps} 
\Bigg[
\frac1{ 4 \eps^4}
+
\frac1{\eps^3}
+
\!\left(\! 4 - \frac{7}{24}\pi^2 \!\right)\!\!\frac1{\eps^2}
+
\!\left(\! 16 - \frac{7}{6}\pi^2 - \frac{14}{3}\zeta_3 \!\right) \!
\frac1{\eps}
\nnb \\
&& \hspace{3cm} + \, 
64
-
\frac{14}{3}\pi^2 
-
\frac{56}{3}\zeta_3 
-
\frac{7}{480}\pi^4
+
\mc O(\eps)
\Bigg]
\, .
\nnb
\eeq
We now discuss the last and most interesting contribution to \eq{eq:int bSi RV}: the 
soft integral proportional to the triple-colour-correlated Born amplitude. 
It is defined by 
\beq
\tilde J_{\rm s}^{\, (i), lmp}
& \equiv &
\Norm
\int d\Phi_{\rm rad}^{(ilm)}\,
\widetilde{\mc I}_{lmp}^{(i)}
\, = \,
\delta_{f_i g} \,
\Norm \,
\frac{(e^{\euler\!}\mu^2)^{\eps}\,\Gamma(1\!+\!\eps)\Gamma^2(1\!-\!\eps)}
     {\eps\,\Gamma(1-2\eps)}
\!
\int d\Phi_{\rm rad}^{(ilm)} \, 
\frac{s_{lm}}{s_{il}s_{im}} \!
\left( \frac{s_{mp}}{s_{im}s_{ip}} \right)^{\!\eps}
\nnb\\
& \equiv &
\delta_{f_i g} \,
\tilde J_{\rm s}^{\, \rm tripole}
\left(\sk{lm}{ilm},\frac{\sk{lp}{ilm}}{\sk{mp}{ilm}}\right) 
\, .
\eeq
whose kinematic dependence is described by the 
\emph{radiative tripole soft function} $\tilde J_{\rm s}^{\, \rm tripole}$ which in 
our approach turns out to be a function of the invariant $\sk{lm}{ilm}$ and of 
the ratio $\sk{lp}{ilm}/\sk{mp}{ilm}$.
As can be guessed from the more intricate kinematic dependence, this part of the
soft one-loop kernel requires more refined techniques to be analytically integrated:
the reason is its peculiar kinematic structure, involving two eikonal kernels 
linking four particles, which leads to a non-trivial azimuthal dependence. 
With the phase-space mapping $\{k_{\aA},k_{\bB},k_{\cC}\} \to \{k_i,k_l,k_m\}$, we can 
use the results of \appn{app:azimuth} to parametrise the Mandelstam invariants 
present in $\widetilde{\mc I}_{lmp}^{(i)}$ in the form
\beq
s_{il} 
& = & 
y \, \sk{lm}{ilm} 
\, , 
\nnb\\
s_{im} 
& = & 
z (1 - y) \,\sk{lm}{ilm} 
\, ,
\nnb\\
s_{lm} 
& = & 
(1 - z)(1 - y) \, \sk{lm}{ilm} 
\, ,
\nnb\\
s_{mp} 
& = & 
(1 - y) \, \sk{mp}{ilm} 
\, ,
\nnb\\[-1mm]
s_{ip} 
& = & 
y(1-z)\sk{mp}{ilm} 
+ 
z\,\sk{lp}{ilm} 
- 
2(1-2w)\sqrt{y\,z(1-z)\sk{lp}{ilm}\sk{mp}{ilm}}
\, ,
\eeq
which leads to the expression
\beq
\tilde J_{\rm s}^{\, \rm tripole}(s,\xi) 
& = & 
\Norm \,
2^{-2\eps} \,
N (\eps) \,
\frac{(e^{\euler\!}\mu^2)^{\eps}\,\Gamma(1\!+\!\eps)\Gamma^2(1\!-\!\eps)}
     {\eps\,\Gamma(1-2\eps)} \, 
s^{-2\eps} \!\!
\int_0^1\! dw \, dy \,dz \,
\Big[ w(1\!-\!w) \Big]^{-\eps-\frac12} \,
\\
&& \hspace{-1cm}
\times \,
\Big[ y (1 - y)^2 z (1 - z) \Big]^{\!-\eps} (1 \!-\! y) \,\,
\frac{1-z}{y\,z} \,\,
z^{\!-\eps} \,
\bigg[ y(1-z) + z\,\xi - 2(1-2 w)\,\sqrt{y\,z\,(1-z)\,\xi} \bigg]^{-\eps}
\, .
\nnb
\eeq
At this point, we observe that this expression takes the form of the master 
integral defined in \eq{eq:masterallint}, namely $I_{\eps,1+\eps,-1-2\eps,
1-\eps,-1-\eps,1-2\eps}(\xi,1)$, thus it can be computed and expanded in 
powers of $\eps$ following the procedure discussed in~\appn{app:Iab3}.
The final result reads
\beq
\label{tripolint}
\tilde J_{\rm s}^{\, \rm tripole}(s,\xi) 
& = & 
\frac{\as}{2\pi}
\left( \frac{s}{\mu^2} \right)^{\!\!-2\eps} 
\bigg[
\;\;
\frac38
\frac1{\eps^3}
+
\!\left( \frac32 - \frac{1}{4}\ln\xi \right) \! \frac1{\eps^2}
+
\!\left( 7 - \frac{19}{48}\pi^2 - \ln\xi + \frac{1}{4}\ln^2\xi \right) \!
\frac1{\eps}
\\
&&
\hspace{2mm}
+ \,
32
-
\frac{19}{12}\pi^2 
-
10\zeta_3 
-
\!\left(\! 4 - \frac{\pi^2}{24} \right)\! \ln\xi
+
\ln^2\xi
-
\frac{1}{6}\ln^3\xi
-
\Li_3(-\xi)
+
\mc O(\eps)
\;
\bigg]
\, .
\nnb
\eeq
\eq{tripolint} features up to a triple pole, stemming from the combination of the 
double pole arising from the phase-space integration of the radiated soft gluon, 
and the single pole of the one-loop squared matrix element.
There are however solid arguments to expect that there should be {\it no} infrared
poles proportional to colour tripoles at NNLO. A first hint for this is the calculation 
in Ref.~\cite{Catani:2000pi}, showing that, before the factorisation and mapping
of the ($\npo$)-particle phase-space, the pole arising from the squared matrix 
element cancels by colour conservation, when only final-state partons are 
considered, as is the case here.  A stronger argument comes from the observation 
that real-virtual singular contributions proportional to colour tripoles would find no 
double-virtual or double-real counterparts to cancel against: indeed, the structure 
of virtual infrared poles at NNLO~\cite{Catani:1998bh,Aybat:2006wq,Aybat:2006mz,
Gardi:2009qi,Gardi:2009zv,Becher:2009cu,Becher:2009qa} contains only colour 
dipoles, as well as quadrupoles generated by exponentiation, but no tripoles. 
Similarly, as clearly shown by \eq{eq:SSCC1}, singular contributions to 
double-unresolved real radiation do not contain three-particle colour correlations. 
We conclude that all poles generated by \eq{tripolint}, including those that come 
from the phase-space integration of the radiated soft gluon, should cancel when 
performing the appropriate colour sums, whereas non-singular terms will provide 
important finite contributions to subtraction counterterms. To see that this cancellation 
indeed takes place, consider the sums involved in the tripole term in \eq{eq:int bSi RV}, 
\beq
\label{tripsum}
\sum_{\substack{l \neq i,m \neq i,l \\ p \neq i,l,m}} \!\!\!
\tilde J_{\rm s}^{\, (i), lmp} \, 
B_{lmp}
\, .
\eeq
The sum can be simplified using symmetry arguments, for instance exploiting the 
complete antisymmetry of $B_{lmp}$ under label exchange, as well as colour 
conservation. To give an obvious example, terms contributing to pole residues 
but independent of the Mandelstam invariants will cancel in the sum over colours,
using the Born matrix-element property $B_{lmm} + B_{lml} = 0$. This is sufficient
to prove  the cancellation of triple poles. 
Double and single poles, on the other hand,
feature residues that also contain the structures
\beq
\sum_{\substack{l \neq i,m \neq i,l \\ p \neq i,l,m}} \!\!\!\!\!
B_{lmp}
\ln^k\!\frac{\bar{s}_{lm}}{\mu^2}\,
\ln^{2n}\!\frac{\bar{s}_{lp}}{\bar{s}_{mp}}
\, ,
\qquad
\sum_{\substack{l \neq i,m \neq i,l \\ p \neq i,l,m}} \!\!\!\!\!
B_{lmp}
\ln\!\frac{\bar{s}_{lp}}{\bar{s}_{mp}}
\, ,
\qquad
\sum_{\substack{l \neq i,m \neq i,l \\ p \neq i,l,m}} \!\!\!\!\!
B_{lmp}
\ln\!\frac{\bar{s}_{lm}}{\mu^2} \,
\ln\!\frac{\bar{s}_{lp}}{\bar{s}_{mp}}
\, ,
\qquad
\eeq
with $k,n\in\mathbb{N}$. The first structure vanishes because of the symmetry of 
both logarithms for the exchange  $l \leftrightarrow m$. Similarly, the second structure 
can be rewritten as 
\beq
\label{eq:2structtripole}
\sum_{\substack{l \neq i,m \neq i,l \\ p \neq i,l,m}} \!\!\!\!\!
B_{lmp}
\ln\!\frac{\bar{s}_{lp}}{\bar{s}_{mp}}
& = &
\sum_{\substack{l \neq i,m \neq i,l \\ p \neq i,l,m}} \!\!
\left( 
\Bn_{lmp} \, \ln\!\frac{\bar s_{lp}}{\mu^2}
-
\Bn_{lmp} \, \ln\!\frac{\bar s_{mp}}{\mu^2}
\right) \, = \,  0 
\, ,
\eeq
since the first and second terms vanish separately upon summation over the 
indices $m$ and $l$, respectively. For the remaining structure we get 
\beq
\sum_{\substack{l \neq i,m \neq i,l \\ p \neq i,l,m}} \!\!\!\!\!
B_{lmp}
\ln\!\frac{\bar{s}_{lm}}{\mu^2} \,
\ln\!\frac{\bar{s}_{lp}}{\bar{s}_{mp}}
& = &
\sum_{\substack{l \neq i,m \neq i,l \\ p \neq i,l,m}} \!\!
\left( 
\Bn_{lmp}
\ln\!\frac{\bar s_{lp}}{\mu^2} \, \ln\!\frac{\bar s_{lm}}{\mu^2}
- 
\Bn_{lmp} 
\ln\!\frac{\bar s_{mp}}{\mu^2} \, \ln\!\frac{\bar s_{lm}}{\mu^2}
\right) \,=\, 0 \,,
\label{eq:single_pole_simpl}
\eeq
where individual terms vanish thanks to the same symmetry arguments used in \eq{eq:2structtripole}. This completes the proof that colour tripoles do not contribute to infrared
counterterms at NNLO, except for subtraction-scheme-dependent finite contributions.
In our approach, these are given by
\beq
\sum_{\substack{l \neq i,m \neq i,l \\ p \neq i,l,m}} \!\!\!\!
\tilde J_{\rm s}^{\, (i), lmp} \, 
B_{lmp}
& = &
- \,
\delta_{f_i g} \,
\frac{\as}{2\pi}
\sum_{\substack{l \neq i,m \neq i,l \\ p \neq i,l,m}} \!\!\!\!\!
B_{lmp}
\bigg[
\,
\frac{1}{2}
\ln\!\frac{\bar{s}_{lp}}{\bar{s}_{mp}}
\ln^2\!\frac{\bar{s}_{lm}}{\mu^2}
+
\frac{1}{6}\ln^3\!\frac{\bar{s}_{lp}}{\bar{s}_{mp}}
+
\Li_3\!\left(\!-\frac{\bar{s}_{lp}}{\bar{s}_{mp}}\!\right)
+
\mc O(\eps)
\bigg]
\, .
\nnb\\[-4mm]
\eeq

%%%%%%%%%%%%

\subsubsection{One-loop collinear kernel}
\label{colltwolo}

%%%%%%%%%%%%

\noindent
For the one-loop collinear kernel we choose the momenta $\{k_{\aA},k_{\bB},k_{\cC}\}$ 
of the phase-space mapping in \appn{app:map1}, as was done 
for the tree-level collinear kernel with one real emission. Thus we pick
\beq
k_{\aA} \to k_i \, ,
\qquad\qquad
k_{\bB} \to k_j \, ,
\qquad\qquad
k_{\cC} \to k_r \, .
\eeq
We promote the set $\{k\}_{[ij]}$ to the set of on-shell momenta $\kkl{ijr}$ 
of \appn{app:map1}, and get
\beq
\bbC{ij} \, RV 
& \equiv &
\frac{\Norm}{s_{ij}} \,
\Bigg\{ \,
P_{ij}\,V \!\left( \kkl{ijr} \right)
+  
Q_{ij}^{\mu \nu}\,V_{\mu \nu} \!\left( \kkl{ijr} \right)
\\[-2mm]
&&
\hspace{10mm}
+ \,
\frac{\as}{2\pi} \, 
\Bigg[
\bigg(
\widetilde P_{ij}
- 
P_{ij}
\frac{\beta_0}{2\eps}
\bigg) \,
B \!\left( \kkl{ijr} \right)
+  
\bigg(
\widetilde Q_{ij}^{\mu \nu} \!
-
Q_{ij}^{\mu \nu}
\frac{\beta_0}{2\eps}
\bigg) \,
B_{\mu \nu} \!\left( \kkl{ijr} \right)
\Bigg]
\Bigg\}
\, .
\nnb
\eeq
Once again, the integration of the collinear kernels is simplified by the fact that 
terms proportional to $Q_{ij}^{\mu\nu}$ and $\widetilde Q_{ij}^{\mu\nu}$ integrate 
to zero, because of their Lorentz structure. For the remaining pieces, containing 
$P_{ij}$ and $\widetilde P_{ij}$, integration in the $(\npo)$-particle phase-space
leads to 
\beq
\int d\Phi_\npo \,
\bbC{ij} \, RV 
& = &
\int d\Phi_n \big(\kkl{ijr} \big) \, 
\bigg[ \, 
J_{\rm c}^{ijr} \,
V \! \left( \kkl{ijr} \right) 
+ \,
\frac{\as}{2\pi}
\bigg(
\tilde J_{\rm c}^{ijr}
- 
J_{\rm c}^{ijr}
\frac{\beta_0}{2\eps}
\bigg)
B \!\left( \kkl{ijr} \right)
\bigg]
\, .
\nnb\\
\eeq
The integral $ J_{\rm c}^{ijr}$ is defined in \eq{Jcdef}, where it is expressed in
terms of the radiative collinear functions $J_{\rm c}^{(k {\rm g})}$  (with $k = 0,1,2$). 
These must now be computed to $\mc O(\eps^2)$, and yield
\beq
J_{\rm c}^{\zg}(s)
& = & 
\frac{\as}{2\pi}
\left( \frac{s}{\mu^2} \right)^{\! -\eps} \!
T_R 
\bigg[
-
\frac23\,\frac1{\eps}
-
\frac{16}{9}
-
\left(\! \frac{140}{27} - \frac{7}{18}\pi^2 \!\right) \eps
- 
\bigg( \frac{1252}{81} - \frac{28}{27}\pi^2 - \frac{50}{9}\zeta_3 \bigg) 
\eps^2
+
\mc O(\eps^3)
\bigg]
\, ,
\nnb
\\[2mm]
J_{\rm c}^{\og}(s)
& = &
\frac{\as}{2\pi} 
\left( \frac{s}{\mu^2} \right)^{\! -\eps} C_F  \,
\bigg[
\frac2{\eps^2}
+
\frac72\frac1{\eps}
+
11
-
\frac{7}{6}\pi^2
+ 
\bigg( 33 - \frac{49}{24}\pi^2 - \frac{50}{3}\zeta_3 \bigg) \eps
\nnb\\ 
&& 
\hspace{40mm} 
+ \,
\bigg(\! 
99 - \frac{77}{12}\pi^2 - \frac{175}{6}\zeta_3 - \frac{71}{720}\pi^4 
\bigg) \eps^2
+
\mc O(\eps^3)
\bigg] \, ,
\\
J_{\rm c}^{\tg}(s)
& = &
\frac{\as}{2\pi}
\left( \frac{s}{\mu^2} \right)^{\!-\eps}
C_A 
\bigg[ \,
\frac4{\eps^2}
+
\frac{23}{3}\frac1{\eps}
+
\frac{208}{9}
-
\frac{7}{3}\pi^2
+ 
\bigg(
\frac{1874}{27} - \frac{161}{36}\pi^2 - \frac{100}{3}\zeta_3 
\bigg)
\eps
\nnb\\ 
&& 
\hspace{40mm} 
+ \,
\bigg(
\frac{16870}{81} - \frac{364}{27}\pi^2 - \frac{575}{9}\zeta_3 - \frac{71}{360}\pi^4 
\bigg)
\eps^2
+
\mc O(\eps^3)
\;
\bigg]
\, . \nnb
\eeq
Similarly, the expression for the integral $\tilde J_{\rm c}^{ijr}$ is obtained by 
integrating the one-loop kernels $\widetilde P_{ij}$. We define
\beq
\tilde J_{\rm c}^{ijr}
& \equiv &
\Norm
\int d \Phi_{\rm rad}^{(ijr)} \, \, 
\frac{\widetilde P_{ij}}{s_{ij}} 
\\
& \equiv &
\delta_{ \{f_i f_j\} \{q \bar q\} } \, 
\tilde J_{\rm c}^{\zg} \! \Big( \sk{jr}{ijr}\Big)
+ 
\Big(\! 
\delta_{f_i g} \delta_{f_j \{q, \bar q\}} 
+ 
\delta_{f_j g} \delta_{f_i \{q, \bar q\}} 
\!\Big)
\tilde J_{\rm c}^{\og} \! \Big( \sk{jr}{ijr} \Big)
+ \delta_{f_i g} \delta_{f_j g} \, 
\tilde J_{\rm c}^{\tg} \! \Big( \sk{jr}{ijr} \Big)
\, . 
\nnb
\eeq
where the kinematic dependence can be described by the 
\emph{one-loop radiative collinear functions} $\tilde J_{\rm c}^{\zg}$, 
$\tilde J_{\rm c}^{\og}$, $\tilde J_{\rm c}^{\tg}$  with argument $\sk{jr}{ijr}$.
Terms in $\widetilde P_{ij}$ which are proportional to the simple polynomials 
$N_{ij}^{\og}$ and $N_{ij}^{\tg}$ (see \eq{olocollk}) can be integrated easily. 
Less trivial integrals arise from the $P_{ij} M_{ij}$ term in \eq{olocollk}, and 
in particular from structures of the type
\beq
I_F^{m,n}
& = &
\int_{0}^{1}\!dz\,(1-z)^{m-\eps} z^{n-\eps} \, 
{}_2F_1\left(1,-\eps ;1-\eps ;-\frac{z}{1-z}\right)
\, ,
\eeq
where $n, m$ can take only the integer values $-1,0,1$. For these values, the 
integral can be expressed in terms of a generalised hypergeometric function of 
type ${}_3{F}_2$, evaluated at unit argument. More precisely,
\beq
I_F^{m,n}
& = &
\frac{\Gamma (m-\eps +2) \Gamma (n-\eps +1)}{\Gamma(m+n-2\eps+3)} \, 
{}_3{F}_2(1,1,n-\eps +1;m+n-2 \eps +3,1-\eps ;1)
\, .
\eeq
This, in turn, can be expanded in powers of $\eps$, using for example 
the package \texttt{HypExp}~\cite{Huber:2005yg,Huber:2007dx}. The 
integration over the remaining radiative phase-space variables is then
straightforward. The final results for the three contributions to $\tilde 
J_{\rm c}^{ijr}$ are
\beq
\tilde J_{\rm c}^{\zg}(s)
& = &
\frac{\as}{2\pi} \!
\left( \frac{s}{\mu^2} \!\right)^{\!\!\!-2\eps} \!\!
\Bigg\{
\,
N_f T_R^2 \,
\bigg[\,
\frac{4}{9} \frac1{\eps^2}
+
\frac{64}{27} \frac1{\eps}
+
\frac{284}{27}
-
\frac{2}{3}\pi^2 
+
\mc O(\eps)
\bigg]
\\
&&
\hspace{17mm}
+ \,
C_A T_R 
\bigg[\!
-
\frac{1}{3} \frac1{\eps^3}
-
\frac{31}{18} \frac1{\eps^2}
-
\left( \frac{211}{27} - \frac{1}{2}\pi^2 \!\right) \! \frac1{\eps}
-
\frac{5281}{162}
+
\frac{31}{12}\pi^2 
+
\frac{62}{9}\zeta_3 
+
\mc O(\eps)
\bigg]
\nnb\\
&&
\hspace{17mm}
+ \,
C_F T_R 
\bigg[\,
\frac{2}{3}  \frac1{\eps^3}
+
\frac{31}{9} \frac1{\eps^2}
+
\left( \frac{431}{27} - \pi^2 \!\right) \! \frac1{\eps}
+
\frac{5506}{81}
-
\frac{31}{6}\pi^2 
-
\frac{124}{9}\zeta_3 
+
\mc O(\eps)
\bigg]
\Bigg\}
, 
\nnb
\\[2mm]
\tilde J_{\rm c}^{\og}(s)
& = & 
\frac{\as}{2\pi} 
\left( \frac{s}{\mu^2} \right)^{\!\!-2\eps}
\Bigg\{
\;
C_F^2 \,
\bigg[
-
\left( \frac{5}{4} - \frac{\pi^2}{3} \right) \! \frac1{\eps^2}
-
\!\left( \frac{15}{2} - \frac{2}{3}\pi^2 - 10\zeta_3 \right) \! \frac1{\eps}
\nnb
\\ 
&& 
\hspace{45mm} 
- \, 
\frac{141}{4}
+
\frac{109}{24}\pi^2 
+
20\zeta_3 
-
\frac{7}{45}\pi^4
+
\mc O(\eps)
\,\bigg]
\nnb \\
&&
\hspace{20mm}
+ \,\,
C_F C_A \,
\bigg[
-
\frac{1}{2} \frac1{\eps^4}
-
\frac{7}{4} \frac1{\eps^3}
-
\!\left( \frac{15}{2} - \frac{7}{12}\pi^2 \!\right) \! \frac1{\eps^2}
-
\!\left( 31 - \frac{55}{24}\pi^2 - \frac{16}{3}\zeta_3 \right) \! 
\frac1{\eps}
\nnb \\
&&
\hspace{60mm}
- \,
\frac{503}{4}
+
\frac{119}{12}\pi^2 
+
\frac{157}{6}\zeta_3 
-
\frac{67}{720}\pi^4
+
\mc O(\eps)
\bigg]
\;
\Bigg\}
\, ,
\nnb
\\
\tilde J_{\rm c}^{\tg}(s)
& = &
\frac{\as}{2\pi} 
\left( \frac{s}{\mu^2} \right)^{\!\!-2\eps}
\Bigg\{
\;
C_A N_f T_R \,
\bigg[
\frac{1}{3} \frac1{\eps}
+
\frac{25}{9}
+
\mc O(\eps)
\bigg]
\nnb
\\
&&
\hspace{20mm}
+ \,
C_A^2 \,
\bigg[
- \,
\frac1{\eps^4}
-
\frac{23}{6} \frac1{\eps^3}
-
\!\left( \frac{172}{9} - \frac{11}{6}\pi^2 \right) \! \frac1{\eps^2}
-
\!\left( \frac{253}{3} - \frac{77}{12}\pi^2 - \frac{92}{3}\zeta_3 \right) \! 
\frac1{\eps}
\nnb\\
&&
\hspace{58mm}
- \,
\frac{57277}{162}
+
\frac{94}{3}\pi^2 
+
\frac{893}{9}\zeta_3 
-
\frac{179}{360}\pi^4
+
\mc O(\eps)
\bigg]
\;
\Bigg\}
\, . 
\nnb
\eeq
This completes the list of all the integrals associated with factorised soft and collinear
kernels at NNLO. These integrals form the basis for the construction of all integrated
infrared counterterms for single- and double-unresolved real radiation at NNLO.

%%%%%%%%%%%%%%%%%%%%%%%%%%%%%%%%%%%%%%%%%%%%%

\section{Conclusions}
\label{conclu}

%%%%%%%%%%%%%%%%%%%%%%%%%%%%%%%%%%%%%%%%%%%%%

\noindent

In any massless gauge theory, (squared) matrix elements factorise in soft and 
collinear limits, at leading power in the soft energy and in the small transverse 
momentum, yielding universal soft and collinear kernels, which multiply the
(squared) matrix element for the Born process, without the unresolved particles.
Away from the strict limits (or beyond leading power in the resolving variables)
this factorisation is not exact: in particular, the factorised Born matrix element
does not conserve momentum (near the soft limit), or is not on the mass shell 
(near collinear limits). In order to integrate the factorisation kernels over the 
unresolved degrees of freedom in a universal way ({\it i.e.} requiring no information 
on the underlying Born process), one needs to provide a set of phase-space 
mappings, which must re-express the factorised Born process in terms of 
an on-shell, momentum-conserving set of momenta. This amounts to a specific
choice of a set of sub-leading power terms in the factorisation, and such a choice
is a necessary ingredient for any infrared subtraction procedure.

In the present paper, we have presented the complete integration of the QCD
factorisation kernels at NLO and NNLO, with a set of phase-space mappings
selected along the lines suggested in Ref.~\cite{Magnea:2018hab}, chosen with the 
goal of simplifying as much as possible the analytic integration. As a consequence, 
we have been able to give analytic results for all kernels, including non-singular
terms. In particular, all integrals of the double-real counterterms can be
written exactly, to all orders in $\epsilon$, in terms of hypergeometric functions,
with the most intricate cases involving ${}_4F_3$ evaluated at unit
argument. We have however
chosen to give the expansion of these hypergeometrics in powers of $\eps$,
up to and including ${\cal O}  (\eps^0)$ terms, since this is what is required
in practical calculations. All our results have been validated against independent
numerical integration codes based on sector decomposition \cite{Binoth:2000ps,
Heinrich:2008si,Borowka:2017idc}.  The analytic results of this paper are necessary 
(and indeed sufficient) ingredients to build all integrated counterterms in the context
of the local analytic sector subtraction of Ref.~\cite{Magnea:2018hab}. The 
present work shows that this novel approach allows to use standard techniques 
to compute an important class of integrals that appear in all NNLO QCD 
computations, yielding comparatively very simple results.

We believe that achieving the maximum simplicity in the case at hand - NNLO
radiation of massless partons in the final state - is important not only for 
building an efficient and transparent NNLO subtraction algorithm for these
processes, but also for future extensions.
The method presented here is expected to be generalisable
to initial-state QCD radiation without conceptual changes, and the results
are sufficiently simple that a generalisation to massive particles at NNLO  
appears feasible. Furthermore, since the integrations presented in this 
paper have been performed with conventional techniques, one may reasonably 
hope that more advanced techniques, such as those involving differential 
equations for Feynman integrals (see, for example,~\cite{Chetyrkin:1981qh,
  Laporta:2001dd,Anastasiou:2002yz,Anastasiou:2003yy,Smirnov:2006ry,
  Henn:2013pwa,Henn:2014qga}), might be sufficient to tackle the problem even at the next perturbative order.

\section*{Acknowledgements}
This work was partially supported by the Italian Ministry of University and
Research (MIUR), grant PRIN 20172LNEEZ. C.S-S.~was supported by the Deutsche
Forschungsgemeinschaft (DFG, German Research Foundation) under grant
no.~396021762 - TRR 257. G.P.~was supported by the Bundesministerium f\"ur Bildung
und Forschung (BMBF, German Federal Ministry for Education and Research) under
contract no.~05H18WWCA1.

%%%%%%%%%%%%%%%%%%%%%%%%%%%%%%%%%%%%%%%%%%%%%

\newpage

%%%%%%%%%%%%%%%%%%%%%%%%%%%%%%%%%%%%%%%%%%%%%

\appendix

%%%%%%%%%%%%%%%%%%%%%%%%%%%%%%%%%%%%%%%%%%%%%

%%%%%%%%%%%%%%%%%%%%%%%%%%%%%%%%%%%%%%%%%%%%%

\section{Phase-space mappings}
\label{AA}

%%%%%%%%%%%%%%%%%%%%%%%%%%%%%%%%%%%%%%%%%%%%%

\subsection{One unresolved particle}
\label{app:map1}

%%%%%%%%%%%%%%%%%%%%%%%%%

\noindent
Given an on-shell, momentum conserving $(n+1)$-tuple of final-state 
massless momenta $\{k\} = \{k_i\}$, $i=1, \ldots, n+1$, including the momentum 
$k_{\aA}$ of the unresolved parton, we choose two momenta $k_{\bB}$ and 
$k_{\cC}$, with $b, c \neq a$, and construct an on-shell, momentum conserving 
$n$-tuple of massless momenta $\kkl{\aA\bB\cC}$ (without $k_{\aA}$) as 
\beq
\kkl{\aA\bB\cC} 
& = & 
\left\{ 
\{ k \}_{\slashed \aA \slashed \bB \slashed \cC}, \, 
\kk{\bB}{\aA\bB\cC}, \, 
\kk{\cC}{\aA\bB\cC} 
\right\} 
\, ,
\label{precset}
\eeq
with
\beq 
&&
\hspace{-10mm}
\kk{\bB}{\aA\bB\cC} 
\, = \, 
k_{\aA} + k_{\bB} - \frac{s_{\aA\bB}}{s_{\aA\cC} + s_{\bB\cC}} \, k_{\cC} \, , 
\qquad
\kk{\cC}{\aA\bB\cC} 
\, = \, 
\frac{s_{\aA\bB\cC}}{s_{\aA\cC} + s_{\bB\cC}} \, k_{\cC} \, ,
\qquad
\kk{i}{\aA\bB\cC} 
\, = \, 
k_i,
\quad
\mbox{if } i \neq \aA,\bB,\cC,
\label{eq:CSmap}
\eeq
where we have introduced 
$s_{\aA\bB\cC} = s_{\aA\bB} + s_{\aA\cC} + s_{\bB\cC} = \sk{\bB\cC}{\aA\bB\cC}$. 
These momenta satisfy the condition 
$\kk{\bB}{\aA\bB\cC} + \kk{\cC}{\aA\bB\cC} = k_{\aA} + k_{\bB} + k_{\cC} $, 
ensuring momentum conservation, and they are all light-like, as easily checked.
Next, we introduce Catani-Seymour parameters \cite{Catani:1996vz}
\beq
y \, = \, \frac{s_{\aA\bB}}{s_{\aA\bB\cC}} \, , 
\qquad 
z \, = \, \frac{s_{\aA\cC}}{s_{\aA\cC} + s_{\bB\cC}} \, , 
\label{eq:CSparam}
\eeq
which allow us to write
\beq
s_{\aA\bB} \, = \, y \, \sk{\bB\cC}{\aA\bB\cC}
\, , 
\qquad\qquad
s_{\aA\cC} \, = \, z (1 - y) \, \sk{\bB\cC}{\aA\bB\cC} 
\, , 
\qquad\qquad
s_{\bB\cC} \, = \, (1 - z)(1 - y) \, \sk{\bB\cC}{\aA\bB\cC}
\, .
\label{eq:sij NLO}
\eeq
We use these variables to parametrise the $(\npo)$-body phase space as 
\beq
\label{npofact}
d\Phi_\npo (\{k\})
& = & 
d\Phi_n(\kkl{\aA\bB\cC}) \, d \Phi_{\rm rad}^{(\aA\bB\cC)} 
\, ,
\qquad \qquad 
d \Phi_{\rm rad}^{(\aA\bB\cC)} 
\, \equiv  \, 
d \Phi_{\rm rad} \left( \sk{\bB\cC}{\aA\bB\cC}; y, z, \phi \right) 
\, ,
\eeq
leading to the explicit expression
\beq
\hspace{-5mm} 
\int d \Phi_{\rm rad}^{(\aA\bB\cC)} 
\equiv
N (\eps)  \left(\sk{\bB\cC}{\aA\bB\cC}\right)^{\!1 - \eps} \!\!\!
\int_0^\pi\!\!\!d\phi \, \sin^{- 2 \eps}\!\phi \!
\int_0^1 \!\! dy \int_0^1 \!\! dz
\Big[ y (1 - y)^2 z (1 - z) \Big]^{- \eps} \! (1 - y) 
\, ,
\label{parphsp}
\eeq
where we have defined 
\beq
\quad N(\eps) 
\, \equiv \, 
\frac{(4\pi)^{\eps - 2}}{\sqrt\pi\,\Gam\!\left(\frac12-\eps\right)}  
\, .
\eeq
In \eq{npofact}, $d\Phi_n(\kkl{\aA\bB\cC})$ is the $n$-body phase space 
for partons with momenta $\{\kb\}^{(\aA\bB\cC)}$, while, in \eq{parphsp}, 
$\phi$ is the azimuthal angle of $k_{\aA}$, measured in the rest frame
of the $k_a+k_b+k_c$ system, with $\kk{\bB}{\aA\bB\cC}$ pointing along
the $z$-direction, (see \appn{app:azimuth} for full details).

%%%%%%%%%%%%%%%%%%%%%%%%%

\subsection{Parametrisation of the azimuthal angle}
\label{app:azimuth}

%%%%%%%%%%%%%%%%%%%%%%%%%

\noindent
While in NLO computations the integration on the azimuthal angle is always 
trivial, at NNLO the integration of at least one azimuthal variable is significantly 
more complicated, and has to be treated with care. First of all, one needs an 
auxiliary four-momentum $k_{\dD}$, to fix the plane with respect to which the 
azimuthal angle is defined. We take as reference frame the one where 
$p=k_{\aA}+k_{\bB}+k_{\cC}$ is at rest, and the direction of $\kk{\bB}{\aA\bB\cC}$ 
as the axis with respect to which the polar angle $\theta$ is defined. 
The azimuthal angle $\phi$ is then defined as the angle between the plane 
containing $k_{\aA}$ and $\kk{\bB}{\aA\bB\cC}$, and the plane containing 
$\kk{\bB}{\aA\bB\cC}$ and $k_{\dD}$. Using the formulae derived in the 
second section of~\cite{Byckling:1971vca}, in this reference frame we have 
\beq
\label{cosphi}
\cos\phi \, = \,  
\Big[ \Delta_3 \Big( p,\kk{\bB}{\aA\bB\cC},k_{\dD} \Big) \,
\Delta_3 \Big( p,\kk{\bB}{\aA\bB\cC},k_{\aA} \Big) \Big]^{-1/2}
\,
G\left(
\begin{array}{ccc}
p, & \kk{\bB}{\aA\bB\cC}, & k_{\dD} \\
p, & \kk{\bB}{\aA\bB\cC}, & k_{\aA}
\end{array}
\right)
\, ,
\eeq
where
\[
\Delta_n(p_1,\dots,p_n) = 
G\left(
\begin{array}{ccc}
p_1, & \dots, & p_n \\
p_1, & \dots, & p_n
\end{array}
\right),
\qquad
G\left(
\begin{array}{ccc}
p_1, & \dots, & p_n \\
q_1, & \dots, & q_n
\end{array}
\right)
=
\left|
\begin{array}{ccc}
p_1\dt q_1 & \dots & p_1\dt q_n \\
. & . & . \\[-.2cm]
: & : & : \\
p_n\dt q_1 & \dots & p_n\dt q_n
\end{array}
\right|
\, .
\]
Using \eq{cosphi} we get
\beq
\cos\phi 
&=&
\frac{
  2k_{\aA}\dt\kk{\bB}{\aA\bB\cC} \;  2 k_{\dD} \dt \kk{\cC}{\aA\bB\cC}
+ 2k_{\aA}\dt\kk{\cC}{\aA\bB\cC} \; 2k_{\dD} \dt \kk{\bB}{\aA\bB\cC}
- s_{\aA\bB\cC} \, 2k_{\aA}\dt k_{\dD}
}{
2 \,
\Big[
2k_{\aA}\dt\kk{\bB}{\aA\bB\cC} \; 2k_{\aA}\dt\kk{\cC}{\aA\bB\cC} \,
( 
2\kk{\bB}{\aA\bB\cC}\dt k_{\dD} \; 2\kk{\cC}{\aA\bB\cC}\dt k_{\dD} 
- s_{\aA\bB\cC}k_{\dD}^2
)
\Big]^{1/2}
}
\, ,
\eeq
which, in the case $k_{\dD}^2=0$, and using \eq{eq:sij NLO}, leads to
\beq
\cos\phi =
\frac{
  y(1-z)\,\sk{\cC\dD}{\aA\bB\cC}
+ z\, \sk{\bB\dD}{\aA\bB\cC}
- s_{\aA\dD}
}
{2\,[yz(1-z)\sk{\bB\dD}{\aA\bB\cC} \, \sk{\cC\dD}{\aA\bB\cC}]^{1/2}}
\, ,
\qquad\quad
\sin^2\phi =
-\,\frac{
\Lambda
\Big(
y(1-z)\sk{\cC\dD}{\aA\bB\cC},
z\sk{\bB\dD}{\aA\bB\cC},
s_{\aA\dD}
\Big)
}{
4\,yz(1-z)\sk{\bB\dD}{\aA\bB\cC}\sk{\cC\dD}{\aA\bB\cC}
}
\, ,
\quad 
\eeq
where $\Lambda(a,b,c)=a^2+b^2+c^2-2ab-2bc-2ca$ is the K\"all\'en function. 
Having written $\cos\phi$ in terms of invariants, we introduce a new integration 
variable,
\beq
w = 
\frac{1-\cos\phi}{2}, 
\qquad
\cos\phi = 1-2w,
\qquad
\sin^2\phi = 4w(1-w),
\qquad
d\phi = 
\frac{dw}{[w(1-w)]^{1/2}}
\, .
\eeq
The integration over the azimuthal angle then becomes
\beq
\int_0^{\pi}\!\!d\phi\,\sin^{-2\eps}\!\phi =
2^{-2\eps}\int_0^1\!\!dw \, [w(1-w)]^{-\eps-1/2}
\, ,
\eeq
giving, for the radiative phase space, 
\beq
\int d \Phi_{\rm rad}^{(\aA\bB\cC)} 
& = &
2^{-2\eps}  \, 
N (\eps)  \left(\sk{\bB\cC}{\aA\bB\cC}\right)^{\!1 - \eps} \!
\int_0^1dw \int_0^1 dy \int_0^1dz\,
[w(1-w)]^{-\eps-\frac12}\, 
\times
\nnb \\
&& \qquad \times 
\Big[ y (1 - y)^2 z (1 - z) \Big]^{- \eps} \! (1 - y) \, .
\qquad
\eeq
Among the new scalar products $\sk{\cC\dD}{\aA\bB\cC}$,
$\sk{\bB\dD}{\aA\bB\cC}$, and $s_{\aA\dD}$, only the last one involves 
the unresolved parton $k_{\aA}$. Its relation with the other invariants is then
\beq
s_{\aA\dD}
&=&
y(1-z)\sk{\cC\dD}{\aA\bB\cC}
+ 
z\sk{\bB\dD}{\aA\bB\cC}
- 
2\,(1-2w)
\Big[
yz(1-z)\sk{\bB\dD}{\aA\bB\cC}\sk{\cC\dD}{\aA\bB\cC}
\Big]^{1/2}
\, .
\label{eq:azimutinvar}
\eeq

%%%%%%%%%%%%%%%%%%%%%%%%%

\subsection{Two unresolved particles}
\label{app:map2}

%%%%%%%%%%%%%%%%%%%%%%%%%

\noindent
Given an on-shell, momentum conserving $(n+2)$-tuple of final-state 
massless momenta $\{k\} = \{k_i\}$, ${i=1, \ldots, n+2}$, including the momenta 
$k_{\aA}$, $k_{\bB}$ of the two unresolved partons, we construct an 
on-shell, momentum conserving $n$-tuple of massless momenta, applying 
twice the procedure in \appn{app:map1}. We distinguish the cases involving 
four, five, and six momenta in the mapping. 

%%%%%%%%%%%%

\subsubsection{Mapping involving four momenta}
\label{app:map2 4}

%%%%%%%%%%%%

\noindent
In addition to the momenta $k_{\aA}$, $k_{\bB}$, we choose a third 
momentum $k_{\cC}$ to construct the on-shell, momentum conserving 
$(n+1)$-tuple of massless momenta $\kkl{\aA\bB\cC}$ 
\beq
\kkl{\aA\bB\cC} 
& = & 
\left\{ \{ k \}_{\slashed \aA \slashed \bB \slashed \cC}, \, 
\kk{\bB}{\aA\bB\cC}, \, \kk{\cC}{\aA\bB\cC} \right\} 
\, ,
\eeq
and a fourth momentum $k_{\dD}$ to fix the azimuthal angle of $k_{\aA}$ in the 
reference frame where $k_{\aA}+k_{\bB}+k_{\cC}$ is at rest, as described in
\appn{app:azimuth}. Then in $\kkl{\aA\bB\cC}$  we select the three momenta 
$\kk{\bB}{\aA\bB\cC}$, $\kk{\cC}{\aA\bB\cC}$ and $\kk{\dD}{\aA\bB\cC}=k_{\dD}$ 
to construct the on-shell, momentum conserving $n$-tuple of massless momenta 
$\kkl{\aA\bB\cC\dD}$ 
\beq
\kkl{\aA\bB\cC\dD} 
& = & 
\left\{ \{ k \}_{\slashed \aA \slashed \bB \slashed \cC \slashed \dD}, \, 
\kk{\cC}{\aA\bB\cC\dD}, \, \kk{\dD}{\aA\bB\cC\dD} \right\} \, ,
\eeq
with
\beq
\kk{\cC}{\aA\bB\cC\dD} 
\, = \, 
\kk{\bB}{\aA\bB\cC} + \kk{\cC}{\aA\bB\cC} 
- \frac{\sk{\bB\cC}{\aA\bB\cC}}{\sk{\bB\dD}{\aA\bB\cC}+\sk{\cC\dD}{\aA\bB\cC}} \, 
\kk{\dD}{\aA\bB\cC} \, , 
\qquad
\quad 
\kk{\dD}{\aA\bB\cC\dD} 
\, = \, 
\frac{\sk{\bB\cC\dD}{\aA\bB\cC}}{\sk{\bB\dD}{\aA\bB\cC}+\sk{\cC\dD}{\aA\bB\cC}} \, 
\kk{\dD}{\aA\bB\cC} \, ,
\quad 
\eeq
while all other momenta are left unchanged 
($\kk{n}{\aA\bB\cC\dD} = k_n \, , \, n \neq \aA, \bB, \cC, \dD$).
Introducing Catani-Seymour parameters 
\beq
y' \, = \, \frac{s_{\aA\bB}}{s_{\aA\bB\cC}} \, , 
\qquad
z' \, = \, \frac{s_{\aA\cC}}{s_{\aA\cC} + s_{\bB\cC}} \, , 
\qquad 
y \, = \, 
\frac{\sk{\bB\cC}{\aA\bB\cC}}{\sk{\bB\cC\dD}{\aA\bB\cC}}  \, , 
\qquad
z \, = \, 
\frac{\sk{\bB\dD}{\aA\bB\cC}}{\sk{\bB\dD}{\aA\bB\cC}+\sk{\cC\dD}{\aA\bB\cC}} 
\, , 
\label{eq:CSparam NNLO}
\eeq
we can write the six invariants involving $\aA, \bB, \cC, \dD$ in terms 
of the invariant $\sk{\cC\dD}{\aA\bB\cC\dD}=\sk{\bB\cC\dD}{\aA\bB\cC} =
s_{\aA\bB\cC\dD}$:
\beq
s_{\aA\bB} & = & y' \, y \, \sk{\cC\dD}{\aA\bB\cC\dD}
\, , 
\qquad\qquad
s_{\aA\cC} \; = \; z' ( 1 - y' ) \, y \, \sk{\cC\dD}{\aA\bB\cC\dD}
\, , 
\qquad\qquad
s_{\bB\cC}  \; = \; ( 1 - y' ) ( 1  - z' ) \, y \, \sk{\cC\dD}{\aA\bB\cC\dD}
\, ,
\nnb \\[2mm]
s_{\cC\dD}  & = & ( 1 - y' ) ( 1 - y ) ( 1 - z ) \, \sk{\cC\dD}{\aA\bB\cC\dD}
\, ,
\nnb \\
s_{\aA\dD} 
& = & 
(1-y) 
\left[\, y'(1-z')(1-z) + z'z - 2(1-2w')\sqrt{y'z'(1-z')z(1-z)} \,\right] 
\sk{\cC\dD}{\aA\bB\cC\dD}
\, ,
\nnb \\
s_{\bB\dD} 
& = & 
(1-y) 
\left[\, y'z'(1-z) + (1-z')z + 2(1-2w')\sqrt{y'z'(1-z')z(1-z)} \,\right] 
\sk{\cC\dD}{\aA\bB\cC\dD}
\, .
\label{invaproco}
\eeq
We use these variables to parametrise the $(\npt)$-body phase space as 
\beq
&&
d \Phi_{\npt}(\{k\})
\, = \, 
d \Phi_n(\kkl{\aA\bB\cC\dD}) \, d \Phi_{\rm rad, 2}^{\,(\aA\bB\cC\dD)} 
\, ,
\\
&&
d \Phi_{\rm rad, 2}^{\,(\aA\bB\cC\dD)}
\, = \, 
d \Phi_{\rm rad}(\sk{\cC\dD}{\aA\bB\cC\dD};y,z,\phi) \,
d \Phi_{\rm rad}(\sk{\bB\cC}{\aA\bB\cC};y',z',w') 
\, ,
\eeq
where the explicit expression of $d \Phi_{\rm rad, 2}$ in terms of Catani-Seymour 
parameters reads
\beq
\label{phspproco}
\int d \Phi_{\rm rad, 2}^{\,(\aA\bB\cC\dD)} 
& = & 
2^{- 2 \eps} \,
N^2 (\eps) \, 
\left(\sk{\cC\dD}{\aA\bB\cC\dD}\right)^{\!2 - 2\eps}
\!\!
\int_0^1 \!\!\! d w' \!
\int_0^1 \!\!\! d y' \!
\int_0^1 \!\!\! d z' \!
\int_0^\pi \!\!\! d \phi \, (\sin\phi)^{- 2 \eps} \!
\int_0^1 \!\!\! d y \!
\int_0^1 \!\!\! d z \, 
\\
& & \times\,
\Big[ w'(1-w')\Big]^{-1/2-\eps}  
\Big[ y'(1-y')^2\,z'(1-z')\,y^2(1-y)^2\,z(1-z) \Big]^{- \eps} 
(1-y') \, y(1-y) \, .
\nnb
\eeq
Here $w'=(1-\cos\phi')/2$ parametrises the azimuth $\phi'$ of 
$k_{\aA}$ in the reference frame where $k_{\aA}+k_{\bB}+k_{\cC}$ is at rest, 
while $\phi$ is the azimuth of $\kk{\bB}{\aA\bB\cC}$, whose integration is trivial.

%%%%%%%%%%%%

\subsubsection{Mapping involving five momenta}
\label{app:map2 5}

%%%%%%%%%%%%

\noindent
In this case, we first select two momenta $k_{\cC},k_{\dD}$, and build the 
$(n+1)$-tuple of massless momenta $\kkl{\aA\cC\dD}$ 
\beq
\kkl{\aA\cC\dD} 
& = & 
\left\{ \{ k \}_{\slashed \aA \slashed \cC \slashed \dD}, \, 
\kk{\cC}{\aA\cC\dD},\kk{\dD}{\aA\cC\dD} \right\} 
\, .
\eeq
Then in $\kkl{\aA\cC\dD}$  we choose the three momenta 
$\kk{\bB}{\aA\cC\dD}=k_{\bB}$, $\kk{\eE}{\aA\cC\dD}=k_{\eE}$,
and $\kk{\dD}{\aA\cC\dD}$ to construct the on-shell, momentum 
conserving $n$-tuple of massless momenta $\kkl{\aA\cC\dD,\bB\eE\dD}$ 
\beq
\kkl{\aA\cC\dD,\bB\eE\dD}
& = & 
\left\{ 
\{ k \}_{\slashed \aA \slashed \bB \slashed \cC \slashed \dD \slashed \eE}, \, 
\kk{\cC}{\aA\cC\dD,\bB\eE\dD},
\kk{\dD}{\aA\cC\dD,\bB\eE\dD},
\kk{\eE}{\aA\cC\dD,\bB\eE\dD}
\right\} \, ,
\eeq
with
\beq
\kk{\cC}{\aA\cC\dD,\bB\eE\dD} 
&=& 
\kk{\cC}{\aA\cC\dD} 
, 
\qquad
\qquad
\qquad
\kk{\dD}{\aA\cC\dD,\bB\eE\dD} 
\, = \,  
\frac{\sk{\bB\eE\dD}{\aA\cC\dD}}
     {\sk{\bB\dD}{\aA\cC\dD}\!+\!\sk{\eE\dD}{\aA\cC\dD}} \, 
\kk{\dD}{\aA\cC\dD} 
,
\nnb \\
\kk{\eE}{\aA\cC\dD,\bB\eE\dD} 
& = &
\kk{\bB}{\aA\cC\dD} 
+ 
\kk{\eE}{\aA\cC\dD} 
- 
\frac{\sk{\bB\eE}{\aA\cC\dD}}
     {\sk{\bB\dD}{\aA\cC\dD}\!+\!\sk{\eE\dD}{\aA\cC\dD}} \, 
\kk{\dD}{\aA\cC\dD} 
,
\qquad\quad
\eeq
while all other momenta are left unchanged 
($\kk{n}{\aA\cC\dD,\bB\eE\dD} = k_n \, , \, n \neq \aA,\bB,\cC,\dD,\eE$).
Introducing Catani-Seymour parameters 
\beq
y' \, = \, \frac{s_{\aA\cC}}{s_{\aA\cC\dD}} \, , 
\qquad
z' \, = \, \frac{s_{\aA\dD}}{s_{\aA\dD} + s_{\cC\dD}} \, , 
\qquad 
y \, = \, 
\frac{\sk{\bB\eE}{\aA\cC\dD}}{\sk{\bB\eE\dD}{\aA\cC\dD}}  \, , 
\qquad
z \, = \, 
\frac{\sk{\bB\dD}{\aA\cC\dD}}{\sk{\bB\dD}{\aA\cC\dD}+\sk{\eE\dD}{\aA\cC\dD}} 
\, , 
\eeq
we write the six relevant invariants in terms of 
$\sk{\cC\dD}{\aA\cC\dD,\bB\eE\dD}$, and 
$\sk{\eE\dD}{\aA\cC\dD,\bB\eE\dD}=\sk{\bB\eE\dD}{\aA\cC\dD}$, as
\beq
s_{\aA\cC} & = & y' \, ( 1 - y ) \sk{\cC\dD}{\aA\cC\dD,\bB\eE\dD}
, 
\hspace{30.5mm}
s_{\aA\dD} \; = \; z' \, ( 1 - y' ) ( 1 - y ) \sk{\cC\dD}{\aA\cC\dD,\bB\eE\dD}
, 
\nnb \\[2mm]
s_{\bB\eE} & = & y \, \sk{\eE\dD}{\aA\cC\dD,\bB\eE\dD}
, 
\hspace{42.4mm}
s_{\bB\dD} \; = \;  ( 1 - y' ) \, z \, ( 1 - y ) \sk{\eE\dD}{\aA\cC\dD,\bB\eE\dD}
, 
\nnb \\
\qquad
s_{\cC\dD}  & = & ( 1 - y' ) ( 1  - z' ) ( 1 - y ) \sk{\cC\dD}{\aA\cC\dD,\bB\eE\dD}
,
\hspace{11mm}
s_{\eE\dD}  \; = \; ( 1 - y' ) ( 1  - z ) ( 1 - y ) \sk{\eE\dD}{\aA\cC\dD,\bB\eE\dD}
.
\qquad
\qquad 
\eeq
For the $(\npt)$-body phase space we obtain
\beq
d \Phi_{\npt}(\{k\})
= 
d \Phi_n(\kkl{\aA\cC\dD,\bB\eE\dD}) \, d \Phi_{\rm rad, 2}^{\,(\aA\cC\dD,\bB\eE\dD)} 
\, ,
\eeq
where the double radiative phase space 
\beq
d \Phi_{\rm rad, 2}^{\,(\aA\cC\dD,\bB\eE\dD)}
& = & 
d \Phi_{\rm rad}(\sk{\eE\dD}{\aA\cC\dD,\bB\eE\dD};y,z,\phi) \,
d \Phi_{\rm rad}(\sk{\cC\dD}{\aA\cC\dD};y',z',\phi') 
\, ,
\eeq
can be written as
\beq
\int d \Phi_{\rm rad, 2}^{\,(\aA\cC\dD,\bB\eE\dD)} 
& = & 
N^2 (\eps) \, 
\left(
\sk{\cC\dD}{\aA\cC\dD,\bB\eE\dD} \, \sk{\eE\dD}{\aA\cC\dD,\bB\eE\dD}
\right)^{\!1 - \eps}
\!\!
\int_0^\pi \!\!\! d \phi' \, (\sin\phi')^{- 2 \eps} \!
\int_0^1 \!\!\! d y' \!
\int_0^1 \!\!\! d z' \!
\int_0^\pi \!\!\! d \phi \, (\sin\phi)^{- 2 \eps} \!
\int_0^1 \!\!\! d y \!
\nnb\\
& & 
\quad
\times\, 
\int_0^1 \!\!\! d z \, 
\Big[ y'(1-y')^2\,z'(1-z')\,y(1-y)^3\,z(1-z) \Big]^{- \eps} 
(1-y')(1-y)^2 
\, .
\eeq

%%%%%%%%%%%%

\subsubsection{Mapping involving six momenta}
\label{app:map2 6}

%%%%%%%%%%%%

\noindent
Similarly to the mapping with five momenta, we first select two momenta 
$k_{\cC},k_{\dD}$, and build the $(n+1)$-tuple of massless momenta 
$\kkl{\aA\cC\dD}$ 
\beq
\kkl{\aA\cC\dD} 
& = & 
\left\{ \{ k \}_{\slashed \aA \slashed \cC \slashed \dD}, \, 
\kk{\cC}{\aA\cC\dD},\kk{\dD}{\aA\cC\dD} \right\} 
\, .
\eeq
Then in $\kkl{\aA\cC\dD}$  we choose the three momenta 
$\kk{\bB}{\aA\cC\dD}=k_{\bB}$, $\kk{\eE}{\aA\cC\dD}=k_{\eE}$ and 
$\kk{\fF}{\aA\cC\dD}=k_{\fF}$ to construct the on-shell, momentum 
conserving $n$-tuple of massless momenta $\kkl{\aA\cC\dD,\bB\eE\fF}$ 
\beq
\kkl{\aA\cC\dD,\bB\eE\fF}
& = & 
\left\{ 
\{ k \}_{\slashed \aA \slashed \bB \slashed \cC \slashed \dD \slashed \eE \slashed \fF}, \, 
\kk{\cC}{\aA\cC\dD,\bB\eE\fF},
\kk{\dD}{\aA\cC\dD,\bB\eE\fF},
\kk{\eE}{\aA\cC\dD,\bB\eE\fF},
\kk{\fF}{\aA\cC\dD,\bB\eE\fF}
\right\} \, ,
\eeq
with
\beq
&&
\kk{\cC}{\aA\cC\dD,\bB\eE\fF} 
\, = \,  
\kk{\cC}{\aA\cC\dD} 
\, = \, 
k_{\aA} + k_{\cC} - \frac{s_{\aA\cC}}{s_{\aA\dD} + s_{\cC\dD}} \, k_{\dD} 
\, , 
\qquad\quad
\kk{\dD}{\aA\cC\dD,\bB\eE\fF} 
\, = \, 
\kk{\dD}{\aA\cC\dD} 
\, = \, 
\frac{s_{\aA\cC\dD}}{s_{\aA\dD} + s_{\cC\dD}} \, k_{\dD} 
\, ,
\qquad  \quad 
\nnb \\[0.2cm]
&&
\kk{\eE}{\aA\cC\dD,\bB\eE\fF} 
\, = \, 
\kk{\eE}{\bB\eE\fF} 
\, = \, 
k_{\bB} + k_{\eE} - \frac{s_{\bB\eE}}{s_{\bB\fF} + s_{\eE\fF}} \, k_{\fF} 
\, ,
\qquad\quad
\kk{\fF}{\aA\cC\dD,\bB\eE\fF} 
\, = \,  
\kk{\fF}{\bB\eE\fF} 
\, = \,  
\frac{s_{\bB\eE\fF}}{s_{\bB\fF} + s_{\eE\fF}} \, k_{\fF}
\, ,
\eeq
while all other momenta are left unchanged 
($\kk{n}{\aA\cC\dD,\bB\eE\fF} = k_n \, , \, n \neq \aA,\bB,\cC,\dD,\eE,\fF$).
Introducing Catani-Seymour parameters 
\beq
y' \, = \, \frac{s_{\aA\cC}}{s_{\aA\cC\dD}} \, , 
\qquad
z' \, = \, \frac{s_{\aA\dD}}{s_{\aA\dD} + s_{\cC\dD}} \, , 
\qquad 
y \, = \, 
\frac{s_{\bB\eE}}{s_{\bB\eE\fF}}  \, , 
\qquad
z \, = \, 
\frac{s_{\bB\fF}}{s_{\bB\fF}+s_{\eE\fF}} 
\, , 
\eeq
we write the six relevant invariants in terms of 
$\sk{\cC\dD}{\aA\cC\dD,\bB\eE\fF}=\sk{\cC\dD}{\aA\cC\dD}=s_{\aA\cC\dD}$, 
and 
$\sk{\eE\fF}{\aA\cC\dD,\bB\eE\fF}=\sk{\eE\fF}{\bB\eE\fF}=s_{\bB\eE\fF}$, as
\beq
s_{\aA\cC} & = & y' \, \sk{\cC\dD}{\aA\cC\dD,\bB\eE\fF}
\, , 
\qquad
s_{\aA\dD} \; = \; z' \, ( 1 - y' ) \sk{\cC\dD}{\aA\cC\dD,\bB\eE\fF}
\, , 
\qquad
s_{\cC\dD}  \; = \; ( 1 - z' ) ( 1  - y' ) \sk{\cC\dD}{\aA\cC\dD,\bB\eE\fF}
\, ,
\nnb \\[2mm]
s_{\bB\eE} & = & y \, \sk{\eE\fF}{\aA\cC\dD,\bB\eE\fF}
\, , 
\qquad\;
s_{\bB\fF} \; = \;  z \, ( 1 - y ) \sk{\eE\fF}{\aA\cC\dD,\bB\eE\fF}
\, , 
\qquad\;\;
s_{\eE\fF}  \; = \; ( 1  - z ) ( 1 - y ) \sk{\eE\fF}{\aA\cC\dD,\bB\eE\fF}
\, .
\qquad
\eeq
In this case, the double-radiative phase space is exactly the product of two 
factorised single-radiative phase spaces. Indeed
\beq
d \Phi_{\npt}(\{k\})
& = & 
d \Phi_n(\kkl{\aA\cC\dD,\bB\eE\fF}) \, d \Phi_{\rm rad, 2}^{\,(\aA\cC\dD,\bB\eE\fF)} 
\, ,
\eeq
and
\beq
&&
\hspace{-8mm}
d \Phi_{\rm rad, 2}^{\,(\aA\cC\dD,\bB\eE\fF)}
\, = \,
d \Phi_{\rm rad}(\sk{\eE\fF}{\aA\cC\dD,\bB\eE\fF};y,z,\phi) \,
d \Phi_{\rm rad}(\sk{\cC\dD}{\aA\cC\dD};y',z',\phi') 
\, ,
\nnb \\[2mm] 
&&
\hspace{-8mm}
\int d \Phi_{\rm rad, 2}^{\,(\aA\cC\dD,\bB\eE\fF)} 
\, = \, 
N^2 (\eps) \, 
\left(
\sk{\cC\dD}{\aA\cC\dD,\bB\eE\fF} \, \sk{\eE\fF}{\aA\cC\dD,\bB\eE\fF}
\right)^{\!1 - \eps}
\!\!
\int_0^\pi \!\!\! d \phi' \, (\sin\phi')^{- 2 \eps} \!
\int_0^1 \!\!\! d y' \!
\int_0^1 \!\!\! d z' \!
\int_0^\pi \!\!\! d \phi \, (\sin\phi)^{- 2 \eps} \!
\int_0^1 \!\!\! d y \!
\nnb\\
& & 
\hspace{20mm}
\times\, 
\int_0^1 \!\!\! d z \, 
\Big[ y'(1-y')^2\,z'(1-z')\,y(1-y)^2\,z(1-z) \Big]^{- \eps} 
(1-y')(1-y) 
\, .
\eeq

%%%%%%%%%%%%%%%%%%%%%%%%%%%%%%%%%%%%%%%%%%%%%

\section{Azimuthal integrals}
\label{app:Iw}

In this appendix we show the details of the integration of structures (combinations of 
Lorentz invariants) that feature a non-trivial dependence on the azimuthal variable $w'$.
Such structures appear both in the integration of one-loop kernels over a single-radiation 
phase-space (see \secn{olokore}), and in the integration of tree-level kernels over the 
double-unresolved radiation phase-space (see \secn{treetworeal}). Three master 
integrals are presented, ordered with increasing complexity. In particular the basic 
integral over the azimuthal variable is presented in \secn{app:Iw1}. One and two 
further integrations of the result give rise respectively to the master integrals considered 
in \secn{app:Iab2} and in \secn{app:Iab3}.

%%%%%%%%%%%%%%%%%%%%%%%%%

\subsection{The master integral $I_{a,b}(A,B)$} 
\label{app:Iw1}

%%%%%%%%%%%%%%%%%%%%%%%%%

\noindent 
The basic building block for azimuthal integrals is the
master integral $I_{a,b}(A,B)$, that is defined as
\beq
I_{a,b}(A,B)
&\equiv& 
\int_0^1 \!\!\! d w' \, 
\frac{\left[ w'(1-w')\right]^{\frac12-b}}{{\left[A^2+B^2+2(1-2w')AB\right]}^{a}}
\, ,
\eeq
with $A,B \in \mathbb{R}$ and, in the cases we are interested in, $A,B \ge 0$. 
Notice that $I_{a,b}(A,B)$ is manifestly symmetric under the exchange 
$A \leftrightarrow B$. Defining
\beq
\eta = \frac{4AB}{(A+B)^2},
\eeq
we have
\beq
I_{a,b}(A,B)
& = & 
\frac{1}{(A+B)^{2a}}\,
\int_0^1\!\!dw'\,\frac{\left[w'(1-w')\right]^{\frac12-b}}{\left(1-\eta w'\right)^{a} }
\nnb\\
&=& 
\frac{1}{(A+B)^{2a}}\,
\frac{\Gamma^2(3/2-b)}{\Gamma(3-2b)}\,\,
_2F_1\big(a,3/2-b,3-2b,\eta\big) \, .
\eeq
The hypergeometric functions of this kind satisfy
\beq
_2F_1(\alpha,\beta,2\beta,x)
& = &
\left(\frac{1+\sqrt{1-x}}{2}\right)^{\!\!-2\alpha} \!\!
_2F_1\!\left( 
\alpha , \alpha - \beta + \frac12 , \beta + \frac12 ,
\left(\frac{1-\sqrt{1-x}}{1+\sqrt{1-x}}\right)^2 
\right)\,,
\eeq
so that one can rewrite the master integral as
\beq
I_{a,b}(A,B)
&=& 
\left[ \frac{(1+\sqrt{\rho})^2}{(A+B)^2} \right]^{a}\,
\frac{\Gamma^2(3/2-b)}{\Gamma(3-2b)}\,
_2F_1(a,a+b-1,2-b,\rho)\,,
\eeq
where
\beq
\rho \, \equiv \,  
\left(\frac{1-\sqrt{1-\eta}}{1+\sqrt{1-\eta}}\right)^2 
\, = \,   
\left\{
\begin{array}{cc}
\displaystyle{
\frac{A^2}{B^2} 
}
& 
\mbox{if}\quad A^2 \le B^2 
\\[4mm]
\displaystyle{
\frac{B^2}{A^2} 
}
& 
\mbox{if}\quad A^2 \ge B^2 
\end{array}
\right. \,, 
\quad
\frac{(1+\sqrt{\rho})^2}{(A+B)^2}
\, = \,  
\left\{
\begin{array}{cc}
\displaystyle{
\frac{1}{B^2}
}
& 
\mbox{if}\quad A^2 \le B^2 
\\[4mm]
\displaystyle{
\frac{1}{A^2}
}
& 
\mbox{if}\quad A^2 \ge B^2 
\end{array}
\right. \,,
\eeq
and we used 
\beq
\frac{2}{1+\sqrt{1-\eta}} = 
1 + \frac{1-\sqrt{1-\eta}}{1+\sqrt{1-\eta}} =
1 + \sqrt{\rho}\,. 
\eeq
The final result reads
\beq
I_{a,b}(A,B)
& = & 
\frac{\Gamma^2(3/2-b)}{\Gamma(3-2b)}\,
\Bigg[
(B^{2})^{-a}\,{}_2F_1\bigg(a,a+b-1,2-b,\frac{A^2}{B^2}\bigg) \,
\Theta \big(B^2\!-\!A^2\big)
\nnb\\
&&
\qquad\qquad\qquad
+ \,
(A^{2})^{-a}\,{}_2F_1\bigg(a,a+b-1,2-b,\frac{B^2}{A^2}\bigg) \,
\Theta \big(A^2\!-\!B^2 \big)
\Bigg] \, .
\eeq
For the specific case where $a=1$ we find
\beq
I_{b}(A,B)
& \equiv & 
I_{1,b}(A,B)
\, =\, 
\int_0^1 \!\!\! d w' \, 
\frac{\left[ w'(1-w')\right]^{\frac12-b}}{A^2+B^2+2(1-2w')AB}
\\
&=& 
\frac{\Gamma^2(3/2-b)}{\Gamma(3-2b)}\,
\bigg[\,
\frac{1}{B^2}\,{}_2F_1\bigg(1,b,2-b,\frac{A^2}{B^2}\bigg)
\Theta(B^2\!-\!A^2)\nnb\\
&&\hspace*{1.85cm}+\frac{1}{A^2}\,{}_2F_1\bigg(1,b,2-b,\frac{B^2}{A^2}\bigg)
\Theta(A^2\!-\!B^2) \,
\bigg] \, . \nnb
\label{eq:I1bAB}
\eeq

%%%%%%%%%%%%%%%%%%%%%%%%%

\subsection{The master integral $I_{a,b,\beta,\gamma}(C,D)$} 
\label{app:Iab2}

%%%%%%%%%%%%%%%%%%%%%%%%%

\noindent 
The master integral $I_{a,b,\beta,\gamma}(C,D)$ is defined by
\beq
\label{eq:IabbgCDin}
I_{a,b,\beta,\gamma}(C,D)
&\equiv& 
\int_0^1 \! d v \, 
\int_0^1 \! d w' \, 
\frac{v^{\beta}(1-v)^{\gamma}[w'(1-w')]^{\frac12-b}}
     {\Big[ C\,v+D(1-v)+2(1-2w')\sqrt{CDv(1-v)} \Big]^a}
\, ,
\eeq
with $C,D \in \mathbb{R}$ and, in the cases we are interested in,  $C,D \ge 0$. 
Notice that $I_{a,b,\beta,\gamma}(C,D)$ is symmetric under the simultaneous 
exchange $C \leftrightarrow D$, $\beta \leftrightarrow \gamma$,
\beq
I_{a,b,\gamma,\beta}(D,C)
& = &
I_{a,b,\beta,\gamma}(C,D)
\, ,
\eeq
and the $w'$ integration has the structure of the master integral of \appn{app:Iw1}, 
with $A^2 = Cv$ and $B^2 = D(1-v)$. Thus one may write
\beq
I_{a,b,\beta,\gamma}(C,D)
& = & 
\int_0^1 \! d v \, 
v^{\beta} (1\!-\!v)^{\gamma} \,
I_{a,b} \Big(\sqrt{Cv},\sqrt{D(1-v)} \Big)
\\
& = &
\frac{\Gamma^2(3/2-b)}{\Gamma(3-2b)}\,
\int_0^1 \! d v \, 
v^{\beta}(1\!-\!v)^{\gamma}
\nnb\\
&&
\times \, \Bigg[
\big( D (1\!-\!v) \big)^{-a} \, 
{}_2F_1\bigg(a,a+b-1,2-b,\frac{Cv}{D(1\!-\!v)}\bigg) \, 
\Theta \bigg(1-\frac{Cv}{D(1\!-\!v)} \bigg)
\nnb \\
&&\quad
+ \,
\big( C v \big)^{-a} \,
{}_2F_1 \bigg(a,a+b-1,2-b,\frac{D(1\!-\!v)}{Cv}\bigg) \, 
\Theta \bigg(\frac{Cv}{D(1\!-\!v)}-1\bigg) \,
\Bigg] \, .
\label{eq:inAppC}
\eeq
The step functions appearing in \eq{eq:inAppC} modify the $v$ integration 
domain as
\beq
1-\frac{Cv}{D(1\!-\!v)} \, \gtrless \, 0
\quad \longleftrightarrow \quad
v \, \lessgtr \, \frac{D}{C+D} \, .
\eeq
Since $C,D>0$, then $0<\frac{D}{C+D}<1$ and we get
\beq
\label{mast4oint}
I_{a,b,\beta,\gamma}(C,D) 
& = & 
\frac{\Gamma^2(3/2\!-\!b)}{\Gamma(3-2b)}\,
\Bigg[\;
\frac{1}{D^a} \!
\int_0^{\frac{D}{C\!+\!D}} \!\!\! d v \, 
v^{\beta}(1\!-\!v)^{\gamma-a}\,
{}_2F_1\bigg(a,a+b-1,2-b,\frac{Cv}{D(1\!-\!v)}\bigg)
\nnb\\
&&
\hspace{18mm}
+ \,
\frac{1}{C^a}  \!
\int_{\frac{D}{C\!+\!D}}^1 \!\!\! d v \, 
v^{\beta-a}(1\!-\!v)^{\gamma}\,
{}_2F_1\bigg(a,a+b-1,2-b,\frac{D(1\!-\!v)}{Cv}\bigg)
\Bigg] .
\qquad\quad
\eeq
Next, we restore the integration region to the unit interval $[0,1]$, 
with the following changes of variables: 
\beq
  v \, \to\, \frac{\frac{D}{C}\,v}{1\!+\!\frac{D}{C}\,v} \quad
  {\textrm{(first integral in~\ref{mast4oint})}}\, ,
  \qquad
  v \,\to\, \frac{1}{1\!+\!\frac{C}{D}\,v} \quad 
  {\textrm{(second integral in~\ref{mast4oint})}} \, .  
\eeq
The master integral becomes
\beq
&& \hspace*{-1cm}I_{a,b,\beta,\gamma}(C,D)
= 
\frac{\Gamma^2(3/2\!-\!b)}{\Gamma(3-2b)}\,
\Bigg[
\;
\frac{D^{1+\beta-a}}{C^{1+\beta}} \!
\int_0^1 \!\! d v \, 
v^{\beta} \!
\left(\! 1\!+\!\frac{D}{C}v \!\right)^{\!\!a-\beta-\gamma-2} \!\!
{}_2F_1 \big(a,a+b-1,2-b,v\big)
\nnb\\
&&
\hspace{33mm}
+ \,\,
\frac{C^{1+\gamma-a}}{D^{1+\gamma}} \!
\int_0^1 \!\! d v \, 
v^{\gamma} \!
\left(\! 1\!+\!\frac{C}{D}v \!\right)^{\!\!a-\beta-\gamma-2} \!\!
{}_2F_1 \big(a,a+b-1,2-b,v \big)
\Bigg] 
.
\quad
\label{eq:C7}
\eeq	
In the integration of the two-unresolved tree-level kernels the integration over 
the azimuthal angle gives rise to the master integral $I_{a,b,\beta,\gamma}(C,D)$ 
with $a=1$ (see \secn{subsec:yxp}), which deserves a separate analysis.
We define
\beq
\label{eq:appIv}
\hspace*{-0.7cm}I_{b,\beta,\gamma}(C,D) \, \equiv \,  I_{1,b,\beta,\gamma}(C,D)
& = & 
\int_0^1 \!\!\! d v \, 
\int_0^1 \!\!\! d w' \, 
\frac{v^{\beta}(1-v)^{\gamma}[w'(1-w')]^{\frac12-b}}
     {C\,v+D(1-v)+2(1-2w')\sqrt{CDv(1-v)}}
\, ,
\eeq
with $C,D \in \mathbb{R}$ and $C,D \ge 0$. The $w'$ integration can be 
performed using \eq{eq:I1bAB}, with $A^2 = Cv, B^2 = D(1-v)$, with the result
\beq
I_{b,\beta,\gamma}(C,D)
& = & 
\int_0^1 \!\!\! d v \, 
v^{\beta}(1\!-\!v)^{\gamma} \,
I_{b}\Big(\sqrt{Cv},\sqrt{D(1-v)}\Big) \, .
\eeq
The hypergeometric functions with the first argument set to unity satisfy
\beq\label{eq:hypprop1}
{}_2F_1 \big(1,b,c,x\big) 
&=& 
-\,\frac{c-1}{b-1}\,\frac1x \,\, {}_2F_1 \bigg(1,2-c,2-b,\frac1x \bigg)
+
\frac{\Gamma(c)\Gamma(1-b)}{\Gamma(c-b)} 
\left(-\frac1x\right)^{\!b} \!
\left(1-\frac1x\right)^{\!c-b-1}\!
,
\qquad
\label{eq:invF}
\eeq
which leads to
\beq
I_{b,\beta,\gamma}(C,D)
& = &
\frac{1}{C}\,
\frac{\Gamma^2(3/2-b)}{\Gamma(3-2b)}\,
\int_0^1 \! d v \, 
\Bigg\{\;
v^{\beta-1}(1\!-\!v)^{\gamma} \, 
{}_2F_1\bigg(1,b,2-b,\frac{1\!-\!v}{\alpha\,v}\bigg)
\nnb\\
&&
\hspace{35mm}
- \,
( -\alpha)^{b}\,
\frac{\Gamma(2-b)\Gamma(1-b)}{\Gamma(2-2b)}\,
v^{\beta+b-1}(1\!-\!v)^{\gamma+b-1}
\, 
\nnb\\
&&
\hspace{45mm}
\times
\Big[ 1 - v(1\!+\!\alpha) \Big]^{1-2b}
\Theta\!\left(1-\frac{\alpha\,v}{1\!-\!v}\right) \!
\Bigg\} \, ,
\eeq
where we have defined $\alpha=C/D$. Upon making the substitution 
$v\to {v}/({1\!+\!\alpha})$ in the second term, it can be integrated, giving 
yet another hypergeometric function. Thus
\beq
I_{b,\beta,\gamma}(C,D)
& = & 
\frac{1}{C}\,
\frac{\Gamma^2(3/2-b)}{\Gamma(3-2b)}\,
\bigg[\;
\int_0^1 \! d v \, 
v^{\beta-1} (1\!-\!v)^{\gamma} \, 
{}_2F_1\bigg(1,b,2-b,\frac{1\!-\!v}{\alpha\,v} \bigg)
\nnb\\
&&
\hspace{24mm}
- \,
\frac{( -\alpha)^{b}}{(1\!+\!\alpha)^{\beta+b}} \, 
\frac{\Gamma(2-b)\Gamma(1-b)\Gamma(\beta+b)}{\Gamma(\beta-b+2)}
\nnb\\
&&
\hspace{35mm}
\times
{}_2F_1 \bigg(1-\gamma-b,\beta+b,\beta-b+2,\frac{1}{1\!+\!\alpha}\bigg)
\;
\bigg] 
\, .
\eeq
Though the integral $I_{b,\beta,\gamma}(C,D)$ is well defined for real 
positive $C$ and $D$, in order to properly keep track of the imaginary 
parts we give a small imaginary part to $\alpha$, according  to
\beq
\alpha \to \alpha \pm {\rm i} \delta,
\qquad
(- \alpha)^s \, \to \, \big( \! - \alpha \mp {\rm i} \delta \big)^s 
\, = \, 
\alpha^s \,e^{\mp{\rm i} s \pi}
\, ,
\qquad
\delta \to 0^+
\, .
\eeq
Then we can write the first hypergeometric function using its integral 
representation, as 
\beq
{}_2F_1\bigg(1,b,2-b,\frac{1\!-\!v}{\alpha\,v}\bigg)
& = & 
- \,
\alpha\,v \,
\frac{\Gamma(2-b)}{\Gamma(b)\Gamma(2-2b)}\,
\int_0^1 \!\!\! d t \, 
t^{b-2}(1\!-\!t)^{1-2b}
\left[1-\frac{t+\alpha}{t}\,v\right]^{-1} \, ,
\eeq
and integrate in $v$, with the result
\beq
I_{b,\beta,\gamma}(C,D)&=&
\frac{1}{C}\,
\frac{\Gamma^2(3/2-b)}{\Gamma(3-2b)}\,\\
&&
\times
\Bigg[
- 
\frac{\alpha\,\Gamma(2\!-\!b)}{\Gamma(b)\Gamma(2\!-\!2b)}
\frac{\Gamma(\beta\!+\!1)\Gamma(\gamma\!+\!1)}{\Gamma(\beta\!+\!\gamma\!+\!2)}
\!\!
\int_0^1 \!\!\! d t \, 
t^{b-2}(1\!-\!t)^{1-2b}
{}_2F_1\bigg(1,\beta\!+\!1,\beta\!+\!\gamma\!+\!2,\frac{t\!+\!\alpha}{t}\bigg)
\nnb\\
&&
\hspace{6mm}
- \,
\frac{\alpha^{b} \, e^{\mp{\rm i} b \pi}} {(1\!+\!\alpha)^{\beta+b}}
\frac{\Gamma(2\!-\!b)\Gamma(1\!-\!b)\Gamma(\beta\!+\!b)}{\Gamma(\beta\!-\!b\!+\!2)}
\,
{}_2F_1\bigg(
1\!-\!\gamma\!-\!b,\beta\!+\!b,\beta\!-\!b\!+\!2,\frac{1}{1\!+\!\alpha}
\bigg)
\,
\Bigg] 
\, .
\nnb
\eeq
Using simple hypergeometric identities (similar to \eq{eq:hypprop1}), we obtain 
then the expression 
\beq
\label{eq:tempC16}
I_{b,\beta,\gamma}(C,D)
&=&
\frac{1}{C}
\frac{\Gamma^2(3/2\!-\!b)}{\Gamma(3\!-\!2b)}\,
\Bigg\{\,
\alpha \,
\frac{\Gamma(2-b)}{\Gamma(b)\Gamma(2-2b)}\,
\frac{\Gamma(\beta+1)\Gamma(\gamma+1)}{\Gamma(\beta+\gamma+2)}\,
\int_0^1 \! \! d t \, 
t^{b-2}(1\!-\!t)^{1-2b}
\\
&&
\hspace{35mm}
\times 
\Bigg[\,
\frac{t}{\alpha}\,
\frac{\beta+\gamma+1}{\beta}
{}_2F_1\bigg(1,\gamma+1,1-\beta,-\frac{t}{\alpha}\bigg)
\nnb\\
&&
\hspace{41mm}
- \,
\frac{\Gamma(\beta+\gamma+2)\Gamma(-\beta)}{\Gamma(\gamma+1)} 
\left(-\frac{\alpha}{t}\right)^{\!-\beta-1} \!
\left(1+\frac{t}{\alpha}\right)^{\!-\beta-\gamma-1}
\;
\Bigg]
\nnb\\
&&
\hspace{24mm}
- \,
\alpha^{-\beta} e^{\mp{\rm i} b \pi}\,
\frac{\Gamma(2-b)\Gamma(1-b)\Gamma(\beta+b)}{\Gamma(\beta-b+2)} 
\nnb\\
&&
\hspace{35mm}
\times
{}_2F_1\bigg(\beta+\gamma+1,\beta+b,\beta-b+2,-\frac{1}{\alpha}\bigg)
\;
\Bigg\}
.
\nnb
\eeq
The second term of the integral over $t$ in \eq{eq:tempC16} can be now 
integrated, giving the same hypergeometric function that appears in the 
last line. Recalling now that
\beq\label{gammaimaginary}
\Gamma(z)\,\Gamma(1-z) \, = \, \frac{\pi}{\sin(\pi z)}, 
\qquad
e^{\mp{\rm i} z \pi}\,\Gamma(z)\,\Gamma(1-z) 
\, = \, 
\frac{\pi\cos(\pi z)}{\sin(\pi z)} \mp i\,\pi
\, , 
\eeq
using straightforward trigonometric identities, and inserting back $\alpha=C/D$, 
we obtain
\beq
I_{b,\beta,\gamma}(C,D)
& = & 
\frac{1}{C}\,
\frac{\Gamma^2(3/2-b)\Gamma(2-b)}{\Gamma(3-2b)\Gamma(b)}\,\\
&&
\hspace*{-1.8cm}\times\,\Bigg\{
\frac{\Gamma(\beta)\Gamma(\gamma+1)}
     {\Gamma(2-2b)\Gamma(\beta+\gamma+1)}\,
\int_0^1 \!\!\! d t \, 
t^{b-1}(1\!-\!t)^{1-2b}
{}_2F_1\bigg(1,\gamma+1,1-\beta,-\frac{D}{C}\,t\bigg)
\nnb\\
&&
\hspace*{-1.1cm}-
\left(\frac{C}{D}\right)^{\!-\beta}\,
\frac{\Gamma(\beta+b)}{\Gamma(\beta-b+2)}\,
\frac{\pi\,\sin(\pi(\beta+b+1))}{\sin(\pi(\beta+1))\sin(\pi b)}\,
{}_2F_1\bigg(\beta+\gamma+1,\beta+b,\beta-b+2,-\frac{D}{C}\bigg)
\Bigg\} \, .\nnb
\label{eq:Iv 1}
\eeq
We notice that the imaginary part of \eq{gammaimaginary} drops
out of the latter expression, as it does not depend on $z$.
In the special case where $\beta=1-b$, the second hypergeometric in \eq{eq:Iv 1} 
does not contribute, since its prefactor vanishes. We then obtain
\beq
I_{b,1-b,\gamma}(C,D)
& = & 
\frac{1}{C}\,
\frac{\Gamma^2(3/2-b)\Gamma(2-b)}{\Gamma(3-2b)\Gamma(b)}\,
\frac{\Gamma(1-b)\Gamma(\gamma+1)}
     {\Gamma(2-2b)\Gamma(\gamma-b+2)}\,\nnb\\
     &&
\times \int_0^1 \!\!\! d t \, 
t^{b-1}(1\!-\!t)^{1-2b}
{}_2F_1\bigg(1,\gamma+1,b,-\frac{D}{C}\,t\bigg) \, . \qquad
\eeq
In this case, the integral yields again a simple hypergeometric function, so that
we get the compact result
\beq
I_{b,1-b,\gamma}(C,D)
& = & 
\frac{1}{C}\,
\frac{\Gamma^2(3/2-b)}{\Gamma(3-2b)}\,
\frac{\Gamma(1-b)\Gamma(\gamma+1)}{\Gamma(\gamma-b+2)}\,
{}_2F_1\bigg(1,\gamma+1,2-b,-\frac{D}{C}\bigg)
\, .
\label{eq:Iv s1}
\eeq
Using the symmetry of the original master integral under the simultaneous 
exchange $C \leftrightarrow D$, $\beta \leftrightarrow \gamma$, we similarly 
get the result
\beq
I_{b,\beta,1-b}(C,D)
& = & 
\frac{1}{D}\,
\frac{\Gamma^2(3/2-b)}{\Gamma(3-2b)}\,
\frac{\Gamma(1-b)\Gamma(\beta+1)}{\Gamma(\beta-b+2)}\,
{}_2F_1\bigg(1,\beta+1,2-b,-\frac{C}{D}\bigg)
\, .
\label{eq:Iv s2}
\eeq

%%%%%%%%%%%%%%%%%%%%%%%%%

\subsection{The master integral $I_{a,b,\beta,\gamma,\delta,\sigma}(P,Q)$}
\label{app:Iab3}

%%%%%%%%%%%%%%%%%%%%%%%%%

\noindent
In the integration of the colour-tripole contributions to the one-loop
single-soft kernel (see \secn{softwolo}), the integral of
\eq{eq:IabbgCDin} needs to be integrated over one further variable.
We then define the master integral $I_{a,b,\beta,\gamma,\delta,\sigma}(P,Q)$
as follows,
\beq
\label{eq:masterallint}
I_{a,b,\beta,\gamma,\delta,\sigma}(P,Q)
&\equiv& 
\int_0^1 \!\!\! d u 
\int_0^1 \!\!\! d v 
\int_0^1 \!\!\! d w' \,
\frac{u^{\delta}(1-u)^{\sigma}v^{\beta}(1-v)^{\gamma} \big[w'(1-w')\big]^{\frac12-b}}
     {\Big[P\,v+Q\,u(1-v)+2(1-2w')\sqrt{P\,Q\,uv(1-v)}\Big]^a}
 \nnb\\
 & = & 
\int_0^1 \!\!\! d u \, 
u^{\delta}(1-u)^{\sigma}
I_{a,b,\beta,\gamma}\Big(P,Q\,u\Big)
\, .
\eeq
According to the result in \eq{eq:C7} we can write
\beq
I_{a,b,\beta,\gamma,\delta,\sigma}(P,Q)
& = & 
\frac{\Gamma^2(3/2\!-\!b)}{\Gamma(3-2b)}\,
\frac{Q^{1+\beta-a}}{P^{1+\beta}}
\int_0^1 \!\!\! d u 
\int_0^1 \!\!\! d v \,
u^{\beta+\delta-a+1}(1-u)^{\sigma} \, 
{}_2F_1 \big(a,a+b-1,2-b,v\big)\nnb
\\
&&
\hspace{20mm}
\times\,
\Bigg[ \, 
v^{\beta} \!
\left(\! 1 \!+\! \frac{Q}{P}\,u\,v \!\right)^{\!\!a-\beta-\gamma-2} 
\!\!\!\!
+ 
v^{a-\beta-2} \!
\left(\! 1 \!+\! \frac{Q}{P}\,\frac{u}{v} \!\right)^{\!\!a-\beta-\gamma-2}
\Bigg] 
\, .
\eeq
The integration over $u$ gives another hypergeometric function,
\beq
I_{a,b,\beta,\gamma,\delta,\sigma}(P,Q)
& = & 
\frac{\Gamma^2(3/2\!-\!b)}{\Gamma(3-2b)}\,
\frac{\Gamma(\beta+\delta-a+2)\Gamma(\sigma+1)}
     {\Gamma(\beta+\delta+\sigma-a+3)}\,
\frac{Q^{1+\beta-a}}{P^{1+\beta}}
\int_0^1 \!\! d v \,\,
{}_2F_1 \big(a,a+b-1,2-b,v\big)
\nnb\\
&&
\times
\bigg[ \; 
v^{\beta} 
{}_2F_1\!\left(
\beta\!+\!\gamma\!-\!a\!+\!2,
\beta\!+\!\delta\!-\!a\!+\!2,
\beta\!+\!\delta\!+\!\sigma\!-\!a\!+\!3,
-\frac{Q}{P}\,v 
\right) 
\nnb\\
&&
\hspace{4mm}
+ \,
v^{a-\beta-2}
{}_2F_1\!\left( 
\beta\!+\!\gamma\!-\!a\!+\!2,
\beta\!+\!\delta\!-\!a\!+\!2,
\beta\!+\!\delta\!+\!\sigma\!-\!a\!+\!3,
-\frac{Q}{P}\frac1v 
\right) 
\bigg] 
\, .
\eeq
The expansion of these hypergeometric functions in powers of $\epsilon$ is 
simpler if the integer part of the first index is $0$. Since this quantity is positive
for the cases of interest, we can lower the first index (taking care that in the 
generated hypergeometric functions $b > 0$ and $c-b > 0$) using the identities
\beq
{}_2F_1(a,b,c,x)
& = &
-\,
\frac{c-1}{a-1} \,
\frac1x \,
\Big[\,
{}_2F_1(a-1,b-1,c-1,x)
-
{}_2F_1(a-1,b,c-1,x)
\,\Big]
\, ,
\\
{}_2F_1(a,b,c,x)
& = &
\frac{b}{a-1} \,
{}_2F_1(a-1,b+1,c,x)
+
\frac{a-b-1}{a-1} \,
{}_2F_1(a-1,b,c,x)
\, ,
\nnb\\
{}_2F_1(a,b,c,x)
& = &
\frac1{1-x} \,
\left[\,
\frac{c-b}{a-1} \,
{}_2F_1(a-1,b-1,c,x)
+
\frac{a-c+b-1}{a-1} \,
{}_2F_1(a-1,b,c,x)
\,\right]
\, .
\nnb
\eeq
Once the integer part of the first index is $0$, we can then expand in powers of
$\eps$ using
\beq
{}_2F_1(\alpha\eps,b,c,x)
& = &
1
+
\frac{\Gamma(c)}{\Gamma(b)\Gamma(c-b)}
\sum_{n=1}^{+\infty}
\frac{(-\alpha\eps)^n}{n!}
\int_0^1\!\!dt \,
t^{b-1}(1-t)^{c-b-1}
\ln^n(1-tx)\,,
\eeq
and then easily perform the remaining integrations. 

%%%%%%%%%%%%%%%%%%%%%%%%%%%%%%%%%%%%%%%%%%%%%

%%%%%%%%%%%%%%%%%%%%%%%%%%%%%%%%%%%%%%%%%%%%%

\bibliographystyle{JHEP}
\bibliography{integration}

%%%%%%%%%%%%%%%%%%%%%%%%%%%%%%%%%%%%%%%%%%%%%

\end{document}